\documentclass{lmcs}

\usepackage[utf8]{inputenc}
\usepackage{color}

\usepackage[T1]{fontenc}
\usepackage{lmodern}

\usepackage{xspace} 

\DeclareMathSymbol{:}{\mathpunct}{operators}{"3A}

\definecolor{BlueViolet}{rgb}{0, 0, 0.55}
\definecolor{RubineRed}{rgb}{0.88, 0.07, 0.37}
\definecolor{ForestGreen}{rgb}{0.13, 0.55, 0.13}
\definecolor{Blue}{rgb}{0.0, 0.0, 1.0}
\definecolor{NavyBlue}{rgb}{0.0, 0.0, 0.5}
\definecolor{Black}{rgb}{0.02, 0.02, 0.02}
\definecolor{MidnightBlue}{rgb}{0.0, 0.2, 0.4}
\definecolor{Gray}{rgb}{0.41, 0.41, 0.41}
\definecolor{TealBlue}{rgb}{0.212,0.459,0.533}
\definecolor{Plum}{rgb}{0.6,0.25,0.6}

\usepackage{centernot}
\usepackage{hhline}
\usepackage{mdframed}

\usepackage{tensor}

\usepackage{bm} 

\usepackage{mathbbol} 

\usepackage{siunitx}
\usepackage{textcomp}
\usepackage{wasysym} 
\usepackage{caption}
\usepackage{pst-func}
\usepackage{amssymb}
\usepackage{graphicx}

\usepackage{environ}
\usepackage{pgfplots}
\usepackage{tikz}
\usetikzlibrary{shadows}
\usepackage{mathtools}
\usepackage{array,multirow}
\usepackage{url}
\usepackage{cite}

\usepackage{etoolbox}
\usepackage{arydshln}

\usepackage{stmaryrd} 

\usepackage[inline,shortlabels]{enumitem} 
\newlist{inlinelist}{enumerate*}{1}
\setlist*[inlinelist,1]{%
  label=(\roman*),
}

\usepackage{xifthen}  
\usepackage{ifthen}

\usepackage{nicefrac}

\usepackage{hyperref}
\usepackage{cleveref}
\usepackage{bigints}

\usetikzlibrary{pgfplots.dateplot}
\usetikzlibrary{matrix,chains,positioning,decorations.pathreplacing}
\usetikzlibrary{patterns,calc,fit,arrows}
\usetikzlibrary{shapes.geometric}

\usepackage[final,nomargin,inline,index]{fixme} 
\fxusetheme{color}

\FXRegisterAuthor{bart}{anbart}{\color{magenta} {\underline{bart}}}
\FXRegisterAuthor{enrico}{anenrico}{\color{blue} {\underline{enrico}}}
\FXRegisterAuthor{alberto}{analberto}{\color{ForestGreen} {\underline{alberto}}}

\hypersetup{
  breaklinks   = true,
  colorlinks   = true, 
  urlcolor     = blue, 
  linkcolor    = blue, 
  citecolor    = red   
}
\hypersetup{final} 

\newcommand{\ifempty}[3]{%
  \ifthenelse{\isempty{#1}}{#2}{#3}%
}

\newcommand{\ifdots}[3]{%
  \ifthenelse{\equal{#1}{...}}{#2}{#3}%
}

\newcommand{\hidden}[1]{}

\newcommand{\keyterm}[1]{\emph{#1}}%

\newcommand{\mypar}[1]{\paragraph*{#1}}

\crefname{appendix}{appendix}{appendices}
\Crefname{appendix}{Appendix}{Appendices}
\crefname{notation}{notation}{notations}
\Crefname{notation}{Notation}{Notations}

\definecolor{LightGrey}{rgb}{0.95,0.95,0.95}
\definecolor{keyword}{HTML}{7F0055}

\newcommand{\qedex}{\ensuremath{\diamond}}

\renewcommand{\vec}[1]{\boldsymbol{#1}}

\def\etc{etc.\@\xspace}

\newcommand{\eg}{e.g.\@\xspace}
\newcommand{\ie}{i.e.\@\xspace}
\newcommand{\wrt}{w.r.t.\@\xspace}

\newcommand{\eqdef}{\triangleq}

\newcommand{\irule}[2]{\dfrac{#1}{#2}}

\newcommand{\nrule}[1]{{\scriptsize \textsc{#1}}}

\newcommand{\R}{\mathbb{R}}
\newcommand{\RNN}{\mathbb{R}_0^{+}}
\newcommand{\RP}{\mathbb{R}^{+}}

\newcommand{\RNZeroB}{\R \setminus \{0\}}

\newcommand{\setenum}[1]{\{#1\}}

\renewcommand{\epsilon}{\varepsilon}

\newenvironment{thmproof}[2][]{%
  \ifempty{#1}
  {\subsection*{Proof of Theorem~\ref{#2}}}
  {\subsection*{Proof of Theorem~\ref{#2} ({#1})}}
  \label{#2-proof}
  }%
  {}

\newenvironment{lemproof}[2][]{%
  \ifempty{#1}
  {\subsection*{Proof of Lemma~\ref{#2}}}
  {\subsection*{Proof of Lemma~\ref{#2} ({#1})}}
  \label{#2-proof}
  }%
  {}

\newenvironment{corproof}[2][]{%
  \ifempty{#1}
  {\subsection*{Proof of Corollary~\ref{#2}}}
  {\subsection*{Proof of Corollary~\ref{#2} ({#1})}}
  \label{#2-proof}
  }%
  {}

  {}

\newcommand{\BTC}{\textup{%
  \leavevmode
  \vtop{\offinterlineskip 
    \setbox0=\hbox{B}%
    \setbox2=\hbox to\wd0{\hfil\hskip-.03em
    \vrule height .3ex width .15ex\hskip .08em
    \vrule height .3ex width .15ex\hfil}
    \vbox{\copy2\box0}\box2}}\xspace}

\def\pmvColor{\color{ForestGreen}}

\newcommand{\pmvFmt}[1]{{\pmvColor{\sf{#1}}}} 

\newcommand{\PmvU}{\pmvFmt{\mathbb{A}}\xspace} 

\newcommand{\pmv}[2][]{\pmvFmt{#2}_{\pmvColor{#1}}\xspace}

\newcommand{\pmvA}[1][]{\pmv[{#1}]{A}}
\newcommand{\pmvAi}[1][]{\pmv[{#1}]{A'}}
\newcommand{\pmvB}[1][]{\pmv[{#1}]{B}}
\newcommand{\pmvC}[1][]{\pmv[{#1}]{C}}

\def\tokColor{\color{magenta}}
\newcommand{\tokFmt}[1]{{\tokColor{\mathsf{#1}}}}

\newcommand{\tok}[2][]{\tokFmt{#2}_{\tokColor{#1}}\xspace}

\newcommand{\tokT}[1][]{{\tok[#1]{T}}}    
\newcommand{\tokTi}[1][]{\tok[#1]{T'}}
\newcommand{\TokU}{\tokFmt{\mathbb{T}}} 

\newcommand{\mintedT}[1][]{{\tokColor \tensor*{\tokT}{^\CreditTag_{#1}}}}
\newcommand{\debtT}[1][]{{\tokColor \tensor*{\tokT}{^\DebtTag_{#1}}}}

\newcommand{\base}[1]{{\tokColor{#1}}}
\newcommand{\minted}[1]{#1^{{\tokColor c}}}
\newcommand{\debt}[1]{{\tokColor{#1^{d}}}}

\newcommand{\TokUM}{\minted{\TokU}} 
\newcommand{\TokUD}{\debt{\TokU}} 

\def\lpColor{\color{blue}}

\newcommand{\LpL}[1][]{{\lpColor{\Lambda_{#1}}}}
\newcommand{\LpLi}[1][]{{\lpColor{\Lambda'_{#1}}}}

\newcommand{\projRes}[1]{{#1}}

\newcommand{\lp}[2]{[{#1}\ifempty{#2}{}{,{#2}}]}

\newcommand{\supplyWal}[2][]{\mathit{S}_{#1}{\ifempty{#2}{}{(#2)}}} 

\newcommand{\supply}[2][]{\mathit{S}_{#1}{\ifempty{#2}{}{(\minted{#2})}}} 

\newcommand{\supplyDebt}[2][]{\mathit{S}_{#1}{\ifempty{#2}{}{(\debt{#2})}}} 

\newcommand{\reserves}[2][]{{#1}{\ifempty{#2}{}{(\base{#2})}}} 

\newcommand{\liqThreshold}{\mathit{T}_{\liquidateOp}}
\newcommand{\liqReward}{\mathit{R}_{\liquidateOp}}
\newcommand{\rLiq}{\liqReward}

\newcommand{\X}[2][]{\ifempty{#2}{{\mathit{XR}}}{{\mathit{XR}}_{#1}({#2})}}

\newcommand{\CreditTag}{c}
\newcommand{\DebtTag}{d}

\newcommand{\baseval}[2][]{W_{#1}\ifempty{#2}{}{({#2})}}
\newcommand{\walVal}[2][]{\baseval[{#1}]{#2}}

\newcommand{\wealth}[2]{W_{#1}\ifempty{#2}{}{(#2)}}

\newcommand{\creditval}[2][]{W^\CreditTag_{#1}\ifempty{#2}{}{(#2)}}

\newcommand{\debtval}[2][]{W^\DebtTag_{#1}\ifempty{#2}{}{(#2)}}

\newcommand{\Coll}[2]{\mathit{C}_{#1}\ifempty{#2}{}{(#2)}}

\newcommand{\netpos}[2]{\mathit{W}^{c-d}_{#1}\ifempty{#2}{}{(#2)}}
\newcommand{\netCredit}[2]{\netpos{#1}{#2}}

\newcommand{\Health}[2]{\mathit{H}_{#1}\ifempty{#2}{}{(#2)}} 

\newcommand{\Util}[2][]{\mathit{U}_{#1}{\ifempty{#2}{}{({#2})}}} 
\newcommand{\Intr}[1]{\mathit{I}_{#1}} 

\newcommand{\WalW}[1][]{{\pmvColor{\omega_{#1}}}}
\newcommand{\WalWi}[1][]{{\pmvColor{\omega'_{#1}}}}

\newcommand{\statePrice}[2][]{\price_{\tokColor #2}{\ifempty{#1}{}{(#1)}}}
\newcommand{\statePricei}[2][]{\pricei_{\ifempty{#2}{}{\!\!\tokColor #2}}{\ifempty{#1}{}{(#1)}}}

\newcommand{\stateBase}[3][]{\ifempty{#1}{\WalW}{#1}(\base{#2}, #3)}
\newcommand{\stateDebt}[3][]{\ifempty{#1}{\LpL}{#1}(\debt{#2}, #3)}
\newcommand{\stateCredit}[3][]{\ifempty{#1}{\LpL}{#1}(\minted{#2}, #3)}

\newcommand{\stateBasei}[2]{\stateBase[\WalWi]{#1}{#2}}
\newcommand{\stateDebti}[2]{\stateDebt[\LpLi]{#1}{#2}}
\newcommand{\stateCrediti}[2]{\stateCredit[\LpLi]{#1}{#2}}

\newcommand{\wal}[2]{{#1}[{#2}]}
\newcommand{\walA}[2][]{\wal{\pmvA[#1]}{#2}}
\newcommand{\walB}[2][]{\wal{\pmvB[#1]}{#2}}

\newcommand{\tokBal}[1][]{w_{#1}}

\newcommand{\confG}[1][]{\Gamma_{#1}}
\newcommand{\confGi}[1][]{\Gamma'_{#1}}
\newcommand{\confGii}[1][]{\Gamma''_{#1}}

\newcommand{\price}[1][]{{\tokColor\pi}{\ifempty{#1}{}{(#1)}}}
\newcommand{\pricei}[1][]{{\tokColor\pi'}{\ifempty{#1}{}{(#1)}}}

\def\txColor{\color{MidnightBlue}}

\newcommand{\txFmt}[1]{{\txColor{\sf #1}}}

\newcommand{\tx}[2][]{\txFmt{#2}_{\txColor{#1}}}
\newcommand{\txT}[1][]{\tx[#1]{X}}
\newcommand{\txTi}[1][]{\txFmt{X'_{\txColor{{#1}}}}}

\newcommand{\TxTS}[1][]{\vec{\tx[#1]{\mathcal{X}}}} 
\newcommand{\TxTiS}[1][]{\vec{\tx[#1]{\mathcal{X'}}}} 
\newcommand{\TxYS}[1][]{\vec{\tx[#1]{\mathcal{Y}}}} 

\newcommand{\val}[2][]{#2_{#1}} 
\newcommand{\valV}[1][]{\val[#1]{v}}
\newcommand{\valVi}[1][]{\val[#1]{v'}}

\newcommand{\prVal}{p}

\newcommand{\prIncr}{\delta}

\newcommand{\gainSt}[3]{\ensuremath{g_{{#1}}({#2}, {#3})}}
\newcommand{\gain}[3]{\ensuremath{g_{{#1}}({#2},{#3})}}

\newcommand{\accrueIntOp}{{\txColor{\sf int}}} 
\newcommand{\exchUpdateOp}[1]{{\txColor{\sf px}}\ifempty{#1}{}{({#1})}}
\newcommand{\pxOp}[1]{{\txColor{\sf px}}\ifempty{#1}{}{({#1})}}

\newcommand{\projTok}[2][]{\ensuremath{{#2}\!\mid_{\ifempty{#1}{\tokT}{#1}}}}
\newcommand{\projToki}[1]{\projTok[\tokTi]{#1}}

\newcommand{\ltsLabel}[1][]{{\txT_{#1}}}
\newcommand{\ltsLabeli}[1][]{{\txTi_{#1}}}

\newcommand{\depositOp}{{\txColor{\sf dep}}}
\newcommand{\redeemOp}{{\txColor{\sf rdm}}}
\newcommand{\borrowOp}{{\txColor{\sf bor}}} 
\newcommand{\repayOp}{{\txColor{\sf rep}}} 
\newcommand{\liquidateOp}{{\txColor{\sf liq}}} 
\newcommand{\swapOp}{{\txColor{\sf swp}}}

\newcommand{\actDeposit}[2]{\ifempty{#1}{}{{#1}:}\depositOp({#2})}
\newcommand{\actRedeem}[2]{\ifempty{#1}{}{{#1}:}\redeemOp({#2})}
\newcommand{\actBorrow}[2]{\ifempty{#1}{}{{#1}:}\borrowOp({#2})}
\newcommand{\actRepay}[2]{\ifempty{#1}{}{{#1}:}\repayOp({#2})}
\newcommand{\actLiquidate}[4]{\ifempty{#1}{}{{#1}:}\liquidateOp({#2},{#3},{#4})}

\newcommand{\actSwap}[3]{\ifempty{#1}{}{{#1}:}\swapOp({#2},{#3})}
\newcommand{\depRule}{%
\irule
  {
  \begin{array}{c}
    \WalW(\base{\tokT},\pmvA) \geq \valV > 0
    \qquad
    \minted{\valV} = \nicefrac{\valV}{\X[\LpL]{\tokT}}
    \qquad
    \LpLi =
    \LpL 
    + \setenum{\base{\tokT} \mapsto \valV} 
    + \setenum{(\minted{\tokT},\pmvA) \mapsto 
    \minted{\valV}}
    \end{array}
  }
  {
    (\WalW,\LpL,\price) 
    \xrightarrow{\actDeposit{\pmvA}{\valV:\tokT}}
    (\WalW - \setenum{(\tokT,\pmvA) \mapsto \valV},
    \LpLi,
    \price) 
  }
  \;\nrule{[Dep]}
}

\newcommand{\borRule}{%
  \irule
  {
  \begin{array}{c}
    \projRes{\LpL}(\tokT) \geq \valV > 0 
    \qquad
    \LpLi =
    \LpL 
    - \setenum{\tokT \mapsto \valV}
    + \setenum{(\debt{\tokT},\pmvA) \mapsto \valV})
    \qquad
    \Health{\LpLi,\price}{\pmvA} \geq 1 
    \end{array}
  }
  {
    (\WalW,\LpL,\price)
    \xrightarrow{\actBorrow{\pmvA}{\valV:\tokT}}
    (\WalW + \setenum{(\tokT,\pmvA) \mapsto \valV},
    \LpLi,
    \price)
  }
  \;\nrule{[Bor]}
}

\newcommand{\repRule}{%
  \irule
  {
  \begin{array}{c}
    \WalW(\tokT,\pmvA) \geq \valV > 0
    \qquad
    \LpL(\debt{\tokT},\pmvA) \geq \valV
    \qquad
    \LpLi = \LpL
    + \setenum{\tokT \mapsto \valV} 
    - \setenum{(\debt{\tokT},\pmvA) \mapsto \valV}    \end{array}
  }
  {
    (\WalW,\LpL,\price)
    \xrightarrow{\actRepay{\pmvA}{\valV:\tokT}}
    (\WalW - \setenum{(\tokT,\pmvA) \mapsto \valV},
    \LpLi,
    \price)    
  }
  \;\nrule{[Rep]}
}

\newcommand{\rdmRule}{%
  \irule
  {
  \begin{array}{c}
    \LpL(\minted{\tokT},\pmvA) \geq \minted{\valV} > 0
    \quad\;\;
    \valV = \minted{\valV} \cdot \X[\LpL]{\tokT} 
    \qquad
    \projRes{\LpL}(\tokT) \geq \valV
    \\[5pt]
    \LpLi = \LpL - \setenum{\tokT \mapsto \valV} 
    - \setenum{(\minted{\tokT},\pmvA) \mapsto \minted{\valV}}
    \qquad
    \Health{\LpLi,\price}{\pmvA} \geq 1
  \end{array}
  }
  {
    (\WalW,\LpL,\price)
    \xrightarrow{\actRedeem{\pmvA}{\minted{\valV}:\minted{\tokT}}}
    (\WalW + \setenum{(\tokT,\pmvA) \mapsto \valV},
    \LpLi,
    \price)
  }
  \;\nrule{[Rdm]}
}

\newcommand{\intRule}{%
  \irule
  {
  \LpLi = \LpL 
  + \sum_{\tokT,\pmvA}
  \LpL(\debt{\tokT},\pmvA) \cdot \Intr{\LpL}(\tokT)
  }
  {
    (\WalW,\LpL,\price)
    \xrightarrow{\accrueIntOp}
    (\WalW,\LpLi,\price)
  }
  \; \nrule{[Int]}
}

\newcommand{\liqRule}{%
  \irule
  {
    \begin{array}{c}
    \WalW(\tokT[0],\pmvA) \geq \valV[0] > 0
    \qquad
    \LpL(\debtT[0],\pmvB) \geq \valV[0]
    \qquad
    \minted{\valV[1]} = \frac{\valV[0]}{\X[\LpL]{\tokT[1]}} \cdot \frac{\price[{\tokT[0]}]}{\price[{\tokT[1]}]} \cdot \rLiq
    \qquad
    \LpL(\mintedT[1],\pmvB) \geq \minted{\valV[1]}
    \\[7pt]
    \LpLi = \LpL 
    + \setenum{\tokT[0] \mapsto \valV[0]}
    + \setenum{(\mintedT[1],\pmvA) \mapsto \minted{\valV[1]}}
    - \setenum{(\mintedT[1],\pmvB) \mapsto \minted{\valV[1]}}
    - \setenum{(\debtT[0],\pmvB) \mapsto \valV[0]}
    \\[5pt]
    \pmvA \neq \pmvB 
    \qquad
    \Health{\LpL,\price}{\pmvB} < 1 
    \qquad
    \Health{\LpLi,\price}{\pmvB} \leq 1
    \end{array}
  }
  {
  (\WalW,\LpL,\price)
  \xrightarrow{\actLiquidate{\pmvA}{\pmvB}{\valV[0]:\tokT[0]}{\mintedT[1]}}
  (\WalW - \setenum{(\tokT[0],\pmvA) \mapsto \valV[0]},
  \LpLi,
  \price)
  }
  \;\nrule{[Liq]}
}

\newcommand{\pxRule}{%
\irule
  {\price(\tokT) + \delta > 0}
  {
    (\WalW,\LpL,\price)
    \xrightarrow{\
    \pxOp{\prIncr:\tokT}
    } 
    (\WalW,\LpL,\price + \{\tokT \mapsto \prIncr\})
  }
  \;\;\;\nrule{[Px]}
}

\newcommand{\swpRule}{%
  \irule
  {
    \WalW(\tokT[0],\pmvA) \geq \valV > 0
    \qquad
    \WalWi = \WalW
    - \setenum{(\tokT[0],\pmvA) \mapsto \valV}
    + \setenum{(\tokT[1],\pmvA) \mapsto \valV \cdot \frac{\price[{\tokT[0]}]}{\price[{\tokT[1]}]}}
  }
  {
    (\WalW,\LpL,\price)
    \xrightarrow{\actSwap{\pmvA}{\valV:\tokT[0]}{\tokT[1]}} 
    (\WalWi,\LpL,\price)
  }
  \;\;\;\nrule{[Swp]}
}

\makeatletter
\renewcommand\paragraph{\@startsection{paragraph}{4}{\z@}%
  {2.25ex \@plus 1ex \@minus .2ex}%
  {-0.75em}%
  {\normalfont\normalsize\bfseries}}
\makeatother

\title{A theory of Lending Protocols in DeFi}

\author[M.~Bartoletti]{Massimo Bartoletti}
\address{University of Cagliari, Cagliari, Italy}
\email{bart@unica.it}

\author[E. Lipparini]{Enrico Lipparini}
\address{University of Cagliari, Cagliari, Italy}
\email{enrico.lipparini@unica.it}

\begin{document}

\maketitle

\begin{abstract}
Lending protocols are one of the main applications of Decentralized Finance (DeFi), enabling crypto-assets loan markets with a total value estimated in the tens of billions of dollars.
Unlike traditional lending systems, these protocols operate without relying on trusted authorities or off-chain enforcement mechanisms.
To achieve key economic goals such as stability of the loan market, they devise instead trustless on-chain mechanisms, 
such as rewarding liquidators who repay the loans of under-collateralized borrowers  by awarding them part of the borrower's collateral.
%
The complexity of these incentive mechanisms, combined with their entanglement in low-level implementation details, makes it challenging to precisely assess the structural and economic properties of lending protocols,
as well as to analyze user strategies and attacks.
Crucially, since participation is open to anyone, any weaknesses in the incentive mechanism may give rise to unintended emergent behaviours, 
or even enable adversarial strategies aimed at making profits to the detriment of legit users, or at undermining the stability of the protocol.
In this work, we propose a formal model of lending protocols that captures the essential features of mainstream platforms, enabling us to identify and prove key properties related to their economic and strategic dynamics.
\end{abstract}

\section{Introduction} \label{section:introduction}

\newcommand{\compoundsupply}{$\sim$\$3B\xspace}
\newcommand{\aavesupply}{$\sim$\$25B\xspace}

\newcommand{\uniswapsupply}{\$5.3B\xspace}
\newcommand{\curvesupply}{\$4.6B\xspace}
\newcommand{\uniswapvolumedaily}{\$1.3B\xspace}
\newcommand{\curvevolumedaily}{\$180M\xspace}
\newcommand{\DAOsupply}{\$6B\xspace}
\newcommand{\statsdate}{June 2025\xspace}

Decentralized Finance (DeFi) refers to a collection of interoperable protocols run on permissionless blockchains that replicate traditional financial services without relying on centralized intermediaries.
In this setting, \emph{lending protocols} have established loan markets of crypto-assets that collectively manage tens of billions of dollars in value:  
as of \statsdate, two of the main lending platforms, Aave and Compound, 
hold respectively \aavesupply and \compoundsupply worth of crypto-assets~\cite{aavestats,compstats}.

At an abstract level, lending protocols can be seen as state transition systems, where the system state keeps track of the credit and debit positions --- abstractly modelled through \emph{tokens} --- associated with each user. 
Such system is partitioned into two components: the \emph{user wallets}, which represent the tokens freely available for user disposal, and \emph{lending pools}, which record the tokens available for lending as well as the outstanding credit and debit positions.
For example, lending can be modelled as a transfer of tokens from the user's wallet to the lending pool, together with a contextual minting of \emph{credit tokens} that represent the user's claim. 
Dually, borrowing can be modelled as a transfer of tokens from the lending pool to the user's wallet, along with the minting of \emph{debt tokens} that record the user's obligation. 

A key distinguishing feature of decentralized lending protocols, compared to traditional lending systems, is the absence of off-chain enforcement mechanisms to prevent loan defaults. 
Instead, lending protocols rely entirely on \emph{on-chain} mechanisms to incentivize, one the one hand, lenders to provide liquidity, and, on the other hand, borrowers to repay their loans.
As in traditional finance, borrowing in decentralized lending protocols requires users to provide a collateral, and debts accrue interests over time. 
However, unlike traditional finance, decentralized lending protocols are open to all users, who can freely participate as \emph{liquidity providers} --- gaining from interests accrued on debts --- and as \emph{liquidators}, by repaying (part of) an under-collateralized loan in exchange for a discounted amount of the seized collateral. 
Another key difference is that all protocol parameters, such as the interest rate function, collateral and token prices, are algorithmically determined by smart contracts.

This openness, combined with the inherent complexity of the emergent behaviour resulting from interactions between users and lending pools, makes them an attractive target for adversaries. By exploiting weaknesses in their economic incentive mechanisms, adversaries can devise sophisticated attack strategies to extract undue profits, harm legitimate users, and, more broadly, undermine the stability of the protocol.
Since real-world implementations of lending protocols are too complex for effective formal analysis, we need at least an abstract model of their behavior that can faithfully analyse such strategic aspects.
Such model and analysis should offer relevant insights about the following research questions:
\vspace{0.6em}
\begin{enumerate}[label={\bf RQ\arabic*:}]
  \setlength\itemsep{0.6em}
\item \label{RQ1}
What structural properties and invariants are enjoyed by lending pools?

\item \label{RQ2}
What is the economic effect of each individual interaction with a lending pool?

\item \label{RQ3}
Which strategies can be followed by rational users anticipating a forthcoming action? 

\item \label{RQ4}
Which attacks are possible for adversaries with a large amount of capital? 

\end{enumerate}

\paragraph{Contributions} 

This paper proposes a formal analysis of lending protocols, focussing on the properties that arise through the interaction between users and lending pools.  
To this purpose, we introduce a new operational model that captures the state machine behavior of lending protocols by synthesizing the common features of leading implementations such as Aave and Compound.

\vspace{0.6em}
More specifically, our contributions can be summarized as follows: 
\vspace{0.6em}
\begin{enumerate}
  \setlength\itemsep{0.6em}
\item A formal model of lending protocols that precisely captures their behavior as transitions in a state machine. Our model encompasses all typical interactions between users and lending pools, along with key economic features such as collateralization, exchange rates, token prices, and interest accruals (\Cref{sec:lending-pools}).
 
\item An analysis of the fundamental \emph{structural properties} of lending protocols, in the form of invariants on the machine states.
(\Cref{sec:struct-properties}).
In particular, we prove in Lemma~\ref{lem:ER-increasing} that exchange rates are preserved by all actions except interest accruals --- which always increase them --- and for the corner case where all the credits of a token are redeemed.
Another crucial invariant is given by Theorem~\ref{th:net-worth-preservation}, which establishes that the total net worth of all users is preserved by all actions except price updates.

\item An analysis of the economic effects of individual interactions with a lending protocol, both in terms of the change to the users' net worth (\ie, their \emph{gain}) and collateralization (\Cref{sec:econ-properties}).
In particular, Lemma~\ref{lem:gain-base} measures the gain of actions performed by users: it shows that the only action that affects a user's gain is the liquidation, which yields a positive gain for the liquidator and an equivalent loss for the liquidated borrower. 
Lemma~\ref{lem:gain-px} quantifies the effect of price updates on the users' net worth: it shows that users with lots of debts in a given token type would benefit from a price decrease, while those with lots of credits would suffer losses.
Lemma~\ref{lem:gain-int} shows that a dual situation occurs for interest accruals.
Lemma~\ref{lem:health-tx} measures the impact of user actions on the health factor.
More specifically, Lemma~\ref{lem:health-dep-rep} compares the effectiveness of adding more collateral versus repaying debts in order to increase one's health factor.

\item An analysis of \emph{strategic users} --- those who aim at increasing their gain by leveraging partial knowledge of forthcoming actions in the lending protocol (\Cref{sec:strategic-properties}). 
In particular, we investigate the strategies such users should follow when they anticipate future events such as liquidations (Theorem~\ref{th:gameLiq}), price updates (Theorem~\ref{th:gamePx} and Lemma~\ref{lem:gameOptions}), and interest accruals (Theorem~\ref{th:X-int-vs-X}).
For liquidations and price updates, we show that a user can always front-run the impending action with their own transaction in order to achieve a higher gain.
For interest accruals, instead, we show that --- except in the simple case where the interest rate function is constant --- there is no simple front-running strategy that guarantees a higher gain. 

\item An analysis of attacks to lending protocols, where adversaries use their capital to temporarily manipulate token prices or their utilization to obtain an advantage in further interactions with a lending pool (\Cref{sec:attacks}).
More specifically, 
Theorem~\ref{th:gameUndercollLoanAttck} shows an attack in which an adversary manipulates prices in order to borrow more tokens than what they should be allowed to.
Theorem~\ref{th:gameLiqAttck} shows another price manipulation attack,
where the adversary causes other users to become under-collateralized and profits from their subsequent liquidation.
In the other attacks, the adversary manipulates the \emph{utilization} of some token --- roughly defined as the ratio between the total debt in that token and its overall supply --- to induce a variation in the interest rate applied to debts in that token. 
This variation is then exploited by the adversary in order to make a profit.
Theorem~\ref{th:underutilizationAttack} shows an attack where an adversary deposits tokens just before an interest accrual, in order to induce a decrease in the utilization, and so pay less interests on their debts.
Theorem~\ref{th:overutilizationAttack} shows instead an attack where the adversary borrows tokens before an interest accrual, in order to induce an increase in the utilization, and so benefit from a higher appreciation of their credits.

\end{enumerate}
\vspace{0.6em}
We discuss some limitations of our work in~\Cref{sec:limitations}, related work in~\Cref{sec:related}, and draw conclusions in~\Cref{sec:conclusions}.
We include detailed proofs of our results in Appendix~\ref{proofs:struct-properties} to~\ref{proofs:attacks}.


\section{Lending Protocols}
\label{sec:lending-pools}
\label{sec:lp}

We introduce a formal model of lending protocols, encompassing the common features implemented by the main lending platforms, and abstracting away some features that are inessential to understand their underlying economic mechanism. 
We discuss these abstractions and the limitations they induce in~\Cref{sec:limitations}.


\subsection{Blockchain model}
\mypar{Users and tokens}
We assume a denumerable set of \keyterm{addresses} $\PmvU$,
ranged over by $\pmvA, \pmvAi, \ldots$.
Each user can participate in a lending protocol by using one or more addresses, which serve as pseudonyms for that user.
Hereafter, at the cost of a little ambiguity we will often identify users with their addresses. 
We also assume a denumerable set of \keyterm{base token types} $\TokU$, ranged over by $\tokT,\tokTi,\ldots$.
The notation $\valV:\tokT$ stands for $\valV$ units of token type $\tokT$, where $\valV$ is a nonnegative real number ($\valV \in \RNN$).
%
When users deposit tokens of type $\tokT$ into a lending pool,
they receive in return a receipt of the deposit, which we model as \emph{credit tokens} $\minted{\tokT}$.
Dually, when users borrow tokens from a LP, we represent 
their debt as \emph{debit tokens} $\debt{\tokT}$.
We denote the universes of credit and debit tokens as $\TokUM$ and $\TokUD$, respectively. 

\mypar{Wallets}

We model the users' \keyterm{wallets} as a function that associates each base token type and each address to the token balance directly available to the user.
Formally:
\[
\WalW \; : (\TokU \times \PmvU) \rightarrow \RNN
\]
Note that $\WalW(\tokT,\pmvA)$ ranges over a continuous domain. 
While this differs from concrete lending protocol implementations, where token balances are discrete, our model abstracts them as nonnegative real numbers, thereby avoiding the need to account for rounding in balance-related operations.
Hereafter, we use $\valV, \valVi, \ldots$ to range over $\RNN$.

\mypar{Lending pools}

A lending pool (in short, LP) is intuitively formed by three components:
\begin{itemize}
\item a map from base token types to the balance of their reserves in the pool; 
\item a map from addresses to their associated credit tokens;
\item a map from addresses to their associated debit tokens.
\end{itemize}
For notational convenience, rather than modelling a LP as a triple of functions, we model it as a function with domain the disjoint union of the domains of the three maps:
\[
\LpL \; : \, (\TokU \uplus (\TokUM \times \PmvU) \uplus (\TokUD \times \PmvU)) \rightarrow \RNN
\]
where we assume that $\LpL$ has finite support.
%
%
We now introduce some notation to manipulate LPs. 
We denote by $\setenum{x \mapsto y}$ a partial function mapping $x$ to $y$.
Pointwise summation of functions is denoted by $+$ and $\sum$.
When $f$ is defined on $x$ but $g$ is not, then 
$(f+g)(x) = f(x)$.
For example, if $\LpL$ maps each element of its domain to $0$, then $\LpL + \setenum{\tokT \mapsto 2}$ is the function that is equal to $\LpL$ in all points but $\tokT$, where it takes value $2$.


\begin{table}[t]
\caption{Summary of notation.}
\label{tab:notation}
\centering
\begin{tabular}{p{40pt}p{140pt}p{105pt}p{110pt}}
\hline
$\pmvA,\pmvB$ & Users 
& $\X[\LpL]{\tokT}$ & Exchange rate
\\
$\tokT,\minted{\tokT},\debt{\tokT}$ & Token type (base,credit,debit) 
& $\supplyWal[\WalW]{\tokT}, \supply[\LpL]{\tokT}, \supplyDebt[\LpL]{\tokT}$ & Token supply
\\
$\WalW,\WalWi$ & Wallet states 
& $\wealth{\confG}{\pmvA}$ & Net worth of $\pmvA$ in $\confG$
\\
$\LpL,\LpLi$ & LP states 
& $\Health{\confG}{\pmvA}$ & Health factor of $\pmvA$ in $\confG$
\\
$\price,\pricei$ & Price functions 
& $\gain{\pmvA}{\confG}{\txT}$ & Gain of $\pmvA$ upon firing $\txT$
\\
$\txT,\txTi$ & Transactions 
& $\liqThreshold$ & Liquidation threshold
\\
$\confG,\confGi$ & Blockchain states 
& $\rLiq$ & Liquidation reward
\\
\hline
\end{tabular}
\end{table}

\mypar{Prices}

A price oracle is a function associating a strictly positive price to each base token:
\[
\price \; : \, \TokU \rightarrow \RP 
\]
We use the previously introduced notation to describe price updates as well: for example, $\price + \setenum{\tokT \mapsto 0.1}$ denotes the price function that coincides with $\price$ for all token types except for $\tokT$, whose price is increased by 0.1.


\mypar{Blockchain states} 

A blockchain state defines all the components that are needed to represent the interactions between users and LPs. 
Formally, we render a blockchain state as a triple $\confG = (\WalW,\LpL,\price)$  containing the users' wallets $\WalW$, the LP state $\LpL$, and a price oracle $\price$.

\subsection{Basic economic definitions}

\mypar{Token supply}

Given a wallet state $\WalW$ and a base token type $\tokT$,
we denote by $\supplyWal[\WalW]{\tokT}$ the number of units of $\tokT$ in $\WalW$.
We refer to $\supplyWal[\WalW]{\tokT}$ as the \emph{supply of $\tokT$ in $\WalW$}.
Similarly, given a LP state $\LpL$, we denote by $\supply[\LpL]{\tokT}$ and $\supplyDebt[\LpL]{\tokT}$ the supply of a credit token $\minted{\tokT}$ and of a debit token $\debt{\tokT}$, respectively.
Formally:
\begin{equation} \label{eq:supply}
\supplyWal[\WalW]{\tokT} = \sum_{\pmvA} \WalW(\tokT,\pmvA) 
\qquad
\supply[\LpL]{\tokT} = \sum_{\pmvA} \LpL(\minted{\tokT},\pmvA) 
\qquad
\supplyDebt[\LpL]{\tokT} = \sum_{\pmvA} \LpL(\debt{\tokT},\pmvA) 
\end{equation}

\mypar{Exchange rate}

The \keyterm{exchange rate} of a token type $\tokT$ in a LP state $\LpL$
represents the share of deposited units of $\tokT$ (\ie, reserves plus debts) over the units of the associated credit tokens.
Formally, we define the exchange rate $\X[\LpL]{\tokT}$ as:
\begin{equation} 
  \label{eq:LP:ER}
  \X[\LpL]{\tokT}
  \; = \;
  \begin{cases}
    \dfrac
    {
      \LpL(\base{\tokT})
      + 
      \supplyDebt[\LpL]{\tokT}
    }
    {\supply[\LpL]{\tokT}}
    & \text{if $\supply[\LpL]{\tokT} \neq 0$}
    \\[10pt]
    1 & \text{otherwise} 
  \end{cases}
\end{equation}

\noindent
The intuition, which will be more clear once we define the rules for depositing and redeeming tokens, is to define the price $\price[\mintedT]$ of a credit token type $\minted{\tokT}$ in $\LpL$ as:
\begin{equation} \label{eq:LP:price-credit}
\price[\mintedT] \; = \; \X[\LpL]{\tokT} \cdot \price[\tokT]
\end{equation}
Then, when a user deposits $\valV:\tokT$ into a LP, they will receive in exchange an amount $\minted{\valV}:\mintedT$ such that $\valV \cdot \price[\tokT] = \minted{\valV} \cdot \price[\mintedT]$. 
We will also see in Lemma~\ref{lem:ER-increasing} that 
the exchange rate increases upon interest accruals. 
Since this leads to a proportional increase of the price of credit tokens as per~\eqref{eq:LP:price-credit}, users have a direct incentive to providing liquidity to the LP.

\mypar{Net worth}

We define the value of tokens in $\pmvA$'s wallet as the sum of the amounts of the tokens in $\stateBase{\cdot}{\pmvA}{}$ weighted by their price:
\begin{equation} \label{eq:baseVal}
  \walVal[\WalW,\price]{\pmvA}
  \; = \;
  \sum_{\tokT} \stateBase{\tokT}{\pmvA}{} \cdot \price[\tokT]
\end{equation}

\noindent
The value of $\pmvA$'s credits in a LP is the sum of the amounts of all the credit tokens owned by $\pmvA$ weighted by their price:
\begin{equation} \label{eq:creditVal}
  \creditval[\LpL,\price{}]{\pmvA}
  \; = \;
  \sum_{\tokT \in \TokU}
  \LpL(\minted{\tokT},\pmvA) \cdot \X[\LpL]{\tokT} \cdot \price[\tokT]
\end{equation}

\noindent
Similarly, the value of $\pmvA$'s debts in a LP is the sum 
of the amount of $\pmvA$'s debit tokens weighted by the price of the underlying base token:
\begin{equation} \label{eq:debtVal}
  \debtval[\LpL,\price{}]{\pmvA}
  \; = \;
  \sum_{\tokT \in \TokU} \LpL(\debt{\tokT},\pmvA) \cdot \price[\tokT]
\end{equation}

\noindent
We then define the \keyterm{net worth} of $\pmvA$ in a blockchain state $\confG = (\WalW,\LpL,\price)$ as the value of base tokens in $\pmvA$'s wallet, plus the value of credits in the LP, minus the value of $\pmvA$'s debt:
\begin{equation} \label{eq:networth}
  \wealth{\confG}{\pmvA}
  \; = \;
  \walVal[\WalW,\price]{\pmvA}
  + \creditval[\LpL,\price]{\pmvA}
  - \debtval[\LpL,\price]{\pmvA}
\end{equation}

We will show in Theorem~\ref{th:net-worth-preservation} a fundamental preservation property, \ie the net worth is preserved by all LP actions (except for price updates).

In certain cases, it will be useful to refer to the net worth of a user \emph{restricted} to a specific base token type $\tokT$. 
We will write $\projTok{\wealth{\confG}{\pmvA}}$ to denote the quantity obtained by removing from $\wealth{\confG}{\pmvA}$ all the expressions that do not mention $\tokT$, \ie:
\begin{equation} 
\label{eq:wealth-restricted}
\projTok{\wealth{\confG}{\pmvA}}
\; = \;  
\Big(
\stateBase{\tokT}{\pmvA}{} 
+ 
\LpL(\minted{\tokT},\pmvA) \cdot \X[\LpL]{\tokT}
-  \LpL(\debt{\tokT},\pmvA)
\Big) 
\cdot \price[\tokT] 
\end{equation} 

Of course, the overall net worth of $\pmvA$ is given by the sum of her restricted net worth over all the base token types:
\begin{equation}
\label{eq:sum-wealth-restricted}
\wealth{\confG}{\pmvA}
\; = \;
\sum_{\tokT \in \TokU} \projTok{\wealth{\confG}{\pmvA}}
\end{equation}


%

\mypar{Net position}

The net worth $\walVal[\confG]{\pmvA}$ does not perfectly reflect the financial position of $\pmvA$.
On the one hand, $\pmvA$ may have tokens deposited in a LP that she cannot redeem due to insufficient liquidity in the LP: as a result, her disposable wealth is lower than her net worth.
On the other hand, $\pmvA$ may owe debts to the LP without the LP being able to enforce their repayment: in this case, her disposable wealth is actually higher than her net worth. 
This situation arises when $\pmvA$ has more debts than credits, \ie her net position is negative.
Formally, we define the \emph{net position} of $\pmvA$ as: 
\begin{equation}
\label{eq:net-position}
\netpos{\LpL,\price}{\pmvA} 
\; = \;
{\creditval[\LpL,\price]{\pmvA}}
-
{\debtval[\LpL,\price]{\pmvA}}
\; = \;
\wealth{\confG}{\pmvA}
-
\walVal[\WalW,\price]{\pmvA}
\end{equation}

\noindent
A negative net position represents the amount of debts that a user can \emph{default}, \ie  that the LP cannot be guaranteed to recover.


\mypar{Collateralization and health factor}

Collateralization is a measure of a user's ability to repay their debts, defined as the ratio between the values of the user's credits and debts:
\begin{equation} \label{eq:coll}
  \Coll{\LpL,\price}{\pmvA}
  \; = \;
  \begin{cases}
    \dfrac
    {\creditval[\LpL,\price]{\pmvA}}
    {\debtval[\LpL,\price]{\pmvA}}
    & \text{if } \debtval[\LpL,\price]{\pmvA} > 0 \\
    + \infty 
    & \text{otherwise}  
  \end{cases}
\end{equation}

\noindent
The idea is that credit tokens are not directly transferable by users, but rather are kept by the LP as a guarantee in case a borrower fails to repay her debt.
As we will see, LPs allow users to borrow only if they are over-collateralized, featuring an incentive mechanism for borrowers to keep their debt sufficiently collateralized over time. 
More specifically, borrowers must maintain their collateralization above a given value $1/\liqThreshold \geq 1$ in order to avoid that their credit tokens are seized and distributed to other users in exchange for repaying their debt.
The value $\liqThreshold < 1$ is a protocol parameter, called \emph{liquidation threshold}.

The requirement that a user $\pmvA$ has sufficient collateralization can be equivalently expressed by requiring that their \keyterm{health factor} $\Health{\LpL,\price}{\pmvA}$ is at least $1$, where:
\begin{equation} \label{eq:healthfactor}
  \Health{\LpL,\price}{\pmvA}
  \; = \;
  \Coll{\LpL,\price}{\pmvA}
  \cdot 
  {\liqThreshold}
\end{equation}




\mypar{Interest rates}

As in traditional finance, loans in lending protocols accrue interest over time. 
We keep our model parametric with respect to interest rates, by introducing a function $\Intr{\LpL}(\tokT)$, which depends only on the LP state $\LpL$ and the token type $\tokT$.
Coherently to actual lending protocols~\cite{Gudgeon20aft}, 
we assume that interest rates are strictly positive, and that the interest rate for a token $\tokT$ depends solely on the total reserves, credits, and debits denominated in $\tokT$, independently of their distribution across user addresses.
Formally, we require the interest rate function to respect the following constraints, for all~$\tokT$:
\begin{equation} 
\label{eq:interest-rate-gt-zero}
    \Intr{\LpL}(\tokT) > 0
    \qquad
    \qquad
    \LpL \sim_{\tokT} \LpLi 
    \implies
    \Intr{\LpL}(\tokT) = \Intr{\LpLi}(\tokT)
\end{equation}
where we define the relation $\sim_{\tokT}$ between two LP states as:
\[
\LpL \sim_{\tokT} \LpLi
\quad \eqdef \quad
\LpL(\tokT) = \LpLi(\tokT)
\;\land\;
\supply[\LpL]{\tokT} = \supply[\LpLi]{\tokT}
\;\land\;
\supplyDebt[\LpL]{\tokT} = \supplyDebt[\LpLi]{\tokT}
\]

Although most of our results do not depend on the actual choice for of the interest rate function, in examples and in some specific results  
(\eg, Theorems~\ref{th:X-int-vs-X}, \ref{th:underutilizationAttack}, and~\ref{th:overutilizationAttack})
we will consider a concrete instantiation, inspired by actual lending protocols such as Aave and Compound. 
There, the interest rate for a token $\tokT$ in a LP state $\LpL$ is a function of the \keyterm{utilization} of $\tokT$,
which measures the fraction of units of $\tokT$ currently lent to users. 
Formally, the utilization of $\tokT$ in $\LpL$ is defined as $0$ when $\supplyDebt[\LpL]{\tokT} = 0$, and otherwise:
\begin{equation} \label{eq:util}
    \Util[\LpL]{\tokT} 
    \; = \; 
    \dfrac
    {\supplyDebt[\LpL]{\tokT}}
    {\LpL(\tokT) + \supplyDebt[\LpL]{\tokT}}
\end{equation}

\noindent
We then define the \emph{linear utilization interest rate} as a linear function of the token utilization:
\begin{equation}
\label{eq:intr-util}
\Intr{\LpL}(\tokT) 
\; = \;
\alpha \cdot \Util[\LpL]{\tokT} + \beta
\qquad\text{where $\alpha\geq 0, \ \beta > 0$}
\end{equation}
The idea is that if a token $\tokT$ is under-utilized, \ie there are many available reserves in the LP compared to the debts in $\tokT$, then the interest rate for $\tokT$ should be low, in order to incentivize users to borrow it.
Instead, if a token is over-utilized, \ie there are many debts in $\tokT$ compared to the available reserves, a higher interest rate discourages additional loans~\cite{Gudgeon20aft}.

\subsection{Semantics} \label{section:LPrules}

\begin{table}[t]
  \centering
  \caption{Transactions.}
  \label{tab:lending-pool:tx}
  \begin{tabular}{ll}
    \hline
    $\actDeposit{\pmvA}{\valV:\tokT}$ 
    & $\pmvA$ deposits $\valV$ units of $\tokT$, 
      receiving back units of credit token $\minted{\tokT}$
    \\
    $\actBorrow{\pmvA}{\valV:\tokT}$ 
    & $\pmvA$ borrows $\valV$ units of $\tokT$ 
    \\
    $\actRepay{\pmvA}{\valV:\tokT}$ 
    &  $\pmvA$ repays $\valV$ units on $\pmvA$'s debt in $\tokT$ 
    \\
    $\actRedeem{\pmvA}{\valV:\minted{\tokT}}$ 
    & $\pmvA$ redeems $\valV$ units of $\minted{\tokT}$, 
      receiving back units of $\tokT$ 
    \\
    $\actLiquidate{\pmvA}{\pmvB}{\valV:\tokT[0]}{\minted{\tokT[1]}}$ 
    & $\pmvA$ repays $\valV$ units of $\pmvB$'s debt in $\tokT[0]$, 
      seizing units of $\minted{\tokT[1]}$ from $\pmvB$
    \\
    $\accrueIntOp$ 
    & All loans accrue interests
    \\   
    $\exchUpdateOp{\delta:\tokT}$
    & Price of tokens $\tokT$ is increased/decreased by $\delta$
    \\
    \hline
  \end{tabular}
\end{table}

We formalise the interaction between users and LPs
as a labelled transition system between blockchain states.
Labels $\ltsLabel, \ltsLabeli, \ldots$ represent \keyterm{transactions}, which define the actions performed by users and by the environment. 
Transactions have the form displayed in~\Cref{tab:lending-pool:tx}.
In the rest of the section we present the rules that define the state transitions.
An extended example of the application of these rules follows in~\Cref{sec:lp:example}.

The rules below define state transitions of the form $\confG \xrightarrow{\txT} \confGi$.
When such a transition exists, we say that $\txT$ is \emph{enabled} in $\confG$.
We extend this relation to sequences of transactions:
for an empty sequence $\epsilon$ we have $\confG \xrightarrow{\epsilon} \confG$, and for a sequence $\TxTS = \txT \TxYS$ made of a head $\txT$ and a tail $\TxYS$ we define:
\[
\confG \xrightarrow{\TxTS} \confGi
\qquad
\text{iff}
\qquad
\confG \xrightarrow{\txT} \confGii \text{ and } 
\confGii \xrightarrow{\TxYS} \confGi
\]
We say that a blockchain state $\confG[0]$ is \emph{initial} when its LP state has no reserves, no credit tokens, and no debit tokens.
We then say that a state $\confG$ is \emph{reachable} when there exists some initial $\confG[0]$ and some sequence of transactions $\TxTS$ such that $\confG[0] \xrightarrow{\TxTS} \confG$.

\mypar{Deposit} 

Any user $\pmvA$ can deposit $\valV$ units of a base token type $\tokT$ by performing the transaction $\actDeposit{\pmvA}{\valV:\tokT}$.
For each deposit of $\tokT$, the LP mints $\minted{\valV}$ units of the credit token $\minted{\tokT}$.
The amount $\minted{\valV}$ is computed in such a way that the value in credit tokens obtained by $\pmvA$ is equal to the value of the deposited base tokens, \ie, according to~\eqref{eq:LP:price-credit}, 
$\minted{\valV} \cdot \price[\minted{\tokT}] = \valV \cdot \price[\tokT]$.
\begin{equation*}
\depRule
\end{equation*}

\medskip
The premise $\WalW(\base{\tokT},\pmvA) \geq \valV$ ensures that $\pmvA$'s wallet contains at least $\valV$ units of $\tokT$.
In the new blockchain state, the wallet state
$\WalW - \setenum{(\tokT,\pmvA) \mapsto \valV}$
records that $\valV:\tokT$ have been subtracted from $\pmvA$'s wallet.
In the premises, $\minted{\valV}$ is the amount of credit tokens assigned to $\pmvA$ upon the deposit.
In the new LP state $\LpLi$, the reserves of $\tokT$ are increased by $\valV$ units, and the credits of $\pmvA$ are increased by $\minted{\valV}$ units.
We refer to users holding credit tokens as \emph{creditors}.



\mypar{Borrow} 

Any user can borrow units of a base token type $\tokT$ from an LP,
provided that the LP has sufficient reserves of $\tokT$,
and that the user has enough collateral. In the rule premises, this is rendered by requiring that the borrower's health factor is at least $1$ after the action.
\begin{equation*}
\borRule
\end{equation*}

\medskip
In the new blockchain state, $\valV:\tokT$ are added to $\pmvA$'s wallet and removed from the LP reserves. 
Furthermore, the LP records $\valV:\debt{\tokT}$ additional debit tokens for the borrower $\pmvA$.

\mypar{Repay} 

Any user with a loan in tokens $\tokT$ can repay it (in part or as whole) by paying base tokens $\tokT$ to the LP.
In exchange, the LP cancels part of the users' debt, by removing a number of debit tokens $\debt{\tokT}$ equivalent to the number of base tokens paid to the LP.
\begin{equation*}
\repRule
\end{equation*}


\mypar{Redeem} 

Any debt-free user can redeem credit tokens $\minted{\tokT}$ for an equal value of the underlying base tokens, provided that enough reserves of $\tokT$ are available in the LP. 
As in the deposit rule, the number $\valV$ of units of the base token is computed in such a way to have  
$\minted{\valV} \cdot \price[\minted{\tokT}] = \valV \cdot \price[\tokT]$, according to~\eqref{eq:LP:price-credit}.
Any user with non-zero debts can redeem credit tokens as long as it remains over-collateralized. 
This constraint does not apply to users without loans, as credit tokens are not used as collateral.

\begin{equation*}
\rdmRule
\end{equation*}

\medskip
The premise $\LpL(\minted{\tokT},\pmvA) \geq \minted{\valV}$
requires that $\pmvA$ has at least the amount of credit tokens that they want to redeem.
The premise $\LpL(\tokT) \geq \valV$ requires that the LP has enough reserves of base tokens $\tokT$ to give in return.
%
%
The premise $\Health{\LpLi,\price}{\pmvA} \geq 1$
requires that $\pmvA$ remains over-collateralized after the action.

\mypar{Interest Accrual}

Interest accrual models the application of interest to loans. 
The action applies an interest to each loan, 
updating the debt of \emph{all} users with a non-zero debt.
\begin{equation*}
\intRule
\end{equation*}

\medskip
Formally, for each base token type $\tokT$, the number of debit tokens $\debt{\tokT}$ of each $\pmvA$ is increased by $\LpL(\debt{\tokT},\pmvA) \cdot \Intr{\LpL}(\tokT)$, which is strictly greter than zero by~\eqref{eq:interest-rate-gt-zero}.
Note that this action may either increase or decreases the health factor of users with debts, since both $\debtval{\pmvA}$ and $\creditval{\pmvA}$ increase upon the action.
Unlike the previous actions, the label $\accrueIntOp$ omits the name of the address who signs the transaction.
This is because interest accruals are meant to be triggered in a time-dependent fashion, \eg once for each block.

\mypar{Liquidation}

When the health factor of a borrower $\pmvB$ is below $1$, any other user $\pmvA$ with sufficient tokens can \emph{liquidate} part of $\pmvB$'s loan, in return for a discounted amount of credit tokens seized from $\pmvB$. 
The maximum seizable amount is bounded by $\pmvB$'s balance of the credit token and by the ex-post health factor of $\pmvB$,
which cannot exceed $1$ after the action. 
The protocol parameter $\rLiq > 1$ represents the \emph{reward factor}, which implies that the value of tokens obtained by the liquidator $\pmvA$ is greater than the value of $\pmvB$'s debt repaid.

\begin{equation*}
\liqRule
\end{equation*}

\medskip
Note that the amount $\minted{\valV[1]}$ of credit tokens received by $\pmvA$ is computed in such a way to equal the value of repaid debt, multiplied by the reward factor.
That is, according to~\eqref{eq:LP:price-credit}:
\[
\minted{\valV[1]} \cdot \price(\mintedT[1])
\; = \;
\minted{\valV[1]} \cdot \X[\LpL]{\tokT[1]} \cdot \price(\tokT[1])
\; = \;
\valV[0] \cdot \price(\tokT[0]) \cdot \rLiq
\; > \;
\valV[0] \cdot \price(\tokT[0])
\]



\mypar{Price updates} 

The price of any base token can be increased/decreased by an amount $\delta \in \RNZeroB$, provided that the new price is still strictly positive:
\begin{equation*}
\pxRule
\end{equation*}

\medskip
Similarly to $\accrueIntOp$, also the transition label $\exchUpdateOp{\delta:\tokT}$ is not linked to any address. 
This is because while in other actions the address in the label is the transaction signer, in a price update transaction we assume that the action can be performed only by a special user, acting as a price oracle.

\mypar{Token swap} 

The actions considered so far fully characterise the behaviour of lending protocols. 
However, in order to be able to analyse the economic impact of strategies where users can also interact with the environment, we extend our transition system with an additional \emph{swap} action, allowing users to exchange base tokens of type $\tokT[0]$ with a price-equivalent amount of tokens of another type $\tokT[1]$:
\begin{equation*} 
\swpRule
\end{equation*}

In practice, swap actions can be executed through centralized or decentralized exchange services. 
For example, Automated Market Makers (AMMs) are decentralized protocols that allow users to swap between two token types at an algorithmically determined exchange rate, and also serve as decentralized price oracles~\cite{Angeris20aft,BCL22lmcs}.  
In real-world settings, token swaps --- especially when involving large amounts --- typically result in price adjustments. 
For example, a large sale of a token is usually accompanied by a decrease of its price (\eg, in AMMs this price update is applied automatically as part of the swap action).
In our \nrule{[Swp]} transition we assume that token prices are preserved:
when necessary, we can still represent a price-updating swap action as an atomic sequence of \nrule{[Swp]} and \nrule{[Px]}.

\subsection{An example}
\label{sec:lp:example}

We now illustrate our semantics through a simple example involving users $\pmvA$ and $\pmvB$. 
We display their interactions in~\Cref{fig:lending-pool:ex1}, using an alternative representation of blockchain states for readability.
Namely, we write a blockchain state as:
\[
  \wal{\pmvA[1]}{\tokBal[1]} \mid \cdots \mid \wal{\pmvA[n]}{\tokBal[n]}
  \mid
  \lp{\valV[1]:\tokT[1],\cdots,\valV[k]:\tokT[k]}{}
  \mid
  \price
\]

In this representation, a term $\wal{\pmvA[i]}{\tokBal[i]}$
includes the tokens of \emph{all} kinds (base, credit, and debit) associated to $\pmvA[i]$, while the term $\lp{\valV[1]:\tokT[1],\cdots,\valV[k]:\tokT[k]}{}$ describes the reserves of base tokens deposited in the LP.
So, for example, the blockchain state $(\WalW,\LpL,\price)$ where:
\[
\WalW = \setenum{(\tokT[0],\pmvA) \mapsto 1}
\qquad
\LpL =  \setenum{\tokT[0] \mapsto 2, \tokT[1] \mapsto 3, (\mintedT[0],\pmvA) \mapsto 4, (\mintedT[1],\pmvB) \mapsto 5, (\debtT[0],\pmvB) \mapsto 6}
\]
would be represented in our sugared syntax as follows:
\[
\wal{\pmvA}{1:\tokT[0],4:\mintedT[0]} \mid
\wal{\pmvB}{5:\mintedT[1],6:\debtT[0]} \mid
\lp{2:\tokT[0],3:\tokT[1]}{} \mid
\price
\]


We now discuss the state transitions in~\Cref{fig:lending-pool:ex1}.
In the initial blockchain state, $\pmvA$ has $100$ units of $\tokT[0]$, $\pmvB$ has $50$ units of $\tokT[1]$, the LP has no reserves, and the price of both token types is $1$.
We assume that the protocol parameters are as follows:
the liquidation threshold is $\liqThreshold = \nicefrac{2}{3}$,  
the liquidation reward is $\liqReward = 1.1$, and the interest rate function is utility-based and has parameters $\alpha=0$ and $\beta = 0.12$, meaning that there is a constant interest factor $\Intr{\LpL}(\tokT) = 12\%$ for all token types $\tokT$.
\begin{itemize}

\item In steps 1 and 2, $\pmvA$ and $\pmvB$ deposit 50 units of $\tokT[0]$ and $\tokT[1]$, respectively, for which they receive equal amounts of credit tokens $\mintedT[0]$ and $\mintedT[1]$. 

\item In step 3, $\pmvB$ borrows \mbox{$30:\tokT[0]$}, using his credit tokens $\mintedT[1]$ as collateral for the loan.
The loan is permitted because the $\pmvB$'s health factor after the action is above the safety threshold $1$.
Although $\pmvB$ could have borrowed up to $\creditval{\pmvB} \cdot \liqThreshold = 50 \cdot \nicefrac{2}{3} = 33.3$ units of $\tokT[0]$, given the collateral of $50:\minted{\tokT[1]}$, here we assume that $\pmvB$ decides to leave some margin to manage future price volatility and the accrual of interest, which could decrease $\pmvB$'s health factor. 

\item In step 4, interest accrues on $\pmvB$'s debt.
Since the interest rate is 12\%, $\pmvB$'s debt on $\tokT[0]$ grows from $30$ to $33.6$.

\item In step 5, $\pmvB$ repays part of her debt, by paying $5:\tokT[0]$ to the LP. In this way, $\pmvB$'s health factor grows from $0.99$ to $1.16$,  avoiding the risk of being immediately liquidated by $\pmvA$.

\item In step 6, the price of $\tokT[0]$ is increased by $0.3$: since the debt value is at the denominator in the formula of collateralization~\eqref{eq:coll}, this yields a decrease of $\pmvB$'s health factor. This value drops to $0.89$, crossing the threshold for liquidations.

\item In step 7, $\pmvA$ liquidates \mbox{$11:\tokT[0]$} of $\pmvB$'s debt, obtaining in exchange $\minted{\valV[1]}:\mintedT[1]$, where:
\[
\minted{\valV[1]} = \frac{11}{\X{\tokT[1]}} \cdot \frac{\pricei[{\tokT[0]}]}{\pricei[{\tokT[1]}]} \cdot \rLiq
= 
\frac{11}{1} \cdot \frac{1.3}{1} \cdot 1.1 = 15.73
\]
Since the liquidation reward $\liqReward > 1$, the value in credit tokens obtained by $\pmvA$ is greater than the value in base tokens she paid, making the liquidation profitable:
\[
11 \cdot \price[{\tokT[1]}] = 11 \cdot 1.3 = 14.3
\quad < \quad
15.73 \cdot \X{\tokT[1]} \cdot \price[{\tokT[1]}] = 15.73
\]
After the liquidation, $\pmvB$'s health factor is increased (to $0.99$), since $\pmvB$'s debt value has decreased while the credit value has been preserved.   
Note that $\pmvA$ could have not liquidated, \eg, $12:\tokT[0]$, since doing so would have made $\pmvB$'s health factor exceed the safety threshold. 

\item In step 8, $\pmvA$ redeems $10: \mintedT[0]$, receiving $10.72:\tokT[0]$ in exchange. 
Here, each unit of $\mintedT[0]$ is exchanged for 
$\X{\tokT[0]} = \nicefrac{36+17.6}{15.73+34.27} = 1.072$ units of $\tokT[0]$, due to accrued interests. 

\end{itemize}

\begin{figure}[t]
  \small
  \scalebox{0.95}{\parbox{\textwidth}{%
    \begin{align*}
    & \;
    \walA{100:\tokT[0]} \mid
    \walB{50:\tokT[1]} \mid 
    \lp{0:\tokT[0],0:\tokT[1]}{} \mid
    \price = \setenum{\tokT[0] \mapsto 1, \tokT[1] \mapsto 1}
    \\
    \xrightarrow{1.\:\actDeposit{\pmvA}{50:\tokT[0]}}
    & \;
    \walA{50:\tokT[0],50:\mintedT[0]} \mid
    \walB{50:\tokT[1]} \mid 
    \lp{50:\tokT[0]}{} \mid
    \cdots
    \\
    \xrightarrow{2.\:\actDeposit{\pmvB}{50:\tokT[1]}}
    & \;
    \walA{\cdots} \mid
    \walB{50:\mintedT[1]} \mid
    \lp{50:\tokT[0], 50:\tokT[1]}{} \mid
    \cdots
    \\
    \xrightarrow{3.\:\actBorrow{\pmvB}{30:\tokT[0]}}
    & \;
    \walA{\cdots} \mid
    \walB{30:\tokT[0],30:\debtT[0]} \mid
    \lp{20:\tokT[0], 50:\tokT[1]}{} \mid
    \cdots
    && (\Health{}{\pmvB} = 1.11)
    \\
    \xrightarrow{4.\:\accrueIntOp}
    & \;
    \walB{30:\tokT[0],33.6:\debtT[0]} \mid
    \lp{20:\tokT[0], 50:\tokT[1]}{} \mid
    \cdots
    && (\Health{}{\pmvB} = 0.99)
    \\
    \xrightarrow{5.\:\actRepay{\pmvB}{5:\tokT[0]}}
    & \;
    \walB{25:\tokT[0],28.6:\debtT[0]} \mid
    \lp{25:\tokT[0],50:\tokT[1]}{} \mid
    \cdots
    && (\Health{}{\pmvB} = 1.16)
    \\
    \xrightarrow{6.\:\pxOp{0.3:\tokT[0]}}
    & \;
    \cdots \mid
    \pricei = \setenum{\tokT[0] \mapsto 1.3, \tokT[1] \mapsto 1}
    && (\Health{}{\pmvB} = 0.89)
    \\
    \xrightarrow{7.\:\actLiquidate{\pmvA}{\pmvB}{11:\tokT[0]}{\mintedT[1]}}
    & \;
    \walA{39:\tokT[0],50:\mintedT[0],15.73:\mintedT[1]} \mid
    \walB{25:\tokT[0],34.27:\mintedT[1],17.6:\debtT[0]} \mid
    && (\Health{}{\pmvB} = 0.99)
    \\
    & \; 
    \lp{36:\tokT[0],50:\tokT[1]}{} \mid
    \pricei
    \\
    \xrightarrow{8.\:\actRedeem{\pmvA}{10:\mintedT[0]}}
    & \;
    \walA{49.72:\tokT[0],40:\mintedT[0],15.73:\mintedT[1]} \mid
    \walB{\cdots} \mid 
    \lp{25.28:\tokT[0],50:\tokT[1]}{} \mid
    \pricei
    && 
    \end{align*}}}%
  \caption{Interactions between two users and a lending pool.}
  \label{fig:lending-pool:ex1}
\end{figure}
\section{Structural properties of lending protocols}
\label{sec:struct-properties}

We establish in this section some structural properties of lending protocols, such as relevant invariants on their reachable states.
As usual, free variables in statements are meant to be universally quantified; furthermore, blockchain states in the hypotheses are always assumed to be reachable.
For simplicity, we will just write, for example, $\txT \neq \pxOp{}$ to mean that there exist no $\prIncr$ and $\tokT$ such that $\txT = \pxOp{\prIncr:\tokT}$, and similarly for other transaction types.

First, we establish that the transition system is deterministic. 
This follows directly from the fact that, given a blockchain state $\confG$ and a transaction $\txT$, there is at most one applicable rule.
Determinism is a key property for blockchains, since it ensures that all the
blockchain nodes can reconstruct the same state from a sequence of transactions. 

\begin{lem}[Determinism]
\label{lem:determinism}
If $\confG \xrightarrow{\txT} \confGi$ and
$\confG \xrightarrow{\txT} \confGii$,
then $\confGi = \confGii$.
\end{lem}

Lemma~\ref{lem:constant-token-supply}
establishes that the amount of any base token is preserved by state transitions.
The only exception is the \nrule{[Swp]} transition, which however does not ``morally'' break the invariant, since it represents the exchange of tokens between the user and an external service. 
By applying Lemma~\ref{lem:constant-token-supply} inductively, it follows that base tokens are preserved along arbitrary sequences of transitions (not containing swaps).

\begin{lem}[Preservation of base tokens]
  \label{lem:constant-token-supply}
  Let \mbox{$(\WalW,\LpL,\price) \xrightarrow{\ltsLabel} (\WalWi,\LpLi,\pricei)$} with $\txT \neq \swapOp$.
  Then, for all $\tokT$:
  \[
  \supplyWal[\WalW]{\tokT} + \LpL(\tokT)
  \; = \;
  \supplyWal[\WalWi]{\tokT} + \LpLi(\tokT)
  \]
\end{lem}

The following lemma gives a useful invariant on reachable LP states: if the LP has no credit tokens $\minted{\tokT}$, then it has neither reserves of $\tokT$ nor debit tokens $\debt{\tokT}$.   
Another invariant relating base, credit and debit tokens will be established later in~\eqref{eq:credit-leq-base-plus-debit}.

\begin{lem}
\label{lem:cred0-imp-debt0}
If $\supply[\LpL]{\tokT} = 0$, then 
$\LpL(\tokT) = 0 = \supplyDebt[\LpL]{\tokT}$.
\end{lem}

The exchange rate of any base token type $\tokT$ is preserved by all state transitions, except interest accruals and, in the case there are no debts in $\tokT$, a redeem that reclaims the entirety of the credits (bringing back the exchange rate to $1$). 
When interests accrue, the exchange rate of $\tokT$
strictly increases whenever users have loans in $\tokT$. 
By~\Cref{eq:coll}, this guarantees that the credit token $\minted{\tokT}$
will gain value (whenever the price of the underlying base token $\tokT$ is not decreased by a price update transition). 

\begin{lem}[Monotonicity of exchange rate]
\label{lem:ER-increasing}
Let \mbox{$(\WalW,\LpL,\price) \xrightarrow{\ltsLabel} (\WalWi,\LpLi,\pricei)$}.
Then, for all $\tokT$:
\begin{enumerate}[(a)]

\item if $\ltsLabel = \accrueIntOp$  
and $\supplyDebt[\LpL]{\tokT} > 0$,
then 
\[
\X[\LpLi]{\tokT}
\; = \;
\X[\LpL]{\tokT}
+   
\dfrac
{\supplyDebt[\LpL]{\tokT}}
{\supply[\LpL]{\tokT}} \cdot \Intr{\LpL}(\tokT) 
\; > \;
\X[\LpL]{\tokT}
\]

\item if $\ltsLabel = \redeemOp$, and $\supply[\LpLi]{\tokT} = 0$, then 
$\X[\LpLi]{\tokT} = 1$ 

\item otherwise, 
$\X[\LpLi]{\tokT} = \X[\LpL]{\tokT}$.

\end{enumerate}
\end{lem}




By~\eqref{eq:LP:ER}, in initial blockchain states the exchange rate of each token is $1$.
Therefore, from the previous lemma it follows that in any reachable state the exchange rate of any token is always greater than or equal to $1$.
This is formalised by the following:

\begin{cor} 
  \label{lem:ER:geq1}
  $\X[\LpL]{\tokT} \geq 1$.
\end{cor}

Together with \eqref{eq:LP:ER}, this corollary gives an upper bound  to the supply of credit tokens in each reachable LP state.
More precisely, the supply of $\minted{\tokT}$ is bounded by the amount of reserves of $\tokT$ in the LP, plus the overall debt on $\tokT$: 
\begin{equation}
\label{eq:credit-leq-base-plus-debit}
\supply[\LpL]{\tokT} 
\; \leq \;
\LpL(\tokT)
+ 
\supplyDebt[\LpL]{\tokT} 
\end{equation}

Note that, in the specific case where 
$\LpL(\tokT) = 0 = \supplyDebt[\LpL]{\tokT}$,  
Equation~\eqref{eq:credit-leq-base-plus-debit} shows that 
also the inverse of Lemma~\ref{lem:cred0-imp-debt0} holds, \ie under that hypothesis, we have that $\supply[\LpL]{\tokT} = 0$.

\medskip
The following theorem establishes that the total net worth of all users remains constant throughout executions, except possibly when token prices are updated.

\begin{thm}[Preservation of net worth]
\label{th:net-worth-preservation}
For all
$\confG \xrightarrow{\txT} \confGi$
such that $\txT \neq \exchUpdateOp{}$:
\[
\sum_{\pmvC\in \PmvU} \wealth{\confGi}{\pmvC}
=
\sum_{\pmvC\in \PmvU} \wealth{\confG}{\pmvC}
\]
\end{thm}



\

\section{Economic analysis of single transactions}
\label{sec:econ-properties}

In this section we exhaustively analyze how each action affects the net worth of  users and their health factor. 
We will see that, among \emph{user actions} (\ie, all the actions except $\accrueIntOp$ and $\pxOp{}$), the only action that can change the net worth of a user is the liquidation (positively if fired by the user, negatively if suffered by the user) --- Lemma~\ref{lem:gain-base}.
Environment actions such as interest accruals and price updates, on the other hand, affect the net worth of users exposed to the relevant tokens --- that is, users holding debt or credit tokens whose price or interest is modified by these actions. 
In particular, $\accrueIntOp$ always penalizes \emph{debtors} (users who have debts), and benefits \emph{creditors} (users who hold credit tokens)  --- Lemma \ref{lem:gain-int}. 
The action $\pxOp{}$, on the contrary, benefits debtors and penalizes creditors if the price goes down, while it penalizes debtors and benefits creditors if the price goes up --- Lemma~\ref{lem:gain-px}.

Besides maximizing gains, users are also compelled with reducing the risk of incurring in losses. In particular, a user should avoid being the subject of a liquidation. The risk of being liquidated depends on the health factor: indeed, to avoid liquidations, one's health factor should not fall below $1$. 
We will prove that deposits, repayments and liquidations increase the health factor of the user who fires them, while borrows and redeems decrease it --- Lemma~\ref{lem:health-tx}.
In particular, we will see that, while in general repayments improve the health factor more than deposits, for users that are severely indebted it is better to deposit rather than repay --- Lemma \ref{lem:health-dep-rep}.

%
%


\subsection{Effect of transactions on net worth}

In order to study the economic impact of actions on users' net worth, we first define the \keyterm{gain} of an address $\pmvA$ upon firing a sequence of transactions $\TxTS$ from a state $\confG$:
\begin{equation} \label{eq:gainSt}
    \gainSt{\pmvA}{\confG}{\TxTS} 
    \; = \; 
    \wealth{\confGi}{\pmvA}
    -
    \wealth{\confG}{\pmvA}
    \qquad
    \text{if }
    \confG \xrightarrow{\TxTS} \confGi
\end{equation}

\noindent
Note that the gain is well-defined, because  if $\confG \xrightarrow{\TxTS} \confGi$ and $\confG \xrightarrow{\TxTS} \confGii$, then by determinism (Lemma~\ref{lem:determinism}) we must have $\confGi = \confGii$.   
When $\gainSt{\pmvA}{\confG}{\TxTS} < 0$, we will use the term \emph{loss} to denote the value $|\gainSt{\pmvA}{\confG}{\TxTS}|$.
Note that  $\TxTS$ must not necessarily be performed by $\pmvA$, as it may include other users' actions, or environment actions. 

Note that when some of the transactions in $\TxTS$ are not enabled in $\confG$, then the gain $\gainSt{\pmvA}{\confG}{\TxTS}$ is not {well-defined}. In this case, with a slight abuse of notation, we will write  $\gainSt{\pmvA}{\confG}{\TxTS}$ to mean  $\gainSt{\pmvA}{\confG}{\TxTiS}$, where $\TxTiS$ is the sequence of transactions obtained from $\TxTS$ by removing all the non-enabled transactions.

The following lemma shows that the only user action (\ie all actions except $\accrueIntOp$ or $\pxOp{}$) that can change the net worth of users is $\liquidateOp$. 
Such action increases the net worth of the liquidator and correspondingly decreases that of the liquidated address.
In particular, the gain of the liquidator is proportional to the amount liquidated, and it coincides with the loss of the user being liquidated.

\begin{lem}[Gain from user actions]
\label{lem:gain-base}
    Let $\txT$ be enabled in $\confG$, 
    with $\txT \not\in \setenum{\accrueIntOp,\pxOp{}}$. 
    Then:
    \begin{enumerate}
    \item $\gainSt{\pmvA}{\confG}{\txT} = 0$ $\iff$ $\txT$ is not a liquidation involving $\pmvA$.
    \item $\gainSt{\pmvA}{\confG}{\txT} > 0$ $\iff$ 
    $\txT$ is a liquidation performed by $\pmvA$;
    \item $\gainSt{\pmvA}{\confG}{\txT} < 0$ $\iff$ $\txT$ is a liquidation suffered by $\pmvA$;
    \end{enumerate}
    In particular, if $\txT = \actLiquidate{\pmvA}{\pmvB}{\valV:\tokT[0]}{\mintedT[1]}$, we have that:
    \begin{enumerate}
    
    \item[(4)] \(
    \gainSt{\pmvA}{\confG}{\txT} 
    = - \gainSt{\pmvB}{\confG}{\txT} 
    = \valV \cdot \price[{\tokT[0]}] \cdot \rLiq
    \).
    \end{enumerate}
\end{lem}





Note that the gain of $\pmvA$ only depends on the value $\valV \cdot \price[{\tokT[0]}]$ being liquidated. This implies that the optimal strategy for a \emph{non-strategic} user $\pmvA$ --- \ie, one who just wants to maximize their istantaneous gain --- is to liquidate as much as possible, regardless of the users being liquidated and of the token types of the collateral received.

The following lemma shows how a user $\pmvA$ benefits from (or gets damaged by) a price update of a token $\tokT$. 
Specifically, the gain of $\pmvA$ is given by the product between her wealth restricted to $\tokT$ and the ratio between the price variation and the old price of $\tokT$. 
Recalling from \eqref{eq:wealth-restricted} the definition of restricted wealth, we see that this gain is proportional to the amount of base tokens owned, plus the credits (adjusted by the exchange rate), and minus the debts, all multiplied by the price variation. 


\begin{lem}[Gain from price updates]
\label{lem:gain-px}
The gain of $\pmvA$ upon a transaction $\exchUpdateOp{\prIncr:\tokT}$ in $\confG = (\WalW,\LpL,\price)$ is given by:
\begin{align*}
\gain{\pmvA}{\confG}{\exchUpdateOp{\prIncr:\tokT}}
& \;\; = \;\;
\projTok{\wealth{\confG}{\pmvA}} \cdot \frac{\prIncr}{\price(\tokT)}
\;\; = \;\;
    \left(
    \WalW(\tokT,\pmvA) 
    +     
    \LpL(\minted{\tokT},\pmvA) \cdot \X[{\LpL}]{\tokT}
    -
    \LpL(\debt{\tokT},\pmvA)  
    \right)                  
    \cdot 
    \prIncr
\end{align*}
\end{lem}

Note that, when the price decreases (\ie $\prIncr<0$), the proportionality is of opposite sign,
\ie a user with lots of debts in $\tokT$ would benefit from the price decrease, while a user with lots of credits or base tokens would suffer losses.

The following lemma quantifies the effect of interest accruals on the users' gain.
For each address $\pmvA$ and token $\tokT$, the gain of $\pmvA$ is proportional to the interest accrued, to the price of $\tokT$, and to the difference between the user credits in $\tokT$ (weighted by the ratio between the supply of debts and credits in $\tokT$) and the user debts in $\tokT$.


\begin{lem}[Gain from interest accruals]
\label{lem:gain-int}
The gain of $\pmvA$ upon a transaction $\accrueIntOp$ in $\confG = (\WalW,\LpL,\price)$ is given by:
\[
\gainSt{\pmvA}{\confG}{\accrueIntOp}
\; =
\sum_{\supply[\LpL]{\tokT} > 0}
\left(
\dfrac
{\stateCredit{\tokT}{\pmvA}{}}
{\supply[{\LpL}]{\tokT}}
\cdot 
{\supplyDebt[\LpL]{\tokT}}
- 
\LpL(\debt{\tokT},\pmvA)
\right)
\cdot 
\Intr{\LpL}(\tokT)
\cdot \price[\tokT]	
\]
\end{lem}

We additionally observe that the overall gain of $\pmvA$ is given by the summation of the pointwise gains, as per~\eqref{eq:sum-wealth-restricted}.
From that, we infer that $\pmvA$'s gain \emph{restricted} to a given token type $\tokT$ is positive if and only if 
$\pmvA$ has credits in $\tokT$ and the ratio of $\pmvA$'s credits in $\tokT$ over the total credits of $\tokT$ exceeds the ratio of $\pmvA$'s debts over the total debit in $\tokT$, \ie:
\[
\projTok[\tokT]{{\gainSt{\pmvA}{\confG}{\accrueIntOp}}}
\; > \; 
0
\quad \iff \quad
\dfrac
{\stateCredit{\tokT}{\pmvA}{}}
{\supply[{\LpL}]{\tokT}}
\; > 
\dfrac
{\LpL(\debt{\tokT},\pmvA)}
{\supplyDebt[{\LpL}]{\tokT}}
\]
In particular, this implies that pure creditors always have a gain from interest accruals.

%

\subsection{Effect of transactions on health factor}

We now study how user actions impact the health factor.
The following lemma shows that deposits, repayments and liquidations increase users' health factor, while borrows and redeems decrease it.

\begin{lem}[Health factor from user actions]
\label{lem:health-tx}
Let $\confG \xrightarrow{\txT} \confGi$, 
with $\txT = \pmvA:\ell(\cdots)$.
Then:
\begin{enumerate}

\item \label{lem:health-tx:dep-rep-liq}
$\ell \in \setenum{\depositOp,\repayOp, \liquidateOp} \;\implies
\Health{\confGi}{\pmvA} \geq \Health{\confG}{\pmvA}$

\item \label{lem:health-tx:bor-rdm}
$\ell \in \setenum{\borrowOp,\redeemOp} \;\implies
\Health{\confGi}{\pmvA} \leq \Health{\confG}{\pmvA}$ 

\item \label{lem:health-tx:swp}
$\ell \in \setenum{\swapOp} \;\implies
\Health{\confGi}{\pmvA} = \Health{\confG}{\pmvA}$ 
\end{enumerate}
Moreover, the inequalities in \eqref{lem:health-tx:dep-rep-liq} and \eqref{lem:health-tx:bor-rdm} are strict if and only if $\debtval[\confG]{\pmvA}>0$.
\end{lem}

We have shown how actions performed by a user impact her health factor. 
It remains to study the effect of transactions that are not performed by the user, \ie price updates, interest accruals, and liquidation suffered. 

Price updates and interest accruals, depending on the state, can arbitrarily  increase and decrease both the credits and the debts,
hence it is quite clear that the health factor after these transactions can either increase or decrease.
 
For liquidations suffered, however, the ratio between the value of the liquidated debts and that of the seized credits is fixed, given by $\liqReward$, hence it is not that straightforward to conclude whether the health factor of the liquidated user always increases, always decreases, or can either increase or decrease.
%
While the previous lemma showed that the health factor of liquidators always increases, here we show that that of the liquidated address may either increase or decrease.
It is not difficult to quantify the variation in the health factor of a borrower $\pmvB$ who is suffering a liquidation, even though this is not particularly insightful.
For a liquidation $\actLiquidate{\pmvA}{\pmvB}{\valV:\tokT[0]}{\mintedT[1]}$
fired in $\confG$, the difference between the new and the old health factor of $\pmvB$ is given by: 
\[
\Health{\confGi}{\pmvB} - \Health{\confG}{\pmvB}  
\; = \; 
    \frac{
		\Big(
			\creditval[\confG]{\pmvB} 
			- \debtval[\confG]{\pmvB}
			\cdot
			 \nicefrac{\liqReward}{\X[\confG]{\tokT[0]}}
		\Big)
            \cdot
   		\valV \cdot \price(\tokT[0])
		\cdot
		\liqThreshold
		}
		{
				\debtval[\confG]{\pmvB}
				\cdot
				(\debtval[\confG]{\pmvB} - \valV \cdot
				\price[{\tokT[0]}])
		}
\]

The following example shows concrete cases where the health factor of the borrower increases or decreases upon a liquidation. 
\begin{exa}
\label{ex:hf-borrower}
Consider a lending protocol with parameters 
$\liqThreshold = 2/3$, $\liqReward = 1.3$, and an utility-based interest rate function with $\alpha = 0$ and $\beta = 1/2$.
Let $\confG$ be an initial blockchain state where the following sequences of transactions are enabled:
\begin{align*}
    \TxTS 
    & = 
    \actDeposit{\pmvB}{50:\tokT} \;\;
    \actBorrow{\pmvB}{30:\tokT} \;\;
    \accrueIntOp \;\;
    \actLiquidate{\pmvA}{10:\tokT}{\pmvB}{\tokT}
    \\
    \TxYS
    & = 
    \actDeposit{\pmvA}{90:\tokT} \;\;
    \TxTS
\end{align*}
In $\TxTS$, the health factor of the borrower $\pmvB$ increases from $0.96$
to $0.99$ with the liquidation while in $\TxYS$ is decreases from $0.82$ to $0.80$.
This is because $\pmvA$'s deposit (which is not present in $\TxTS$) has affected the exchange rate of $\tokT$ after the interest accrual: in $\TxTS$, such exchange rate is $1.3$, while in $\TxYS$ it is $1.1$. 
This decrease in the exchange rate, caused by the reduced impact of $\pmvB$'s debts on the ratio in~\eqref{eq:LP:ER}, makes the value of $\pmvB$'s credits decrease compared to $\TxTS$.
Since the value of $\pmvB$'s debts is preserved, this explains the decrease in $\pmvB$'s health factor.
See: \url{https://github.com/bitbart/lp-model/tree/main/examples-lmcs}.
\hfill\qedex
\end{exa}
%

\medskip\noindent
A user at risk of liquidation should try to immediately improve her health factor in order to avoid the losses coming from being liquidated. 
The following lemma compares the improvements that repays and deposits bring to the health factor.

\begin{lem}[Health factor: deposit \emph{vs.} repay]
\label{lem:health-dep-rep}
Let
$\confG \xrightarrow{\actDeposit{\pmvA}{\valV : \tokT}} \confG[\it dep]$
and
$\confG \xrightarrow{\actRepay{\pmvA}{\valV : \tokT}} \confG[\it rep]$.
Then:
\[
\Health{\confG[\it rep]}{\pmvA} \geq \Health{\confG[\it dep]}{\pmvA} 
\quad \iff \quad
\valV \cdot \price[\tokT] 
\geq 
{\debtval[{\confG}]{\pmvA}}
-
{\creditval[{\confG}]{\pmvA}}
\]
\end{lem}

From the previous lemma we see that repayments increase the health factor more than deposits if and only if the value transferred to the LP is greater than the difference between the value of debts of the user and the value of credit tokens held.
In practice, this means that, for users with a positive net position --- 
\ie, when ${\debtval[{\confG}]{\pmvA}} < {\creditval[{\confG}]{\pmvA}}$ ---
it is more beneficial to repay instead of deposit.
Instead, for users with negative net position, it could be more convenient to deposit, especially when the transferred value is small. 
This contradicts the statement contained in the Aave FAQs%
\footnote{\url{https://web.archive.org/web/20240914031752/https://docs.aave.com/faq/liquidations}}
for which ``\emph{By default, repayments increase your health factor more than deposits}''. 

\section{Economic analysis of strategic players}
\label{sec:strategic-properties}

All results in the previous section pertain to actions that have an immediate impact --- either positive or negative --- on a user. 
In contrast, in this section we consider more complex scenarios in which a user foresees a future event (\eg $\accrueIntOp$, $\pxOp{}$, or a liquidation against them).
We study the \emph{strategic} dimension of lending protocols: in particular, which actions should the user fire before the foreseen action takes effect in order to improve their net worth?
Conversely, if the user intends to execute a specific action, would it be better to fire it \emph{before} or \emph{after} the foreseen event? 

We start by considering a game where a borrower $\pmvA$ foresees that she is going to be liquidated.
In order to avoid the liquidation, the only possibly helpful actions are those which increase $\pmvA$'s health factor --- by Lemma~\ref{lem:health-tx} --- $\depositOp$, $\repayOp$ and $\liquidateOp$. 
We already know from Lemma~\ref{lem:gain-base} that liquidations performed by $\pmvA$ increase her gain, so we limit our analysis to $\depositOp$ and $\repayOp$. 
Theorem~\ref{th:gameLiq} shows that these actions are only helpful if they disable the liquidation fired against $\pmvA$; otherwise, they do not have any effect. 

\begin{thm}[Strategy for impending liquidations]
\label{th:gameLiq}	
Let $\pmvA$ and $\confG = (\WalW,\LpL,\price)$ be such that $\Health{\confG}{\pmvA}<1$, and let $\liquidateOp$ be a shorthand for an arbitrary liquidation on $\pmvA$ enabled in $\confG$.

Let $\txT = \pmvA:\ell(\valV: \tokT)$ 
with $\ell \in \setenum{\depositOp,\repayOp}$, 
and let $\confG \xrightarrow{\txT} \confGi$.
Then:
\begin{enumerate}
 
\item If $\ell = \depositOp$, then $\gain{\pmvA}{\confG}{\txT \, \liquidateOp} 
 \; > \; 
 \gain{\pmvA}{\confG}{\liquidateOp}
 \iff
 \valV 
 \geq 
 \frac{\X[\LpL]{\tokT}}{\price(\tokT)}
 \cdot \left(
 \dfrac
 {\debtval[\LpL,\price]{\pmvA}}
 {\liqThreshold}
 -
 \creditval[\LpL,\price]{\pmvA}
 \right)
 $
 
\item If $\ell = \repayOp$, then $\gain{\pmvA}{\confG}{\txT \, \liquidateOp} 
 \; > \; 
 \gain{\pmvA}{\confG}{\liquidateOp}
 \iff
 \valV 
 	\geq 
 	\frac{1}{\price(\tokT)}
 	\cdot \left(
 	{\debtval[\LpL,\price]{\pmvA}}
 	-
 	\creditval[\LpL,\price]{\pmvA}
 	\cdot
 	{\liqThreshold}
 	\right)
 $

\end{enumerate}

\end{thm}


The theorem also shows that, if $\pmvA$ wants to minimize the parameter $\valV$,
then she should choose $\depositOp$ if and only if ${\debtval[{\confG}]{\pmvA}}
\cdot 
\left(
\nicefrac{\X[\confG]{\tokT}}{\liqThreshold}
-1
\right)
\geq
\creditval[\confG]{\pmvA}
\cdot 
\left(
\X[\confG]{\tokT}-{\liqThreshold}
\right)
$, 
and choose
$\repayOp$ otherwise.

Note that $\pmvA$ cannot fire valid $\borrowOp$ and $\redeemOp$ transactions in $\Gamma$, since, by hypothesis, $\pmvA$ can be subject to liquidation, and so $\Health{\Gamma}{\pmvA} <1$.

\medskip
The following theorem shows how a user can take advantage of an incoming price update.
It turns out that the only effective action is front-running the price update with a swap of the token affected by the price update.
The rational strategy is to sell the token when its price is going to decrease, and to buy it otherwise.    

\begin{thm}[Strategy for impending price updates]
\label{th:gamePx}
Let $\txT = \pmvA:\ell(\cdots)$ mentioning token $\tokT$,
let $\confG \xrightarrow{\txT} \confGi$,
and let $\pxOp{}$ be a shorthand for $\exchUpdateOp{\prIncr:\tokT}$.
We have that:
\begin{equation*}
\gain{\pmvA}{\confG}{\txT \, \exchUpdateOp{}} 
\; \circ \; 
\gain{\pmvA}{\confG}{\exchUpdateOp{}} 
\; = \; 
\gain{\pmvA}{\confG}{\exchUpdateOp{} \, \txT}
\end{equation*}
where the relation $\circ$ is given by:
\[
\circ
\; = \;
\begin{cases}
=
& \text{if $\ell \in \setenum{\depositOp, \repayOp, \borrowOp, \redeemOp}$}
\\
\,> & \text{if ($\prIncr > 0$ and $\ell = \swapOp(\valV: \tokTi, \tokT)$) or ($\prIncr < 0$ and $\ell = \swapOp(\valV: \tokT, \tokTi)$) }
\\
\,< & \text{if ($\prIncr < 0$ and $\ell = \swapOp(\valV: \tokTi, \tokT)$) or ($\prIncr > 0$ and $\ell = \swapOp(\valV: \tokT, \tokTi)$) }
\end{cases}
\]
More precisely, if $\confG$ has price function $\price$, then:
\[
\gain{\pmvA}{\confG}{\txT \, \exchUpdateOp{}} 
\; = \;
\gain{\pmvA}{\confG}{\exchUpdateOp{}} +    
    \sigma \cdot
    \valV \cdot \prIncr \cdot 
    \left(
    \frac{\price(\tokTi)}{\price(\tokT)}
    \right)^{\sigma}
    \qquad
    \sigma
        \; = \; 
    \begin{cases}
    1
    & \text{$\ell = \swapOp(\valV: \tokTi, \tokT)$}
    \\
    -1
    & \text{$\ell = \swapOp(\valV: \tokT, \tokTi)$}
\end{cases}
\]
\end{thm}

Notice that in Theorem~\ref{th:gamePx} we have not included the case $\ell = \liquidateOp$. Indeed, by Lemma~\ref{lem:gain-base} we already know that $\liquidateOp$ yields a positive gain, but only because of the liquidation reward. It is irrelevant to perform it before or after the price update (\ie $\gain{\pmvA}{\confG}{\liquidateOp \, \exchUpdateOp{}} = \gain{\pmvA}{\confG}{\exchUpdateOp{} \, \liquidateOp}$).


\medskip
The previous theorem shows that, anticipating an increase in the price of  $\tokT$, the only single actions that a user $\pmvA$ can perform to improve her net worth is to sell another token $\tokTi$ to buy $\tokT$. 
%
%
But what if $\pmvA$ is already  fully exposed on $\tokT$ (\ie she possesses only tokens in $\tokT$)?

In traditional finance, traders can increase their exposition to a given asset via \emph{financial options}: the trader buys a contract from a issuer, acquiring the right to buy from the issuer the underlying asset for a fixed  \emph{strike price} in the future. 
Hence, if in the future the market price of the asset exceeds the strike price,  the trader can make a profit by buying the discounted asset from the issuer and resell it at market price.
Some analogies between lending protocols and financial options has been investigated in~\cite{LPOptions2,LPOptions1}.

The following lemma shows how $\pmvA$ can exploit the lending protocol to increase her exposure to token $\tokT$ and
benefit from a foreseen increase in the price of $\tokT$,
analogously what buyers do with financial options. 

\vbox{%
\begin{lem}[Strategy for impending price updates, II]
\label{lem:gameOptions}
Let:
\[
\TxTS[] 
\; = \;
\pmvA:\depositOp(\valV:\tokT)
\;\;
\pmvA:\borrowOp(\valVi:\tokTi)
\;\;
\pmvA:\swapOp(\valVi:\tokTi, \tokT)
\]
be enabled in $\confG$, and
let $\pxOp{}$ be a shorthand for $\pxOp{\prIncr:\tokT}$ with $\prIncr > 0$.
We have that:
\[
\gain{\pmvA}{\confG}{\TxTS[] \exchUpdateOp{}} 
\; > \;
\gain{\pmvA}{\confG}{ \exchUpdateOp{}} 
\]
\end{lem}
}



%
\medskip
The following theorem considers a game in which $\pmvA$ foresees an impending interest accrual.
If we consider an arbitrary interest rate function, then there is no single action that $\pmvA$ can fire before the interest accrual that is guaranteed to benefit her.
Even if we limit to linear utility interest rate functions, \ie 
$\Intr{\LpL}(\tokT) = \alpha \cdot \Util[\LpL]{\tokT}  + \beta$, as
in~\eqref{eq:intr-util}, then $\pmvA$'s strategy is not straightforward:
depending on the state of the lending pool and on the parameters $\alpha$ and $\beta$,  firing a given action before the interest accrual may be beneficial or detrimental for $\pmvA$. 
Indeed, we have that:
\begin{enumerate}[label=(\roman*)]

\item deposits increase $\pmvA$'s credit (which is going to appreciate after the interest accrual), but decrease the utilization (implying that the credit previously held by $\pmvA$ is going to appreciate less);

\item borrows increase $\pmvA$'s debt (which is going to increase after the interest accrual), but increase the utilization (implying that the credit previously held by $\pmvA$ is going to appreciate more);

\item repayments, symmetrically to borrows, decrease $\pmvA$'s debt, and decrease the utilization; 

\item redeems, symmetrically to deposits, decrease $\pmvA$'s credit but increase the utilization;

\item liquidations behave similarly to deposits, with the only difference that the credits received are of a different token type than that of the deposited tokens, and of higher value; but this is not always enough to compensate for the lower credit appreciation.

\end{enumerate}
Note that swaps do not interact with the lending pool whatsoever, hence firing them before or after an interest accrual is not going to have any impact in any case.
In Section \ref{sec:attacks:utilization}, we will show how an adversary can manipulate the utilization in order to increase her gain.

In the special case in which $\alpha=0$, \ie interest rates are constant and do not depend on the utilization, we can conclude that certain actions are surely going to benefit $\pmvA$ (or, at most, have no impact), and other actions are surely going to penalize $\pmvA$ (or, at most, have no impact).
More specifically, deposits and repayments are beneficial, while borrows and redeems are detrimental. 
For liquidations, even in this case, there is no monotonicity.

\begin{thm}[Strategy for impending interest accruals]
\label{th:X-int-vs-X}
Assume that the lending protocol uses the linear utility interest rate function 
$\Intr{\LpL}(\tokT) = \alpha \cdot \Util[\LpL]{\tokT}  + \beta$ 
in~\eqref{eq:intr-util}. 
Let $\txT = \pmvA:\ell(\cdots)$ mentioning token $\tokT$ with transaction parameter $\valV$.
%
\begin{enumerate}

\item If the parameters $\alpha$ and $\beta$ are arbitrary, 
then for every $\ell \in \setenum{\depositOp,\borrowOp,\repayOp,\redeemOp,\liquidateOp}$ and for every $\circ \in \{>,  =, <\}$, there exists $\confG$ and 
$\valV$ such that 
$\gain{\pmvA}{\confG}{\txT \, \accrueIntOp} \circ \gain{\pmvA}{\confG}{\accrueIntOp}$.
	
\item If $\alpha = 0$, then:
\begin{enumerate}

\item if $\ell \in \setenum{\depositOp,\repayOp}$, then
for all $\confG$ and $\valV$,
$\gain{\pmvA}{\confG}{\txT \, \accrueIntOp} \geq \gain{\pmvA}{\confG}{\accrueIntOp}$
	
\item if $\ell \in \setenum{\borrowOp, \redeemOp}$, then
for all $\confG$ and $\valV$,
$\gain{\pmvA}{\confG}{\txT \, \accrueIntOp} \leq \gain{\pmvA}{\confG}{\accrueIntOp}$

\item if $\ell = \liquidateOp$, then for all $\circ \in \{\geq, \leq\}$, there exist $\confG,\valV$ such that $\gain{\pmvA}{\confG}{\txT \, \accrueIntOp} \circ \gain{\pmvA}{\confG}{\accrueIntOp}$


\end{enumerate}

\end{enumerate}


\end{thm}

Note that $\liquidateOp$ does not enjoy any monotonicity, not even in the case where $\alpha = 0$. Indeed, in certain situations it is beneficial to perform a liquidation before interests accrue, while in others, it is better to wait until after the interest accrual. 
This is due to the fact that higher overall amounts of debts imply higher exchange rates increases after $\accrueIntOp$. In particular, given that credit tokens appreciate after an increase in exchange rates, a user who holds a high amount of credit tokens would benefit more from waiting interest rates to increase.

\begin{exa}[Liquidations and interest accruals]
Recall the sequence of transactions in Fig.~\ref{fig:lending-pool:ex1} up to step 6 included. 
Assume that $\pmvA$ anticipates that an $\accrueIntOp$ is going to happen,
and she has to decide whether to liquidate $\pmvB$ before the interests accrual or not. 
The convenience of liquidating $\pmvB$ or not depends on the specific interest rate function. 
Consider \eg, the utility-based interest rate function in~\eqref{eq:intr-util} in the simple case where $\alpha = 0$, that is, there is a constant interest rate $\beta$ for each token type.
If the interest increase is relatively low (\eg, $\beta = 10\%$), 
then the credit tokens held by $\pmvA$ do not appreciate significantly, and so the liquidation reward is sufficiently high to incentivize $\pmvA$ to liquidate $\pmvB$. 
However, if interest rates increase significantly (\eg $\beta = 100\%$) 
or the liquidation bonus is very small,
then the appreciation of the credit tokens held by $\pmvA$ can be so impactful that the best strategy for $\pmvA$ would be wait to liquidate $\pmvB$, 
so that the overall amount of debts will make the appreciation of the credit tokens higher enough to surpass the benefit given by the liquidation bonus.
\hfill\qedex
\end{exa}

\section{Attacks}
\label{sec:attacks}

In this section we illustrate some attacks to lending protocols, which only require the adversary to own sufficient liquidity of certain tokens. 

We start by considering
\emph{price manipulation attacks}, where an adversary uses their capital to trigger a temporary price fluctuation of a token handled by the lending pool (\Cref{sec:attacks:price-manipulation}). 
More specifically, Theorem~\ref{th:gameUndercollLoanAttck} shows an attack where the adversary's goal is to borrow more tokens than what they should be allowed to.
Theorem~\ref{th:gameLiqAttck} shows another attack where the adversary exploits a price manipulation to make a borrower under-collateralized and then liquidate her credit tokens.

Then, we consider \emph{utilization attacks}, where an adversary manipulates the utilization of certain tokens to benefit from a change in the interest accrual (\Cref{sec:attacks:utilization}). 
More specifically, Theorem~\ref{th:underutilizationAttack} shows an \emph{under-utilization} attack in which an adversary deposits some tokens to decrease the utilization in order to pay less interests on her debts (penalizing creditors).
Theorem~\ref{th:overutilizationAttack} shows an \emph{over-utilization attack} in which an adversary borrows some token to increase the utilization in order to gain more from the interest accrual (penalizing debtors).

Although these kinds of attacks are already known in literature  
\cite{Gudgeon20cvcbt,Qin21fc,BCL21wtsc,Mackinga22icbc,Zhou23sp,Arora24asiaccs}, our results are the first to formally establish general conditions under which they can occur. 

In our results, we will make some simplifying assumptions on the credits or debts of the addresses involved in the attack, \eg that the adversary has all the credits of a given token type, or none. 
This allows us to prove that the attacks always succeed, regardless of the actual token amounts invested by the adversary in the attack.
In practice, even when such conditions are not precisely met (\eg, the adversary does not possesses exactly \emph{all} the credit tokens, but most of them), the attack will still succeed for suitable choices of the transaction parameters. 
It is possible to estimate such suitable values by using the formulas to compute the gain provided in~\Cref{sec:econ-properties}.
For the sake of clarity, and to provide a better intuition on the attacks, we will consider the cases in which the hypotheses hold.

\subsection{Price manipulation attacks}
\label{sec:attacks:price-manipulation}

In the decentralized setting, price oracles are usually implemented as smart contracts that determine the price of a token depending on an underlying market on that token.
A typical implementation is given by constant-function Automated Market Makers (AMMs)~\cite{Angeris20aft}, which realize a market on two or more token types, allowing users to swap tokens at an algorithmically-determined exchange rate that depends solely on the offer and supply of the supported tokens.
For example, in the simple case of a constant-product AMM on two tokens $\tokT[0]$, $\tokT[1]$, swaps preserve the product of the reserves of the two tokens in the AMM. Accordingly, the price of $\tokT[0]$ \wrt $\tokT[1]$ is defined as the ratio between the reserves of $\tokT[1]$ and those of $\tokT[0]$ in the AMM. 
When a user swaps units of $\tokT[0]$, since the product of the reserves must remain constant, then after the swap the reserves of $\tokT[1]$ decrease, and so the price of $\tokT[0]$ will decrease as well.
This design, in principle, makes AMMs suitable as price oracles, as users have an economic incentive to perform tokens swap in order to align the AMM token prices to external prices~\cite{BCL22lmcs}.

In practice, relying on instantaneous AMM prices in a lending protocol can be insecure~\cite{werner2021sok}.
Indeed, an adversary with sufficient capital in a given token can induce significant price fluctuations of that token. This manipulated price can then be exploited in interactions with a lending protocol --- as we will demonstrate below --- before the adversary reverses the manipulation to restore the original price on the AMM.
Formally, we model such a price manipulation attack as a sequence of transactions where the adversary fires a transaction $\pxOp{\prIncr:\tokT}$ to manipulate the price, then perform a sequence of interactions with the LP, and finally restores the original price by firing $\pxOp{-\prIncr:\tokT}$. 
We remark that the proposer-builder separation scheme currently adopted by Ethereum~\cite{Heimbach23imc} makes is possible for an adversary to perform such transaction bundles atomically. 

The following theorem shows an attack in which an adversary $\pmvA$ manipulates price updates in order to borrow more tokens than what she should be allowed to.
Specifically, after the attack, although $\pmvA$'s gain remain constant, her net position becomes negative. 
This allows $\pmvA$ to extract value from the LP by effectively defaulting on debt that is no longer backed by a sufficient collateral.
Since the lending protocol cannot enforce repayments, the uncovered debt results in a loss for the pool.



\begin{thm}[Undercollateralized loan attack]
\label{th:gameUndercollLoanAttck}
Let $\confG =  (\WalW,\LpL,\statePrice{})$, and assume that $\pmvA$ has no credits or debts with the LP, \ie, ${\creditval[{\confG}]{\pmvA}=\debtval[\confG]{\pmvA}=0}$.
\noindent
Consider the following sequence of transactions:
\[
\TxTS 
\; = \;
\pmvA:\depositOp(\valV[1]:\tokT[1]) \;\; 
\pxOp{-\prIncr:\tokT[2]} \;\;
\pmvA:\borrowOp(\valV[2]:\tokT[2]) \;\;	
\pxOp{\prIncr:\tokT[2]}
\]
where 
$0  < \prIncr < \price(\tokT[2])$
and
$\valV[2] = \frac{\valV[1]}{\X[{\LpL}]{\tokT[1]}} \cdot \frac{\price[{\tokT[1]}]}{\price[{\tokT[2]}]-\prIncr} \cdot \liqThreshold$.
Let $\confG \xrightarrow{\TxTS} \confGi$.
Then:
\begin{enumerate}

\item $	\gain{\pmvA}{\confG}{\TxTS} = 0$ 

\item $\netCredit{\confGi}{\pmvA} < 0$ if and only if
\(
\prIncr
>
\price(\tokT[2])
\cdot
(1 - \liqThreshold)
\)
\end{enumerate}	
\end{thm}

The theorem gives a lower bound for the price increase $\prIncr$ under which the attack does not have effect (\ie the net position does not go negative). 
Note that, since $\prIncr$ can take values in $(0,\price(\tokT[2]))$ and $0<\liqThreshold<1$, then there always exists a value for $\prIncr$ large enough to make the attack succeed.

\medskip
The following theorem shows another attack, where the attacker $\pmvA$ manipulates prices in order to make another user $\pmvB$ under-collateralized and so make a gain from $\pmvB$'s liquidation.
In this attack, we assume that the collateral of $\pmvB$ relies on a single token type $\tokT[1]$ (hypothesis~\ref{hp:gameLiqAttck:single-token-collateral}), that $\pmvB$ has debts only in $\tokT[2]$ (hypothesis~\ref{hp:gameLiqAttck:single-token-debts}), that the adversary has some tokens $\tokT[2]$ (hypothesis~\ref{hp:gameLiqAttck:A-has-tokens})
and that $\pmvB$ is not liquidatable in the current state (hypothesis~\ref{hp:gameLiqAttck:B-cannot-liquidate}).


\begin{thm}[Liquidation attack]
\label{th:gameLiqAttck}
Let $\confG = (\WalW,\LpL,\price{})$, and let $\pmvA$ and $\pmvB$ be such that:
\begin{enumerate}

\item \label{hp:gameLiqAttck:single-token-collateral}
$\LpL(\mintedT[1],\pmvB) = \valV[c]$ 
and $\stateCredit{\tokT}{\pmvB} = 0$ for all $\tokT \neq \tokT[1]$

\item \label{hp:gameLiqAttck:single-token-debts}
$\LpL(\debtT[2],\pmvB) = \valV[d]$
and $\stateDebt{\tokT}{\pmvB} = 0$ for all $\tokT \neq \tokT[2]$ 

\item \label{hp:gameLiqAttck:A-has-tokens}
$\stateBase{\tokT[2]}{\pmvA} > 0$ 

\item \label{hp:gameLiqAttck:B-cannot-liquidate}
$\Health{\confG[]}{\pmvB} \geq 1$ 

\end{enumerate}

\noindent
Then, for every $\prIncr>0$ sufficiently small,
and for every $\valV[l]>0$ such that $\valV[l] \leq \stateBase{\tokT[2]}{\pmvA}$, $\valV[l] \leq \valV[d]$ and 
$\valV[l]<
\valV[c] 
\cdot
\frac{
	\X[\LpL]{\tokT[1]}
}
{\rLiq}
\cdot
\frac{\prIncr}{\price(\tokT[2])}$,
given the following sequence of transactions:
\[
\TxTS 
\; = \; 
\pxOp{(-\price(\tokT[1])+\prIncr):\tokT[1]} \;\;
\pmvA:\liquidateOp(\pmvB, \valV[l]: \tokT[2], \mintedT[1]) \;\;
\pxOp{(\price(\tokT[1])-\prIncr):\tokT[1]}
\]
we have that $\TxTS$ is enabled in $\confG$, and $\gainSt{\pmvA}{\confG}{\TxTS} > 0$.
\end{thm}

\subsection{Utilization attacks}
\label{sec:attacks:utilization}

Utilization, defined previously in~\eqref{eq:util}, gives an estimate of how much the reserves of a token are valuable, and are hence often used to determine interest rates: the higher the utilization, the higher the interest rate.
This, however, exposes lending protocols to attacks where the adversary manipulates the utilization function in order to increase or decrease the interest rates to their advantage~\cite{BCL21wtsc}.

We formalise in Theorem~\ref{th:underutilizationAttack} an \emph{under-utilization} attack, where an adversary deposits some tokens before an interest accrual in order to decrease the utilization of the token and pay less interests on her debts, and then immediately redeem her credits.
Dually, Theorem~\ref{th:overutilizationAttack} establishes conditions for an \emph{over-utilization} attack, where an adversary borrows some tokens before the interest accrual in order to increase the utilization of the token and hence increase the appreciation of her credit tokens, and then immediately repays her debt. 
In both attacks, the adversary is not fairly participating in the dynamic of the lending protocol, but rather manipulating its intended behavior to increase her net worth.

\medskip

The following theorem shows an under-utilization attack in which
an adversary $\pmvA$ manipulates the utilization of a token $\tokT$.
We assume that $\pmvA$ has no credits (but possibly debts) in $\tokT$. 
We also consider a user $\pmvB$ who has credits (but no debts) in $\tokT$.
The attack consists in 
$\pmvA$ firing a deposit immediately before an interest accrual, thus decreasing the utilization of $\tokT$, and then redeeming her credits.
This strategy benefits $\pmvA$ in two ways:
first, if $\pmvA$ has debts, then a lower utilization implies a lower interest rate, hence reducing the increase of $\pmvA$'s debt;
secondly, given that $\pmvA$ has now acquired credits in $\tokT$, these credits appreciate and can be redeemed for a higher value.
User $\pmvB$, on the contrary, gets penalized by $\pmvA$'s attack, since a lower utilization implies a lower appreciation of her credits.

\begin{thm}[Under-utilization attack]
\label{th:underutilizationAttack}
Assume that the LP uses the linear utility interest rate function in~\eqref{eq:intr-util} with $\alpha>0$. 
Let $\confG[] = (\WalW,\LpL,\price{})$, and let $\pmvA, \pmvB$ and $\tokT$ be such that:
\begin{enumerate}
\item $\stateCredit{\tokT}{\pmvA}=0$ 
\item $\stateCredit{\tokT}{\pmvB}>0$ and  $\stateDebt{\tokT}{\pmvB}=0$
\end{enumerate}
\noindent
Assume that the following sequence of transactions:
\[
\TxTS 
\; = \;
\pmvA:\depositOp{(\valV: \tokT)}
\;\;	
\accrueIntOp
\;\;
\pmvA:\redeemOp{(\minted{\valV}: \tokT)}
\]
is enabled in $\confG$, where $\minted{\valV}$ is the amount of credits held by $\pmvA$ in the intermediate state before $\redeemOp$.
%
Then, we have that:
\[
\gainSt{\pmvA}{\confG[]}{\TxTS} > \gainSt{\pmvA}{\confG[]}{\accrueIntOp} 
\qquad\qquad
\gainSt{\pmvB}{\confG[]}{\TxTS} < \gainSt{\pmvB}{\confG[]}{\accrueIntOp}
\]
\end{thm}
\noindent
If we drop the assumption that $\pmvA$ does not possess any credit, then the attack does not necessarily succeed for \emph{every} state of the lending pool  for \emph{every} value $\valV$,  since the deposit decreases the appreciation of the credits already possessed by $\pmvA$.
In such a case, there is a trade off between the lower appreciation of $\pmvA$'s credits, on one side, and the reduction in the increment of $\pmvA$'s debt, and the gains coming from the credit appreciation of the newly deposited tokens, 
on the other side.
In general, the attack succeeds when $\pmvA$ has plenty of debts in $\tokT$ but few credits.
Precise relations between these values are established in the proof of Theorem~\ref{th:X-int-vs-X}.

\medskip
The following theorem shows an over-utilization attack performed by an adversary $\pmvA$, which possesses \emph{all} the credits in $\tokT$ (but not all the debts in $\tokT$).
We also consider another user $\pmvB$,  which possesses debts in $\tokT$ (but not credit).
Here, the adversary $\pmvA$ borrows some tokens immediately before the interest accrual, thus increasing the utilization of the token $\tokT$, and then immediately repays the loan. 
This strategy benefits $\pmvA$ as a higher utilization implies a higher interest rate, and, since $\pmvA$ is the only credits possessor, then \emph{all} the overall debt increase in $\tokT$ correspond to an appreciation of $\pmvA$ credits.
Even though also the debts of $\pmvA$ increase, since $\pmvA$ is the only creditor, then this loss is matched by the gain coming from the appreciation of her credits.
User $\pmvB$, on the contrary, gets penalized by the attack, since now her debts have increase more than they would have in the absence of the attack.

\begin{thm}[Over-utilization attack]
\label{th:overutilizationAttack}
Assume that the LP uses the linear utility interest rate function in~\eqref{eq:intr-util} with $\alpha > 0$. 
Let $\confG[] = (\WalW,\LpL,\price{})$, and let  $\pmvA, \pmvB$ and $\tokT$ be such that:
\begin{enumerate}
\item $\stateCredit{\tokT}{\pmvA}=\supply[\LpL]{\tokT}$ and  $\stateDebt{\tokT}{\pmvA}
<
\supplyDebt[\LpL]{\tokT}$
\item $\stateDebt{\tokT}{\pmvB}>0$
\end{enumerate}
Assume that the following sequence of transactions:
\[
\TxTS 
\; = \;
\pmvA:\borrowOp{(\valV: \tokT)}
\;\;	
\accrueIntOp
\;\;
\pmvA:\repayOp{(\valV: \tokT)}
\]
is enabled in $\confG$.
Then, we have that:
\[
\gainSt{\pmvA}{\confG[]}{\TxTS} > \gainSt{\pmvA}{\confG[]}{\accrueIntOp} 
\qquad\qquad
\gainSt{\pmvB}{\confG[]}{\TxTS} < \gainSt{\pmvB}{\confG[]}{\accrueIntOp}
\]
\end{thm}

Similarly to the previous theorem, if we drop the assumption that $\pmvA$ possesses \emph{all} the credits in $\tokT$, then the attack is not guaranteed to succeed for \emph{every} state of the lending pool for \emph{every} value $\valV$. 
Indeed, there is now a trade off between the increase in $\pmvA$'s debt, and the increase in the appreciation of $\pmvA$'s credits.
In general, the attack succeeds when $\pmvA$ has plenty of credits but few debts in $\tokT$.


\section{Limitations}
\label{sec:limitations}

For simplicity, our model of lending protocols abstracts away from 
certain functionalities and fine-grained details of real-world lending platforms.
While these simplifications limit the direct applicability of our theory to existing systems, we believe that our results describe ideal properties that lending protocols should satisfy -- such as preservation of global net worth, incentives for liquidity providers and liquidators, \etc

We discuss here some of the key abstractions made in our model, using the Aave protocol implementation~\cite{aaveimpl} for reference. 

\begin{itemize}

\item \emph{Governance}.
In early implementations of Aave and Compond, administrators had the authority to set key economic parameters of the protocol, such as the interest rate function, liquidation threshold, and and liquidation reward. 
To mitigate the risk of administrators engaging in improper behavior, 
more recent versions of these platforms use \emph{governance tokens}, distributed among liquidity providers, to collectively govern and update protocol parameters.
In our model, we assume for simplicity that protocol parameters are fixed.

\item \emph{Deposits and collaterals}.
Our \nrule{[Dep]} rule allows any user to add a new token type to the LP 
by just performing the first deposit of tokens of that type. 
In contrast, adding a new token type to an Aave LP must be authorized by the governance mechanisms. 
Additionally, our \nrule{[Dep]} rule, together with the definition of user collateralization~\eqref{eq:coll}, ensures that every deposit is automatically enabled as collateral.
In contrast, Aave allows users to selectively disable specific deposited token types from being used as collateral.


\item \emph{Liquidations}.
Our \nrule{[Liq]} rule allows liquidators to receive credit tokens in exchange for the repaid debt; if liquidators want to convert the credit tokens in the underlying base tokens, they must send a redeem transaction after the liquidation.
In Aave, liquidators can also choose to receive the underlying base tokens directly.
Furthermore, \nrule{[Liq]} allows liquidators to repay any fraction of the debt, with the only constraint that the borrower's health factor after the liquidation does not exceed $1$.
In Aave, instead, liquidations can repay up-to a 50\% fraction 
of the debt~\cite{aavemaxliq}. 
Since this constraint can be easily bypassed by splitting a liquidation into  multiple actions, we have omitted it.
Another difference is that in our model the protocol parameters
$\liqThreshold$ (liquidation threshold) and $\rLiq$ (liquidation reward) are uniform across all token types.
In contrast, Aave allows these parameters to vary by token, enabling the protocol to fine-tune incentives in selected tokens.

\item \emph{Interest accrual}.
Our \nrule{[Int]} rule models the accrual of interest on loans as an update triggered by a privileged entity. This abstraction reflects the fact that the $\accrueIntOp$ action is not associated with any address, unlike other actions such as deposit or redeem. 
In lending protocol implementations, interest accruals are not executed on demand by privileged entities; instead, they occur automatically whenever a user performs an action that requires up-to-date debt amounts. To reduce execution costs, a single interest rate is applied over the entire period since the last accrual, which can introduce minor inaccuracies.

\item \emph{Price updates}.
Our \nrule{[Px]} rule models the update of a token price performed by a price oracle. 
In contrast, Aave associates each token type with its own price oracle.
This oracle is usually a smart contract that aggregates multiple independent sources, in order to mitigate the risk of price manipulation~\cite{Gansauer25caaw}.

\item \emph{Flash loans}.
Lending platforms typically expose a flash loan functionality, 
which allows users to borrow arbitrary amounts of tokens without a collateral, provided that the borrowed funds are repaid within the same transaction.
This transaction can bundle multiple actions performed by the same user: 
its atomicity guarantees that all operations --- \ie borrowing, using the borrowed tokens, and repaying the loan --- must either complete successfully or be entirely reverted~\cite{aaveflashloan}. 
Our model does not include flash loans, as they are usually meant to be used in combination with other protocols. 
Note that the possibility of obtaining large amounts of funds without providing a collateral is implicitly assumed in the attacks in~\Cref{sec:attacks}, where the adversary needs sufficient capital to obtain the desired manipulation of token prices or utilization. 

\item \emph{Fees}.
Aave requires users to pay fees for certain actions, such as borrowing and executing flash loans. These fees are accumulated in a reserve managed by the protocol's governance mechanisms and are intended to serve as a protection against unforeseen events.

\item \emph{Token amounts}.
In our model, we let token amounts range over the continuous domain of non-negative real numbers.
In contrast, real-world lending protocol implementations operate over a discrete domain, representing token amounts as fixed-size integers. 
As a result, all operations involving token amounts require rounding, which can introduce small inaccuracies and edge cases (and, possibly, attack vectors) abstracted away by our model.

\end{itemize}
\section{Related work}
\label{sec:related}

Even though lending protocols have been extensively studied in recent years, only a few works base their analysis on formal, operational models that capture the interactions in lending protocols at the granularity of individual transactions. 
The work most closely related to ours is~\cite{BCL21wtsc}, whose LP model served as a key inspiration for our model.
Although both models encompass the same types of transactions (except swaps, which are not present in~\cite{BCL21wtsc}), the representation of LP and blockchain states are quite different. 
While, similarly to process algebras, \cite{BCL21wtsc} renders states as parallel compositions of simple terms (\eg, an individual user wallet, a lending pool handling a single token type), in our model we gather all the components of the LP state (reserves, credit and debit maps) into a single function.
Besides that, credits and debits are represented asymmetrically in~\cite{BCL21wtsc}: namely, credits are associated to users' wallets, while debits to LP states.
These differences are not merely aesthetic, however, as they deeply impact the way states predicates are represented, and how states are updates.
Overall, our design choices lead to a substantially clearer LP semantics and to more succinct proofs than~\cite{BCL21wtsc}.
Another key difference lies in the comprehensiveness of the theoretical analysis. 
Compared to our work, the results in~\cite{BCL21wtsc} constitute a narrower subset: they include only simplified versions of 
exchange rate monotonicity (Lemma~\ref{lem:ER-increasing}),
net worth preservation (Theorem~\ref{th:net-worth-preservation}),
and gain from user actions (Lemma~\ref{lem:gain-base}).
Our theoretical framework provides refined versions of these results, along with a thorough analysis of the effects of each individual actions and  more complex strategies, supported by rigorous proofs for all statements.



\medskip
Several works have focused on specific features of lending protocols, such as interest rate functions, price stabilization mechanisms, liquidation strategies, and flash loans.

\paragraph{Interest rate functions}
The impact of interest rate functions on market liquidity and efficiency in lending protocols has been studied by~\cite{Gudgeon20aft}, which provides an empirical analysis of the interest rate models employed by various protocols. 
The two main platforms Aave and Compound rely on static interest rate curves, which often struggle to adapt to rapid market changes such as major price fluctuations.
Dynamic interest rates aiming at stabilizing utilization have been studied in various works~ \cite{econIntRate,DataDrivenAdaptIntRate,AgileRate,AdaptiveCurves,RiskAwareIntRate}.
However, it has been observed that, although dynamic interest rates provide better utilization levels, they also increase the exposure to manipulation attacks~\cite{AttacksDynamicIR,AgileRate}.

\paragraph{Prices}
While, in our model, prices are simply modelled as a function from token types to non-negative real numbers, the literature has explored sophisticated price models and mechanism to mitigate price volatility.
A taxonomy of various price stabilization mechanisms has been proposed in  \cite{moin2019classification}. 
The behaviour of lending protocols in times of high price volatility has been discussed in \cite{Gudgeon20cvcbt}. 
This work also uncovers a vulnerability in the governance design of MakerDAO that allowed attackers to utilize flash loans to steal funds from the contract.
The performance of MakerDAO's oracles has been studied empirically in \cite{gu2020empirical}, which also proposes alternate price feed aggregation models to improve oracle accuracy.
The profitability competition for user deposits between staking in proof-of-stake systems and lending protocols has been studied in~\cite{chitra2019competitive,chitra2020stake}: 
when lending is believed to be more profitable than staking, 
users may shift deposits away from the staking contract of the underlying consensus protocol towards lending pools, thereby endangering the security of the system.
The work~\cite{AMMImplicationsLP} studies the interaction between AMMs and LPs, analysing the impact of transaction costs, arbitrage opportunities, hedging of impermanent losses, and risk management.
Price manipulation attacks, such as those presented in~\Cref{sec:attacks:price-manipulation}, have been formally characterized in~\cite{BMZ24fc}
as instances of \emph{MEV interference} between a contract (the lending pool) and its dependencies (the AMM serving as a price oracle). 
This interference allows the adversary interacting with the AMM to extract more value from the LP than would be possible by interacting with the LP alone.

\paragraph{Liquidations}

A {liquidation model}
intended to simulate interactions between lending pool liquidations 
and token exchange markets in times of high price volatility has been studied in~\cite{Gudgeon20cvcbt}. 
%
The optimal bidding strategy for collateral liquidators in MakerDAO auctions has been studied in~\cite{darlin2020optimal}.
The work \cite{Gudgeon20cvcbt} analyzed what happens when large price drops make many accounts under-collateralized. 
A key observation of this work is that if liquidators sell off collateral at an external market for units of the repaid token type, the limited market demand for collateral tokens may prevent liquidations from being executed. 
In the Compound protocol,
liquidation efficiency  has been studied through historical data ~\cite{Perez21fc}, while  the evolution of {liquidatable} and {undercollateralized} debt has been studied in \cite{kaoanalysis}
and the risk of financial contagion triggering a cascade of defaults in \cite{CompoundDefaultsCascade}.
%
%
%
The impact of different liquidation strategies and protocol designs on the net position of borrowers has been studied in ~\cite{BCJLMV22isola}.
Strategies to liquidate under-collateralized borrowers, such as those 
studied in~\Cref{sec:strategic-properties}, have been formally characterized as instances of Maximal Extractable Value (MEV) in~\cite{BZ25fc}.
In particular, liquidations that do not exploit the knowledge of the mempool are classified by~\cite{BZ25fc} as a benign form of MEV.
This classification aligns with the broader community consensus
\cite{Barczentewicz23ssrn,Ji24fc,Torres24ccs}, which views such liquidations as a necessary incentive mechanism to keep lending protocols aligned with their intended functionality.

\paragraph{Flash loans}

An analysis of flash loans transactions in the main lending platforms has been conducted in \cite{wang2020towards}. 
%
Flash loans have been exploited in several attacks, as they enable attackers to have access to large amount of funds that can be used to initiate attacks
~\cite{valuedefi,harvestfin,origindefi,akryopolisdefi}. 
Attacks such as pump and arbitrage and price manipulation have been studied in ~\cite{Qin21fc}.
A framework for the automated synthesis of attacks that exploit flash loans has been proposed in ~\cite{FlashLoansAutSynt}.


\section{Conclusions}
\label{sec:conclusions}

We have presented a formal model and analysis of decentralized lending protocols, based on common features synthesised from mainstream lending platforms such as Aave and Compound~\cite{aave,comp}. 
Our theoretical investigation of lending protocols has provided answers to the following research questions:
\begin{enumerate}

\item \label{concl:RQ1}
What structural properties and invariants are enjoyed by lending pools?

\item \label{concl:RQ2}
What is the economic effect of each individual interaction with a lending pool?

\item \label{concl:RQ3}
Which strategies can be followed by rational users anticipating a forthcoming action? 

\item \label{concl:RQ4}
Which attacks are possible for adversaries with a large amount of capital? 

\end{enumerate}

Overall, our formal model proved to be sufficiently granular to precisely reproduce and analyse known attacks from the literature, and at the same time streamlined enough to allow for succinct proofs.
To the best of our knowledge, we are the first to have systematically studied the incentive mechanism of lending protocols at that level of granularity, providing a comprehensive analysis of user strategies.
Most notably, we focused on strategies for front-running of impending transactions.
In our analysis, we have observed that the dynamics of interest accruals are among the most complex aspects of the lending protocols, giving rise to different manipulation attacks and non-trivial user strategies.


\paragraph{Future work}

Our analysis shows that lending protocols --- despite the relative simplicity of the rules governing the semantics of individual actions ---
exhibit complex emergent behaviour.
While the strategic properties and the attacks formalised in~\Cref{sec:strategic-properties,sec:attacks} capture relevant aspects of these behaviour, it remains an open question whether more sophisticated strategies or attacks may exist.  
The search of strategies for reaching certain economic goals could be facilitated by specialised automatic tools.  
A relatively light-weight approach towards this goal is statistical model checking, a simulation-based technique that allows to observe the quantitative behavior of complex systems, based on statistical techniques to measure the confidence in the result produced.
Some initial results regarding the application of this approach to lending protocols are in~\cite{BCJLMV22isola}, which studies how different liquidation strategies and choices of the protocol parameters $\liqThreshold$ and $\rLiq$ impact the net position of borrowers.   
A drawback of this approach --- aside not guaranteeing 100\% accuracy in the results --- is that players' (probabilistic) strategies must be given as input to the simulator.
Therefore, this technique does not seem suitable for automatically discovering, or ruling out the existence of, strategies that achieve given economic goals. 

Another approach to inferring strategies and attacks in lending protocols is SMT-based bounded model checking. 
This technique involves encoding the semantics of lending protocols as a set of logical constraints and then querying whether there exists a sequence of transactions --- up to a given length --- that satisfies a specified property over blockchain states. 
While existing tools that apply SMT-based model checking to smart contracts written in real-world languages such as Solidity or Move are generally unable to verify --- or even express --- strategic properties~\cite{BCL25fmbc}, a specialized tool built upon our abstract model could potentially offer greater expressiveness and effectiveness.


\bartnote{proof assistant tipo il lavoro sugli AMM}
\paragraph*{Acknowledgements}
Work partially supported by project SERICS (PE00000014)
under the MUR National Recovery and Resilience Plan funded by the
European Union -- NextGenerationEU, and by PRIN 2022 PNRR project DeLiCE (F53D23009130001).

\bibliographystyle{alpha}
\bibliography{main}

\newpage
\appendix
\begin{figure}[h]
\small
\[
\begin{array}{c}
\depRule
\\[30pt]
\borRule
\\[30pt]
\repRule
\\[30pt]
\rdmRule
\\[30pt]
\intRule
\\[30pt]
\liqRule
\\[30pt]
\pxRule
\\[20pt]
\swpRule
\end{array}
\]
\caption{LP semantics.}
\label{fig:lp-semantics}
\end{figure}

\section{Proofs for~\Cref{sec:struct-properties}}
\label{proofs:struct-properties}

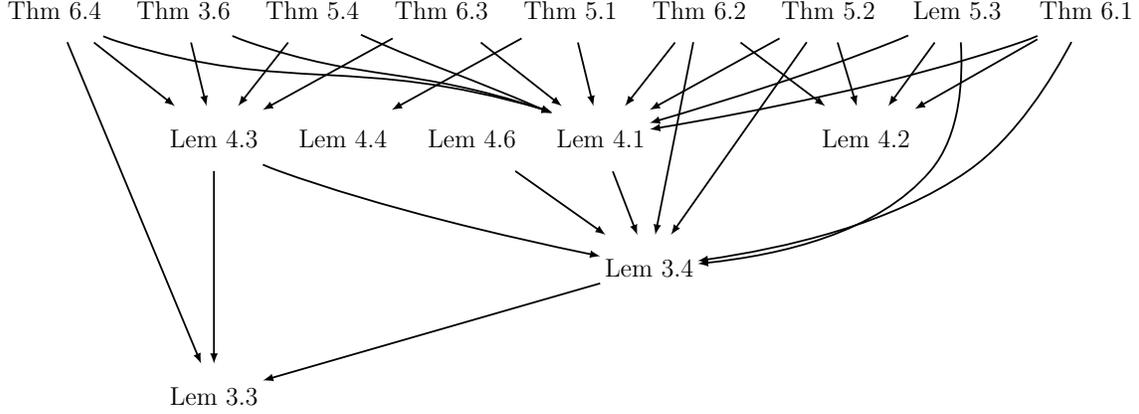
\begin{figure}[t]
  \centering
  \scalebox{0.675}{
    \Large
    \begin{tikzpicture}[>=latex,line join=bevel,]
  \pgfsetlinewidth{1bp}
\begin{scope}
  \pgfsetstrokecolor{black}
  \definecolor{strokecol}{rgb}{1.0,1.0,1.0};
  \pgfsetstrokecolor{strokecol}
  \definecolor{fillcol}{rgb}{1.0,1.0,1.0};
  \pgfsetfillcolor{fillcol}
  \filldraw (0.0bp,0.0bp) -- (0.0bp,252.0bp) -- (630.0bp,252.0bp) -- (630.0bp,0.0bp) -- cycle;
\end{scope}
  \pgfsetcolor{black}
  \draw [->] (331.6bp,81.106bp) .. controls (287.91bp,68.522bp) and (202.4bp,43.89bp)  .. (143.3bp,26.865bp);
  \draw [->] (126.26bp,219.36bp) .. controls (129.17bp,218.14bp) and (132.12bp,216.99bp)  .. (135.0bp,216.0bp) .. controls (201.37bp,193.22bp) and (224.64bp,202.87bp)  .. (304.74bp,176.64bp);
  \draw [->] (103.2bp,215.7bp) .. controls (105.1bp,207.9bp) and (107.38bp,198.51bp)  .. (111.85bp,180.1bp);
  \draw [->] (338.67bp,143.7bp) .. controls (341.71bp,135.81bp) and (345.38bp,126.3bp)  .. (352.4bp,108.1bp);
  \draw [->] (116.0bp,143.87bp) .. controls (116.0bp,119.67bp) and (116.0bp,75.211bp)  .. (116.0bp,36.189bp);
  \draw [->] (143.29bp,147.46bp) .. controls (146.2bp,146.21bp) and (149.14bp,145.03bp)  .. (152.0bp,144.0bp) .. controls (209.91bp,123.13bp) and (279.72bp,107.05bp)  .. (331.71bp,96.254bp);
  \draw [->] (284.22bp,143.88bp) .. controls (296.92bp,134.89bp) and (312.66bp,123.76bp)  .. (334.69bp,108.19bp);
  \draw [->] (198.05bp,220.02bp) .. controls (201.06bp,218.64bp) and (204.09bp,217.28bp)  .. (207.0bp,216.0bp) .. controls (243.41bp,200.05bp) and (255.3bp,197.43bp)  .. (304.95bp,175.98bp);
  \draw [->] (157.4bp,215.7bp) .. controls (150.88bp,207.39bp) and (142.93bp,197.28bp)  .. (129.44bp,180.1bp);
  \draw [->] (319.2bp,215.7bp) .. controls (321.1bp,207.9bp) and (323.38bp,198.51bp)  .. (327.85bp,180.1bp);
  \draw [->] (287.76bp,217.98bp) .. controls (269.17bp,207.74bp) and (244.36bp,194.07bp)  .. (215.2bp,177.99bp);
  \draw [->] (446.94bp,215.87bp) .. controls (429.52bp,191.14bp) and (397.2bp,145.24bp)  .. (371.11bp,108.19bp);
  \draw [->] (431.76bp,217.98bp) .. controls (413.17bp,207.74bp) and (388.36bp,194.07bp)  .. (359.2bp,177.99bp);
  \draw [->] (464.19bp,215.7bp) .. controls (466.53bp,207.9bp) and (469.35bp,198.51bp)  .. (474.87bp,180.1bp);
  \draw [->] (532.96bp,215.78bp) .. controls (534.2bp,196.1bp) and (533.08bp,164.03bp)  .. (516.0bp,144.0bp) .. controls (486.18bp,109.04bp) and (432.45bp,97.069bp)  .. (386.09bp,92.03bp);
  \draw [->] (503.64bp,219.64bp) .. controls (500.75bp,218.36bp) and (497.83bp,217.12bp)  .. (495.0bp,216.0bp) .. controls (452.51bp,199.14bp) and (402.33bp,183.43bp)  .. (359.2bp,170.69bp);
  \draw [->] (518.39bp,215.7bp) .. controls (512.4bp,207.47bp) and (505.12bp,197.48bp)  .. (492.46bp,180.1bp);
  \draw [->] (594.73bp,215.97bp) .. controls (584.1bp,195.85bp) and (563.78bp,162.76bp)  .. (537.0bp,144.0bp) .. controls (494.27bp,114.07bp) and (434.46bp,100.8bp)  .. (386.04bp,93.672bp);
  \draw [->] (575.74bp,219.37bp) .. controls (572.83bp,218.14bp) and (569.88bp,216.99bp)  .. (567.0bp,216.0bp) .. controls (498.98bp,192.59bp) and (416.29bp,176.71bp)  .. (359.32bp,167.16bp);
  \draw [->] (575.7bp,217.46bp) .. controls (558.11bp,207.45bp) and (535.03bp,194.32bp)  .. (507.22bp,178.49bp);
  \draw [->] (383.62bp,215.87bp) .. controls (378.83bp,191.56bp) and (370.01bp,146.82bp)  .. (362.39bp,108.19bp);
  \draw [->] (373.4bp,215.7bp) .. controls (366.88bp,207.39bp) and (358.93bp,197.28bp)  .. (345.44bp,180.1bp);
  \draw [->] (409.99bp,215.7bp) .. controls (421.81bp,206.8bp) and (436.39bp,195.82bp)  .. (457.28bp,180.1bp);
  \draw [->] (265.0bp,215.7bp) .. controls (276.21bp,206.88bp) and (290.0bp,196.03bp)  .. (310.25bp,180.1bp);
  \draw [->] (215.76bp,217.98bp) .. controls (197.17bp,207.74bp) and (172.36bp,194.07bp)  .. (143.2bp,177.99bp);
  \draw [->] (34.133bp,215.85bp) .. controls (49.626bp,178.6bp) and (86.444bp,90.067bp)  .. (108.83bp,36.232bp);
  \draw [->] (54.193bp,219.17bp) .. controls (57.12bp,217.98bp) and (60.092bp,216.89bp)  .. (63.0bp,216.0bp) .. controls (160.16bp,186.24bp) and (193.77bp,209.2bp)  .. (304.81bp,176.83bp);
  \draw [->] (49.0bp,215.7bp) .. controls (60.209bp,206.88bp) and (74.002bp,196.03bp)  .. (94.253bp,180.1bp);
\begin{scope}
  \definecolor{strokecol}{rgb}{0.0,0.0,0.0};
  \pgfsetstrokecolor{strokecol}
  \draw (116.0bp,18.0bp) node {Lem~\ref{lem:cred0-imp-debt0}};
\end{scope}
\begin{scope}
  \definecolor{strokecol}{rgb}{0.0,0.0,0.0};
  \pgfsetstrokecolor{strokecol}
  \draw (359.0bp,90.0bp) node {Lem~\ref{lem:ER-increasing}};
\end{scope}
\begin{scope}
  \definecolor{strokecol}{rgb}{0.0,0.0,0.0};
  \pgfsetstrokecolor{strokecol}
  \draw (99.0bp,234.0bp) node {Thm~\ref{th:net-worth-preservation}};
\end{scope}
\begin{scope}
  \definecolor{strokecol}{rgb}{0.0,0.0,0.0};
  \pgfsetstrokecolor{strokecol}
  \draw (332.0bp,162.0bp) node {Lem~\ref{lem:gain-base}};
\end{scope}
\begin{scope}
  \definecolor{strokecol}{rgb}{0.0,0.0,0.0};
  \pgfsetstrokecolor{strokecol}
  \draw (116.0bp,162.0bp) node {Lem~\ref{lem:gain-int}};
\end{scope}
\begin{scope}
  \definecolor{strokecol}{rgb}{0.0,0.0,0.0};
  \pgfsetstrokecolor{strokecol}
  \draw (480.0bp,162.0bp) node {Lem~\ref{lem:gain-px}};
\end{scope}
\begin{scope}
  \definecolor{strokecol}{rgb}{0.0,0.0,0.0};
  \pgfsetstrokecolor{strokecol}
  \draw (188.0bp,162.0bp) node {Lem~\ref{lem:health-tx}};
\end{scope}
\begin{scope}
  \definecolor{strokecol}{rgb}{0.0,0.0,0.0};
  \pgfsetstrokecolor{strokecol}
  \draw (260.0bp,162.0bp) node {Lem~\ref{lem:health-dep-rep}};
\end{scope}
\begin{scope}
  \definecolor{strokecol}{rgb}{0.0,0.0,0.0};
  \pgfsetstrokecolor{strokecol}
  \draw (171.0bp,234.0bp) node {Thm~\ref{th:X-int-vs-X}};
\end{scope}
\begin{scope}
  \definecolor{strokecol}{rgb}{0.0,0.0,0.0};
  \pgfsetstrokecolor{strokecol}
  \draw (315.0bp,234.0bp) node {Thm~\ref{th:gameLiq}};
\end{scope}
\begin{scope}
  \definecolor{strokecol}{rgb}{0.0,0.0,0.0};
  \pgfsetstrokecolor{strokecol}
  \draw (459.0bp,234.0bp) node {Thm~\ref{th:gamePx}};
\end{scope}
\begin{scope}
  \definecolor{strokecol}{rgb}{0.0,0.0,0.0};
  \pgfsetstrokecolor{strokecol}
  \draw (531.0bp,234.0bp) node {Lem~\ref{lem:gameOptions}};
\end{scope}
\begin{scope}
  \definecolor{strokecol}{rgb}{0.0,0.0,0.0};
  \pgfsetstrokecolor{strokecol}
  \draw (603.0bp,234.0bp) node {Thm~\ref{th:gameUndercollLoanAttck}};
\end{scope}
\begin{scope}
  \definecolor{strokecol}{rgb}{0.0,0.0,0.0};
  \pgfsetstrokecolor{strokecol}
  \draw (387.0bp,234.0bp) node {Thm~\ref{th:gameLiqAttck}};
\end{scope}
\begin{scope}
  \definecolor{strokecol}{rgb}{0.0,0.0,0.0};
  \pgfsetstrokecolor{strokecol}
  \draw (243.0bp,234.0bp) node {Thm~\ref{th:underutilizationAttack}};
\end{scope}
\begin{scope}
  \definecolor{strokecol}{rgb}{0.0,0.0,0.0};
  \pgfsetstrokecolor{strokecol}
  \draw (27.0bp,234.0bp) node {Thm~\ref{th:overutilizationAttack}};
\end{scope}
\end{tikzpicture}
  }
  \caption{Dependencies among the statements.}
  \label{fig:graph-proofs}
\end{figure}

In this and the following appendices we provide detailed proofs for all your statements.
These proofs are presented in the order in which the statements appear in the paper, even though this order does not always reflect their logical dependencies.
To clarify the relationship among our statements, \Cref{fig:graph-proofs} displays a graph of the dependencies: an arrow $a \rightarrow b$ means that the proof of statement $a$ depends on statement $b$.
Note that this graph is acyclic.

For quick reference, we also summarize the semantics of LPs in~\Cref{fig:lp-semantics}.

\begin{lemproof}{lem:constant-token-supply}
  By inspection of the rules in~\Cref{fig:lp-semantics}, it is immediate to 
  observe that all the rules but \nrule{[Swp]} ensure that
  transition preserve the supply of base token types.
  \qed





    




\end{lemproof}

\begin{lemproof}{lem:cred0-imp-debt0}
Let $\confG = (\WalW,\LpL,\price)$ be a reachable state.
We proceed by induction on the length of the trace to reach $\confG$.
The base case holds trivially, since in initial LP states there are no tokens.
For the inductive case, assume that $\confG \xrightarrow{\txT} \confGi = (\WalWi,\LpLi,\pricei)$ and that the statement holds in $\confG$.
There are the following cases:
\begin{itemize}

\item $\actDeposit{\pmvA}{\valV:\tokT}$.
We have that $\supply[\LpLi]{\tokT} > 0$, hence the thesis holds trivially.


\item $\actBorrow{\pmvA}{\valV:\tokT}$.
We have that $\LpL(\tokT) > 0$, so by hypothesis it must be $\supply[\LpL]{\tokT} > 0$. Since the \nrule{[Bor]} rule does not affect credit tokens, then $\supply[\LpLi]{\tokT} > 0$, hence the thesis holds trivially.

\item $\actRepay{\pmvA}{\valV:\tokT}$.
We have that $\supplyDebt[\LpL]{\tokT} > 0$, hence by the induction  hypothesis we must have $\supply[\LpL]{\tokT} > 0$.
Since the \nrule{[Rep]} rule does not affect credit tokens, then $\supply[\LpLi]{\tokT} > 0$, hence the thesis holds trivially.

\item $\actRedeem{\pmvA}{\minted{\valV}:\minted{\tokT}}$.
If $\supply[\LpLi]{\tokT} > 0$, then the thesis holds trivially.
Otherwise, if $\supply[\LpLi]{\tokT} = 0$, then it must be
$\minted{\valV} = \supply[\LpL]{\tokT}$.
Therefore, we have:
\begin{align*}
    \LpLi(\tokT) 
    & = \LpL(\tokT) - \minted{\valV} \cdot \X[\LpL]{\tokT}
    = \LpL(\tokT) - \supply[\LpL]{\tokT} \cdot \X[\LpL]{\tokT}
    && \text{by \nrule{[Rdm]} and hyp.}
\end{align*}
We have now two cases, depending on whether $\supply[\LpL]{\tokT} = 0$ or not.
If $\supply[\LpL]{\tokT} = 0$, then by~\eqref{eq:LP:ER}
we have that $\X[\LpL]{\tokT} = 1$. 
Furthermore, by the induction hypothesis we have that $\LpL(\tokT) = 0 = \supplyDebt[\LpL]{\tokT}$.
Therefore:
\begin{align*}
    \LpLi(\tokT)
    & = \LpL(\tokT) - \supply[\LpL]{\tokT}
    = - \supply[\LpL]{\tokT}
    = 0
    \\
    \supplyDebt[\LpLi]{\tokT}
    & = \supplyDebt[\LpL]{\tokT}
    = 0
\end{align*}
Otherwise, if $\supply[\LpL]{\tokT} > 0$, then by~\eqref{eq:LP:ER}:
\begin{align*}
    \LpLi(\tokT) 
    & =
    \LpL(\tokT) - \supply[\LpL]{\tokT} \cdot \frac{\LpL(\tokT) + \supplyDebt[\LpL]{\tokT}}{\supply[\LpL]{\tokT}}
    = - \supplyDebt[\LpL]{\tokT} 
\end{align*}
Since $\LpLi(\tokT)$ cannot be negative, it must be $\supplyDebt[\LpL]{\tokT} = 0 = \LpLi(\tokT)$.

\item $\actLiquidate{\pmvA}{\pmvB}{\valV[0]:\tokT[0]}{\mintedT[1]}$.
We have two cases, depending on whether $\tokT$ is the repaid token $\tokT[0]$ or the base token underlying the credit token $\minted{\tokT[1]}$.
\begin{itemize}

    \item $\tokT = \tokT[0] \neq \tokT[1]$.
    Assume that $\supply[\LpLi]{\tokT[0]} = 0$.
    Since the \nrule{[Liq]} transition does not affect the amount of $\minted{\tokT[0]}$, then it must be
    $\supply[\LpL]{\tokT[0]} = 0$.
    Then, the thesis follows by the induction hypothesis.

    \item $\tokT = \tokT[1]$.
    In this case thesis holds trivially, since $\supply[\LpLi]{\tokT[1]} \geq \minted{\valV[1]} > 0$ by the \nrule{[Liq]} rule.

\end{itemize}

\item  In all the other cases, the transition does not affect tokens $\tokT$ in the LP, hence the thesis follows directly from the hypothesis.
\qed
\end{itemize}
\end{lemproof}

\begin{lemproof}{lem:ER-increasing}
Let \mbox{$\confG \xrightarrow{\ltsLabel} \confGi$},
where $\confG = (\WalW,\LpL,\price)$
and $\confGi = (\WalWi,\LpLi,\pricei)$.
We must prove that, for all $\tokT$:
\begin{enumerate}[(a)]
\item  if $\ltsLabel = \accrueIntOp$,  
and $\supplyDebt[\LpL]{\tokT} > 0$,
then 
\[
          \X[\LpLi]{\tokT}
          \; = \;
          \X[\LpL]{\tokT}
        +   
            \dfrac
            {\supplyDebt[\LpL]{\tokT}
            }
            {\supply[\LpL]{\tokT}}            
            \cdot \Intr{\LpL}(\tokT) 
        \; > \;
          \X[\LpL]{\tokT}
  \]
    
\item if $\ltsLabel = \redeemOp$, and $\supply[\LpLi]{\tokT} = 0$, then $\X[\LpLi]{\tokT} = 1$ 
    
\item otherwise, $\X[\LpLi]{\tokT} = \X[\LpL]{\tokT}$.
\end{enumerate}
  We proceed by cases on $\ltsLabel$.
  Note that, if $\txT \neq \accrueIntOp$ but $\tokT$ does not appear in the transaction, then its exchange rate does not change. 
  Hence, besides the case $\txT = \accrueIntOp$, we only have to deal with transactions that mention $\tokT$.
  There are the following exhaustive cases:
  \begin{itemize}


  \item \mbox{$\actDeposit{\pmvA}{\valV:\tokT}$}.
    Since the \nrule{[Dep]} rule increases the base tokens $\tokT$, then $\LpLi({\tokT}) > 0$, and so by Lemma~\ref{lem:cred0-imp-debt0} it must be $\supply[\LpLi]{\tokT} > 0$.
    Then:
    \begin{align*}
      \X[\LpLi]{\tokT} 
      & \; = \;
        \dfrac
        {
        \reserves[\LpLi]{\tokT}
        + \supplyDebt[\LpLi]{\tokT}
        }
        {\supply[\LpLi]{\tokT}}
      && \text{by~\eqref{eq:LP:ER}}
      \\
      & \; = \;
        \dfrac
        {
        \reserves[\LpL]{\tokT} 
        + \valV
        + \supplyDebt[\LpL]{\tokT}
        }
        {\supply[\LpL]{\tokT} + \nicefrac{\valV}{\X[\LpL]{\tokT}}}
      && \text{by~\nrule{[Dep]}}
      \\
      & \; = \;
        \dfrac
        {
        \reserves[\LpL]{\tokT} 
        + \valV
        + \supplyDebt[\LpL]{\tokT}
        }
        {\supply[\LpL]{\tokT} \cdot \X[\LpL]{\tokT}
        + 
        \valV}
           \cdot
        \X[\LpL]{\tokT}
      && \text{by arith.}
      \\
      & \; = \;
        \dfrac
        {
        \reserves[\LpL]{\tokT} 
        + \valV
        + \supplyDebt[\LpL]{\tokT}
        }
            {\supply[\LpL]{\tokT}
        \cdot                 
            \dfrac
            {
              \LpL(\base{\tokT})
              + 
              \supplyDebt[\LpL]{\tokT}
            }
            {\supply[\LpL]{\tokT}}
        + 
        \valV}
           \cdot
        \X[\LpL]{\tokT}
      && \text{by ~\eqref{eq:LP:ER}}
      \\
      & \; = \;
        \X[\LpL]{\tokT}
      && \text{by arith.}
    \end{align*}
    
  \item \mbox{$\actBorrow{\pmvA}{\valV:\tokT}$}.
    Since the \nrule{[Bor]} rule increases the debit tokens $\debt{\tokT}$, then $\supplyDebt[\LpLi]{\tokT} > 0$, and so by Lemma~\ref{lem:cred0-imp-debt0} it must be $\supply[\LpLi]{\tokT} > 0$.
    Then:
    \begin{align*}
      \X[\LpLi]{\tokT} 
      & \; = \;
        \dfrac
        {
        \reserves[\LpLi]{\tokT}
        + \supplyDebt[\LpLi]{\tokT}
        }
        {\supply[\LpLi]{\tokT}}
      && \text{by~\eqref{eq:LP:ER}}
      \\
      & \; = \;
        \dfrac
        {
        (\reserves[\LpL]{\tokT}
        - \valV
        )
        + 
        (\supplyDebt[\LpL]{\tokT}
        + \valV
        )
        }
        {\supply[\LpL]{\tokT}}
      && \text{by~\nrule{[Bor]}}
      \\
      & \; = \;
        \dfrac
        {
        \reserves[\LpL]{\tokT}
        + \supplyDebt[\LpL]{\tokT}
        }
        {\supply[\LpL]{\tokT}}
      && \text{by arith.}
      \\
      & \; = \;
        \X[\LpL]{\tokT}
      && \text{by~\eqref{eq:LP:ER}}
    \end{align*}

  \item $\accrueIntOp$. 
    For every $\pmvA$, we have that 
    \begin{align*}
        \stateDebti{\tokT}{\pmvA}
      & =
        \stateDebt{\tokT}{\pmvA}
      +  \stateDebt{\tokT}{\pmvA} \cdot \Intr{\LpL}(\tokT) 
      =
       \stateDebt{\tokT}{\pmvA}
      \cdot (1+\Intr{\LpL}(\tokT))
    \end{align*}
    The supply of credit tokens changes as follows:
    \begin{align*}
    \supplyDebt[\LpLi]{\tokT} 
    & \label{eq:ER:debtSupply} 
    = \sum_{\pmvA \in \PmvU}  \stateDebti{\tokT}{\pmvA}
    && \text{}
    \\
    & = \sum_{\pmvA \in \PmvU} 
       \stateDebt{\tokT}{\pmvA}
      \cdot (1+\Intr{\LpL}(\tokT)
    && \text{}
    \\
    & = \sum_{\pmvA \in \PmvU} 
       \stateDebt{\tokT}{\pmvA}
       +
       \sum_{\pmvA \in \PmvU} 
       \stateDebt{\tokT}{\pmvA}
      \cdot \Intr{\LpL}(\tokT)
    && \text{}
    \\
    & =
    \supplyDebt[\LpL]{\tokT} 
       +
       \sum_{\pmvA \in \PmvU} 
       \stateDebt{\tokT}{\pmvA}
      \cdot \Intr{\LpL}(\tokT)        
    && \text{}
    \end{align*} 


    The exchange rate changes as follows. If ${\supply[\LpL]{\tokT}}=0$, then, by \nrule{[Int]} also ${\supply[\LpLi]{\tokT}} = 0$, and hence, by~\Cref{eq:LP:ER}, $\X[\LpL]{\tokT} = 1 = \X[\LpLi]{\tokT}$. 
    Otherwise, if ${\supply[\LpL]{\tokT}} \neq 0$,
    then $\supply[\LpLi]{\tokT} = \supply[\LpL]{\tokT} \neq 0$, and so:  
        \begin{align*}
          \X[\LpLi]{\tokT} 
          & \; = \;
            \dfrac
            {
            \reserves[\LpLi]{\tokT}
            + 
            \supplyDebt[\LpLi]{\tokT}
            }
            {\supply[\LpLi]{\tokT}}
          && \text{by~\eqref{eq:LP:ER}}
          \\
          & \; = \;
            \dfrac
            {
                \reserves[\LpL]{\tokT}
            +
                \supplyDebt[\LpL]{\tokT} 
           +
                \sum_{\pmvA \in \PmvU} 
               \stateDebt{\tokT}{\pmvA}
              \cdot \Intr{\LpL}(\tokT) 
            }
            {\supply[\LpL]{\tokT}}
          && \text{by~\nrule{[Int]}}            
          \\
          & \; = \;
            \dfrac
            {
                \reserves[\LpL]{\tokT}
            +
                \supplyDebt[\LpL]{\tokT} 
            }
            {\supply[\LpL]{\tokT}}
        +   
            \dfrac
            {
                \sum_{\pmvA \in \PmvU} 
               \stateDebt{\tokT}{\pmvA}
              \cdot \Intr{\LpL}(\tokT) 
            }
            {\supply[\LpL]{\tokT}}
          && \text{by arith.}
          \\          
          & \; = \;
          \X[\LpL]{\tokT}
        +   
            \dfrac
            {
                \sum_{\pmvA \in \PmvU} 
               \stateDebt{\tokT}{\pmvA}
              \cdot \Intr{\LpL}(\tokT) 
            }
            {\supply[\LpL]{\tokT}}
          && \text{by~\eqref{eq:LP:ER}}
          \\          
          & \; = \;
          \X[\LpL]{\tokT}
        +   
            \dfrac
            {\supplyDebt[\LpL]{\tokT}}     
            {\supply[\LpL]{\tokT}}     
            \cdot \Intr{\LpL}(\tokT) 
          && \text{by~\eqref{eq:supply}}
        \end{align*}
    
    \noindent We now have the two following two cases:
    \begin{itemize}
        \item If $\supplyDebt[\LpL]{\tokT} = 0$, the second addend in the previous equation is equal to $0$, hence  $\X[\LpLi]{\tokT}=\X[\LpL]{\tokT}$.
        \item If 
        $\supplyDebt[\LpL]{\tokT} > 0$, 
        since  by~\Cref{eq:interest-rate-gt-zero} $\Intr{\LpL}(\tokT) > 0$,
        then the second addend is strictly positive, hence 
        $\X[\LpLi]{\tokT}>\X[\LpL]{\tokT}$.
        


        \end{itemize}
        
  \item \mbox{$\actRepay{\pmvA}{\valV:\tokT}$}.
    Since the \nrule{[Rep]} rule increases the base tokens $\tokT$, then $\LpLi({\tokT}) > 0$, and so by Lemma~\ref{lem:cred0-imp-debt0} it must be $\supply[\LpLi]{\tokT} > 0$.
    Then:
    \begin{align*}
      \X[\LpLi]{\tokT} 
      & \; = \;
        \dfrac
        {
        \reserves[\LpLi]{\tokT}
        + \supplyDebt[\LpLi]{\tokT}
        }
        {\supply[\LpLi]{\tokT}}
      && \text{by~\eqref{eq:LP:ER}}
      \\
      & \; = \;
        \dfrac
        {
        (\reserves[\LpL]{\tokT}
        + \valV
        )
        + 
        (\supplyDebt[\LpL]{\tokT}
        - \valV
        )
        }
        {\supply[\LpL]{\tokT}}
      && \text{by~\nrule{[Rep]}}
      \\
      & \; = \;
        \dfrac
        {
        \reserves[\LpL]{\tokT}
        + \supplyDebt[\LpL]{\tokT}
        }
        {\supply[\LpL]{\tokT}}
      && \text{by arith.}
      \\
      & \; = \;
        \X[\LpL]{\tokT}
      && \text{by~\eqref{eq:LP:ER}}
    \end{align*}
    
  \item \mbox{$\actRedeem{\pmvA}{\valV:\minted{\tokT}}$}.
    We have two cases. 
    First, consider the case $\supply[\LpLi]{\tokT} \neq 0$, \ie $\valV \neq \supply[\LpL]{\tokT}$.
    Then:
    \begin{align*}
      \X[\LpLi]{\tokT} 
      & \; = \;
        \dfrac
        {
        \reserves[\LpLi]{\tokT}
        + \supplyDebt[\LpLi]{\tokT}
        }
        {\supply[\LpLi]{\tokT}}
      && \text{by~\eqref{eq:LP:ER}}
      \\
      & \; = \;
        \dfrac
        {
        \reserves[\LpL]{\tokT} 
        - \valV \cdot \X[\LpL]{\tokT}
        + \supplyDebt[\LpL]{\tokT}
        }
        {\supply[\LpL]{\tokT} 
        - \valV}
      && \text{by~\nrule{[Rdm]}}
      \\
      & \; = \;
        \dfrac
        {
        \reserves[\LpL]{\tokT}  \cdot \supply[\LpL]{\tokT} 
        - \valV \cdot \X[\LpL]{\tokT} \cdot \supply[\LpL]{\tokT} 
        + \supplyDebt[\LpL]{\tokT} \cdot \supply[\LpL]{\tokT} 
        }
        {
        (\supply[\LpL]{\tokT} 
        - \valV)
            \cdot
        \supply[\LpL]{\tokT} 
        }
      && \text{by arith.}
      \\
      & \; = \;
        \dfrac
        {
        \reserves[\LpL]{\tokT}  \cdot \supply[\LpL]{\tokT} 
        - \valV \cdot 
            (
            \reserves[\LpL]{\tokT}
            + \supplyDebt[\LpL]{\tokT}
            )
        + \supplyDebt[\LpL]{\tokT} \cdot \supply[\LpL]{\tokT} 
        }
        {
        (\supply[\LpL]{\tokT} 
        - \valV)
            \cdot
        \supply[\LpL]{\tokT} 
        }
      && \text{by~\eqref{eq:LP:ER}}
      \\
      & \; = \;
        \dfrac
        {
        \reserves[\LpL]{\tokT}  \cdot (\supply[\LpL]{\tokT} - \valV)
        + \supplyDebt[\LpL]{\tokT} \cdot (\supply[\LpL]{\tokT} - \valV)
        }
        {
        (\supply[\LpL]{\tokT} 
        - \valV)
            \cdot
        \supply[\LpL]{\tokT} 
        }
      && \text{by arith.}
      \\
      & \; = \;
        \X[\LpL]{\tokT}
      && \text{by~\eqref{eq:LP:ER}}
    \end{align*}
    
	Otherwise, if $\supply[\LpLi]{\tokT} = 0$, then, by definition of exchange rate (\ref{eq:LP:ER}), $
	\X[\LpLi]{\tokT}=1$.

  \item In the case of $\liquidateOp$, since there are two tokens that appear in the rule, we will first consider the case in which the token $\tokT$ corresponds to the debt being repayed by the liquidator, and, secondly, the case in which the token $\tokT$ is the token whose associated credit tokens are seized and passed to the liquidator.

  \begin{itemize}
  \item \mbox{$\actLiquidate{\pmvA}{\pmvB}{\valV:\tokT}{\minted{\tokTi}}$}, with $\tokT \neq \tokTi$.
    \begin{align*}
      \X[\LpLi]{\tokT} 
      & \; = \;
        \dfrac
        {
        \reserves[\LpLi]{\tokT}
        + \supplyDebt[\LpLi]{\tokT}
        }
        {\supply[\LpLi]{\tokT}}
      && \text{by~\eqref{eq:LP:ER}}
      \\
      & \; = \;
        \dfrac
        {
        (\reserves[\LpL]{\tokT}
            + \valV)
        + 
        (\supplyDebt[\LpL]{\tokT}
            - \valV
        )
        }
        {\supply[\LpL]{\tokT} }
      && \text{by~\nrule{[Liq]}}
      \\
      & \; = \;
        \dfrac
        {
        \reserves[\LpL]{\tokT}
        + 
        \supplyDebt[\LpL]{\tokT}
        }
        {\supply[\LpL]{\tokT} }
      && \text{by arith.}
      \\
      & \; = \;
        \X[\LpL]{\tokT}
      && \text{by~\eqref{eq:LP:ER}}
    \end{align*}

  \item \mbox{$\actLiquidate{\pmvA}{\pmvB}{\valV:\tokTi}{\minted{\tokT}}$}, with $\tokT \neq \tokTi$.
    \begin{align*}
      \X[\LpLi]{\tokT} 
      & \; = \;
        \dfrac
        {
        \reserves[\LpLi]{\tokT}
        + \supplyDebt[\LpLi]{\tokT}
        }
        {\supply[\LpLi]{\tokT}}
      && \text{by~\eqref{eq:LP:ER}}
      \\
      & \; = \;
        \dfrac
        {
        \reserves[\LpL]{\tokT}
        + 
        \supplyDebt[\LpL]{\tokT}
        )
        }
        {\supply[\LpL]{\tokT} + \valVi - \valVi }
      && \text{by~\nrule{[Liq]}}
      \\
      & \; = \;
        \dfrac
        {
        \reserves[\LpL]{\tokT}
        + 
        \supplyDebt[\LpL]{\tokT}
        }
        {\supply[\LpL]{\tokT} }
      && \text{by arith.}
      \\
      & \; = \;
        \X[\LpL]{\tokT}
      && \text{by ~\eqref{eq:LP:ER}}
    \end{align*}

    \item \mbox{$\actLiquidate{\pmvA}{\pmvB}{\valV:\tokT}{\minted{\tokT}}$}.
    \begin{align*}
    	\X[\LpLi]{\tokT} 
    	& \; = \;
    	\dfrac
    	{
    		\reserves[\LpLi]{\tokT}
    		+ \supplyDebt[\LpLi]{\tokT}
    	}
    	{\supply[\LpLi]{\tokT}}
    	&& \text{by~\eqref{eq:LP:ER}}
    	\\
    	& \; = \;
    	\dfrac
    	{
    		(\reserves[\LpL]{\tokT}
    		+ \valV)
    		+ 
    		(\supplyDebt[\LpL]{\tokT}
    		- \valV
    		)
    	}
    	{\supply[\LpL]{\tokT} + \valVi - \valVi }
    	&& \text{by~\nrule{[Liq]}}
    	\\
    	& \; = \;
    	\dfrac
    	{
    		\reserves[\LpL]{\tokT}
    		+ 
    		\supplyDebt[\LpL]{\tokT}
    	}
    	{\supply[\LpL]{\tokT} }
    	&& \text{by arith.}
    	\\
    	& \; = \;
    	\X[\LpL]{\tokT}
    	&& \text{by~\eqref{eq:LP:ER}}
    \end{align*}

    \end{itemize}
    
    \item \mbox{$\pxOp{\prIncr: \tokT}$} Price updates do not change the number of base/credit/debit tokens in the LP, hence the thesis holds trivially. 
	    
    \item \mbox{$\swapOp(\valV: \tokT[1], \tokT[2])$} Swaps do not change the number of base/credit/debit tokens in the LP, hence the thesis holds trivially.
    \qed
  \end{itemize}
\end{lemproof}

\begin{corproof}{lem:ER:geq1}
We must prove that $\X[\LpL]{\tokT} \geq 1$
for all reachable LP state $\LpL$.
Let $\confG = (\WalW,\LpL,\price)$ be a reachable blockchain state.
If $\confG$ is initial, then $\supply[\LpL]{\tokT} = 0$, and so the second case of~\eqref{eq:LP:ER} applies, giving $\X[\LpL]{\tokT} = 1$.
Otherwise, the statement follows by applying inductively Lemma~\ref{lem:ER-increasing}.
\qed
\end{corproof}

    



    
    
    
    



        

    

\begin{thmproof}{th:net-worth-preservation}
We have to prove that, for every state $\confG$, and
for every
transition:
\[
\confG = (\WalW,\LpL,\price) \; \xrightarrow{\txT} \; \confGi = (\WalWi,\LpLi,\price)
\]
such that $\txT \neq \pxOp{}$, we have that:
\[
\sum_{\pmvC\in \PmvU} \wealth{\confGi}{\pmvC}
\; = \;
\sum_{\pmvC\in \PmvU} \wealth{\confG}{\pmvC}
\]

\noindent 
Since 
${\sum_{\pmvC\in \PmvU} \gain{\pmvC}{\confG}{\txT} = 
    \sum_{\pmvC\in \PmvU} 
       (\wealth{\confGi}{\pmvC}
        - \wealth{\confG}{\pmvC}})$,
we can  equivalently prove that: 
\[
\sum_{\pmvC\in \PmvU} \gain{\pmvC}{\confG}{\txT} = 0
\]
Note that Lemma~\ref{lem:gain-base} implies the thesis for all cases except $\txT = \accrueIntOp$.
Indeed, it states that:
\begin{itemize}
    \item if $\txT$ is not a $\liquidateOp$, then the gain of the user $\pmvA$ who fired the transaction is $\gain{\pmvA}{\confG}{\txT} = 0$, while the gain of the other users does not change.
    \item if  $\txT$ is a $\liquidateOp$, say $\txT = \actLiquidate{\pmvA}{\pmvB}{\valV:\tokT[0]}{\mintedT[1]}$, then the only non-zero gains are those of $\pmvA$ and~$\pmvB$, and  
    $\gain{\pmvA}{\confG}{\txT} = -\gain{\pmvB}{\confG}{\txT}$,
    \ie: 
    \begin{align*}
    \sum_{\pmvC\in \PmvU} \gain{\pmvC}{\confG}{\txT} 
    & = 
    \bigg( 
    \sum_{\pmvC\in \PmvU \setminus \{\pmvA, \pmvB\}} \gain{\pmvC}{\confG}{\txT} 
    \bigg)
    + \gain{\pmvA}{\confG}{\txT}  
    + \gain{\pmvB}{\confG}{\txT} 
    \\
    & =   
    0
    + \gain{\pmvA}{\confG}{\txT}  
    - \gain{\pmvA}{\confG}{\txT}  
    =
    0
    \end{align*}
\end{itemize}

\noindent
We now consider the case $\txT = \accrueIntOp$.
From Lemma~\ref{lem:gain-int}, we have that, for all $\tokT$, if  ${\supply[{\LpL}]{\tokT}} = 0$,
then  for all $\pmvA$, $\projTok{\gainSt{\pmvA}{\confG}{\accrueIntOp}} = 0$;
otherwise,
if ${\supply[{\LpL}]{\tokT}} > 0$,
for all $\pmvA$: 
\[
    \projTok{\gainSt{\pmvA}{\confG}{\accrueIntOp}}
    =
    \left(
        \stateCredit{\tokT}{\pmvA}{}
            \cdot 
        \dfrac                
        {\supplyDebt[\LpL]{\tokT}}
        {\supply[{\LpL}]{\tokT}}
    - 
        \LpL(\debt{\tokT},\pmvA)
    \right)
    \cdot 
    \Big(
        \Intr{\LpL}(\tokT)
    \cdot \price[\tokT]
    \Big)
\]
Let $c = \Intr{\LpL}(\tokT) \cdot \price[\tokT]$.
We have that:
\begin{align*}  
    \sum_{\pmvC\in\PmvU} \projTok{\gain{\pmvC}{\confG}{\accrueIntOp} }
      & \; = \;  
        \sum_{\pmvC\in \PmvU} 
            \left(
                \stateCredit{\tokT}{\pmvC}
                    \cdot   
                \dfrac                
                {\supplyDebt[\LpL]{\tokT}}
                {\supply[{\LpL}]{\tokT}}
            - 
                \LpL(\debt{\tokT},\pmvC)
            \right)
            \cdot c
      && \text{by~\eqref{eq:supply}} 
      \\   
      & \; = \;  
            \left(
                \Big(
                \sum_{\pmvC\in \PmvU} 
                    \stateCredit{\tokT}{\pmvC}
                \Big)
                    \cdot 
                \dfrac                
                {\supplyDebt[\LpL]{\tokT}}
                {\supply[{\LpL}]{\tokT}} 
                - 
                \sum_{\pmvC\in \PmvU} 
                \LpL(\debt{\tokT},\pmvC)
            \right)
            \cdot c
      && \text{by arith.} 
      \\   
      & \; = \;  
            \left(
                 {\supply[\LpL]{\tokT}}
                    \cdot 
                \dfrac    
                {\supplyDebt[\LpL]{\tokT}} 
                {\supply[\LpL]{\tokT}}
                -
                {\supplyDebt[\LpL]{\tokT}}
            \right)
            \cdot c
      && \text{by~\eqref{eq:supply}} 
      \\   
      & \; = \;  
      0
      && \text{by arith.} 
      \\   
 \end{align*} 
 The thesis follows from~\eqref{eq:sum-wealth-restricted}.
 \qed
 \end{thmproof}

\section{Proofs for~\Cref{sec:econ-properties}}
\label{proofs:economic-properties}

\begin{lemproof}{lem:gain-base}
  Let $\txT \not\in \setenum{\accrueIntOp,\pxOp{}}$ be enabled in $\confG$. We have to prove that:
  \begin{enumerate}  
  \item $\gainSt{\pmvA}{\confG}{\txT} = 0$ iff $\txT$ is not a liquidation involving $\pmvA$.
  \item $\gainSt{\pmvA}{\confG}{\txT} > 0$ iff 
  $\txT$ is a liquidation performed by $\pmvA$;
  \item $\gainSt{\pmvA}{\confG}{\txT} < 0$ iff $\txT$ is a liquidation suffered by $\pmvA$;
  \end{enumerate}
    and, in the case $\txT = \actLiquidate{\pmvA}{\pmvB}{\valV:\tokT[0]}{\mintedT[1]}$,  that:
    \begin{enumerate}
    
    \item[(4)] \(
    \gainSt{\pmvA}{\confG}{\txT} 
    = - \gainSt{\pmvB}{\confG}{\txT} 
    = \valV \cdot \price[{\tokT[0]}] \cdot \rLiq
    \).
    \end{enumerate}

  Assume that the state transition is given by:
  \[
  \confG = (\WalW,\LpL,\price) \xrightarrow{\txT} \confGi = (\WalWi,\LpLi,\price)
  \]
  In order to prove the first three points, we proceed by cases on $\txT$. The last point will follow from the cases concerning $\liquidateOp$.

  First, consider the case in which $\pmvA$ does not appear in $\txT$. In this case, the amount of base tokens, credits, and debts held by $\pmvA$ do not change. Since, by hypothesis, $\txT \neq \pxOp{}$,  the token prices do not change too, and hence 
      $\baseval[\WalWi,\price]{\pmvA} = \baseval[\WalWi,\price]{\pmvA}$ and $\debtval[\LpLi,\price]{\pmvA} = \debtval[\LpL,\price]{\pmvA}$.
   For $\creditval[\LpLi,\price]{\pmvA}$,  we have to consider possible changes in the exchange rates. By Lemma \ref{lem:ER-increasing} and the hypothesis that  $\txT \neq \accrueIntOp$,  we know that the exchange rate of a token type $\tokT$ can only change if $\txT$ is a $\redeemOp$ that reclaims the entirety of the credits in $\tokT$.
   If this is the case, since by hypothesis $\pmvA$ does not appear in $\txT$, in means that $\pmvA$ had no credits in $\tokT$, \ie $\projTok{\creditval[\LpL,\price]{\pmvA}}=0=\projTok{\creditval[\LpL,\price]{\pmvA}}$, hence  $\creditval[\LpLi,\price]{\pmvA} = \creditval[\LpL,\price]{\pmvA}$.
    

  Now let's consider the following exhaustive cases in which $\pmvA$ appears in $\txT$:
  \begin{itemize}


  \item \mbox{$\actDeposit{\pmvA}{\valV:\tokT}$}.
    We have that:
    \begin{align}
      \wealth{\confGi}{\pmvA} 
      \nonumber
      & = \baseval[\WalWi,\price]{\pmvA} 
      + \creditval[\LpLi,\price]{\pmvA} 
      - \debtval[\LpLi,\price]{\pmvA}
      && \text{by~\eqref{eq:networth}}
      \\
      \nonumber
      & = 
      \baseval[\WalWi,\price]{\pmvA} 
      +
      \sum_{\tokTi}
      \stateCrediti{\tokTi}{\pmvA}
      \cdot \X[\LpLi]{\tokTi} \cdot \price[{\tokTi}]
      - \debtval[\LpLi,\price]{\pmvA}
      && \text{by \eqref{eq:creditVal}}
      \\
      \nonumber
      & = 
      \baseval[\WalWi,\price]{\pmvA} 
      +
      \sum_{\tokTi}
      \stateCrediti{\tokTi}{\pmvA}
      \cdot \X[\LpL]{\tokTi} \cdot \price[{\tokTi}]
      - \debtval[\LpLi,\price]{\pmvA}
      && \text{by Lem. \ref{lem:ER-increasing}}
      \\
      \nonumber
      & = 
      \Big(
          \baseval[\WalW,\price]{\pmvA} 
          - \valV \cdot \price[\tokT] 
      \Big)
      + \Big(
	      \sum_{\tokTi\neq \tokT}
	      \stateCredit{\tokTi}{\pmvA}
	      \cdot \X[\LpL]{\tokTi} \cdot \price[{\tokTi}]
     \\ & \qquad \nonumber 
          + ( \stateCredit{\tokT}{\pmvA}
            + \minted{\valV}) \cdot \X[\LpL]{\tokT} \cdot \price[\tokT]
      \Big)
      - \debtval[\LpL,\price]{\pmvA}
      && \text{by~\nrule{[Dep]}}
      \\
      \nonumber
      & = \baseval[\WalW,\price]{\pmvA} 
      - \valV \cdot \price[\tokT] 
      + \creditval[\LpL,\price]{\pmvA} 
      + \minted{\valV} \cdot \X[\LpL]{\tokT}  \cdot \price(\tokT) 
      - \debtval[\LpL,\price]{\pmvA}    
      && \text{by arith. + \eqref{eq:creditVal}}
      \\
      \nonumber
      & = \baseval[\WalW,\price]{\pmvA} 
      - \valV \cdot \price[\tokT] 
      + \creditval[\LpL,\price]{\pmvA} 
      + \valV \cdot \dfrac{\X[\LpL]{\tokT}}{\X[\LpL]{\tokT}} \cdot \price(\tokT) 
      - \debtval[\LpL,\price]{\pmvA}    
    && \text{by def.~$\minted{\valV}$}
      \\
      \nonumber
      & = \baseval[\WalW,\price]{\pmvA} 
      - \valV \cdot \price[\tokT] 
      + \creditval[\LpL,\price]{\pmvA} 
      + \valV  \cdot \price(\tokT) 
      - \debtval[\LpL,\price]{\pmvA}
    && \text{by Lem.~\ref{lem:ER-increasing}}
      \\
      \nonumber
      & = \baseval[\WalW,\price]{\pmvA} 
      + \creditval[\LpL,\price]{\pmvA} 
      - \debtval[\LpL,\price]{\pmvA}
    && \text{by arith.}
      \\
      \nonumber
      & = \wealth{\confG}{\pmvA} 
      && \text{by~\eqref{eq:networth}}
    \end{align}


  \item \mbox{$\actBorrow{\pmvA}{\valV:\tokT}$}. We have that:
    \begin{align}
      \wealth{\confGi}{\pmvA} 
      \nonumber
      & = \baseval[\WalWi,\price]{\pmvA} 
      + \creditval[\LpLi,\price]{\pmvA} 
      - \debtval[\LpLi,\price]{\pmvA}
      && \text{by \eqref{eq:networth}}
      \\
      \nonumber
      & = 
      \Big(
          \baseval[\WalW,\price]{\pmvA} 
          + \valV \cdot \price[\tokT] 
      \Big)
      + 
          \creditval[\LpL,\price]{\pmvA}
      - 
        \Big(
            \debtval[\LpL,\price]{\pmvA} 
            +
            \valV \cdot \price[\tokT] 
        \Big)
      && \text{by \nrule{[Bor]}}
      \\
      \nonumber
      & = \baseval[\WalW,\price]{\pmvA} 
      + \creditval[\LpL,\price]{\pmvA} 
      - \debtval[\LpL,\price]{\pmvA}
    && \text{by arith.}
      \\
      \nonumber
      & = \wealth{\confG}{\pmvA} 
      && \text{\eqref{eq:networth}}
    \end{align}

  \item \mbox{$\actRepay{\pmvA}{\valV:\tokT}$}. We have that:
    \begin{align}
      \wealth{\confGi}{\pmvA} 
      \nonumber
      & = \baseval[\WalWi,\price]{\pmvA} 
      + \creditval[\LpLi,\price]{\pmvA} 
      - \debtval[\LpLi,\price]{\pmvA}
      && \text{by \eqref{eq:networth}}
      \\
      \nonumber
      & = 
      \Big(
          \baseval[\WalW,\price]{\pmvA} 
          - \valV \cdot \price[\tokT] 
      \Big)
      + 
          \creditval[\LpL,\price]{\pmvA}
      - 
        \Big(
            \debtval[\LpL,\price]{\pmvA} 
            -
            \valV \cdot \price[\tokT] 
        \Big)
      && \text{by \nrule{[Rep]}}
      \\
      \nonumber
      & = \baseval[\WalW,\price]{\pmvA} 
      + \creditval[\LpL,\price]{\pmvA} 
      - \debtval[\LpL,\price]{\pmvA}
    && \text{by arith.}
      \\
      \nonumber
      & = \wealth{\confG}{\pmvA} 
      && \text{by \eqref{eq:networth}}
    \end{align}

  \item \mbox{$\actRedeem{\pmvA}{\minted{\valV}:\minted{\tokT}}$}.
  By rule~\nrule{[Rdm]}, let $\valV = {\minted{\valV}} \cdot {\X[\LpL]{\tokT}}$.
  We have now two subcases.

  \noindent
  If $\supply[\LpLi]{\tokT} > 0$, then by Lemma~\ref{lem:ER-increasing} it follows that $\X[\LpLi]{\tokT} = \X[\LpL]{\tokT}$, and so we have:
  \begin{align}
      \wealth{\confGi}{\pmvA} 
      \nonumber
      & = \baseval[\WalWi,\price]{\pmvA} 
      + \creditval[\LpLi,\price]{\pmvA} 
      - \debtval[\LpLi,\price]{\pmvA}
      && \text{by \eqref{eq:networth}}
      \\
      \nonumber
      & = 
      \Big(
          \baseval[\WalW,\price]{\pmvA} 
          + \valV \cdot \price[\tokT] 
      \Big)
      + \Big(
          \creditval[\LpL,\price]{\pmvA} 
          - \minted{\valV} \cdot \X[\LpL]{\tokT} \cdot \price[\tokT]
      \Big)
      - \debtval[\LpL,\price]{\pmvA}
      && \text{by \nrule{[Rdm]}}
      \\
      \nonumber
      & = \baseval[\WalW,\price]{\pmvA} 
      + \valV \cdot \price[\tokT] 
      + \creditval[\LpL,\price]{\pmvA} 
      - \valV \cdot \price(\tokT) 
      - \debtval[\LpL,\price]{\pmvA}
      && \text{by def. $\valV$}
    \\
      \nonumber
      & = \baseval[\WalW,\price]{\pmvA} 
      + \creditval[\LpL,\price]{\pmvA} 
      - \debtval[\LpL,\price]{\pmvA}
    && \text{by arith.}
      \\
      \nonumber
      & = \wealth{\confG}{\pmvA} 
      && \text{by \eqref{eq:networth}}
    \end{align}

    \noindent
    Otherwise, if $\supply[\LpLi]{\tokT} = 0$, then 
    it means that $\minted{\valV} = \supply[\LpL]{\tokT} = \LpL(\minted{\tokT},\pmvA)$, and 
    by Lemma~\ref{lem:ER-increasing} it follows that $\X[\LpLi]{\tokT} = 1$.
    Hence, we have that:
    \begin{align}
    \wealth{\confGi}{\pmvA} 
    \nonumber
    & = 
    \baseval[\WalW,\price]{\pmvA} 
    + \valV \cdot \price[\tokT] 
    - \debtval[\LpL,\price]{\pmvA}
    \\
    \nonumber
    & + \sum_{\tokTi \neq \tokT} \LpLi(\minted{\tokTi},\pmvA) \cdot \X[\LpLi]{\tokTi} \cdot \price[\tokTi]
    && \text{by $\LpLi(\minted{\tokT},\pmvA) = 0$}
    \\
    \nonumber
    & = 
    \baseval[\WalW,\price]{\pmvA} 
    + \valV \cdot \price[\tokT] 
    - \debtval[\LpL,\price]{\pmvA}
    \\
    \nonumber
    & + \sum_{\tokTi \neq \tokT} \LpL(\minted{\tokTi},\pmvA) \cdot \X[\LpL]{\tokTi} \cdot \price[\tokTi]
    && \text{by \nrule{[Rdm]}}
    \\
    \nonumber
    & = 
    \baseval[\WalW,\price]{\pmvA} 
    + \valV \cdot \price[\tokT] 
    - \debtval[\LpL,\price]{\pmvA}
    \\
    \nonumber
    & + \sum_{\tokTi} \LpL(\minted{\tokTi},\pmvA) \cdot \X[\LpL]{\tokTi} \cdot \price[\tokTi]
    - \LpL(\minted{\tokT},\pmvA) \cdot \X[\LpL]{\tokT} \cdot \price[\tokT]
    && \text{by arith.}
    \\
    \nonumber
    & = 
    \baseval[\WalW,\price]{\pmvA} 
    + \valV \cdot \price[\tokT] 
    - \debtval[\LpL,\price]{\pmvA}
    + \creditval[\LpL,\price]{\pmvA}
    \\
    \nonumber
    &
    - \LpL(\minted{\tokT},\pmvA) \cdot \X[\LpL]{\tokT} \cdot \price[\tokT]
    && \text{by arith.}
    \\
    \nonumber
    & = \wealth{\confG}{\pmvA} 
    + \valV \cdot \price[\tokT] 
    - \minted{\valV} \cdot \X[\LpL]{\tokT} \cdot \price[\tokT]
    && \text{by \eqref{eq:networth}, $\minted{\valV} = \LpL(\minted{\tokT},\pmvA)$}
    \\
    \nonumber
    & = \wealth{\confG}{\pmvA}
    + \valV \cdot \price[\tokT] 
    - \valV \cdot \price[\tokT]
    && \text{by def. $\minted{\valV}$}
    \\
    \nonumber
    & = \wealth{\confG}{\pmvA} 
    && \text{by arith.}
    \end{align}

  \item \mbox{$\actLiquidate{\pmvA}{\pmvB}{\valV[0]:\tokT[0]}{\mintedT[1]}$}, where $\pmvB \neq \pmvA$.
    We have that:
    \begin{align*}
      \wealth{\confGi}{\pmvA} 
      \nonumber
      & = \baseval[\WalWi,\price]{\pmvA} 
      + \creditval[\LpLi,\price]{\pmvA} 
      - \debtval[\LpLi,\price]{\pmvA}
      && \text{by \eqref{eq:networth}}
      \\
      \nonumber
      & = 
          \baseval[\WalWi,\price]{\pmvA} 
      +
      \sum_{\tokT}
          \stateCrediti{\tokT}{\pmvA}
        \cdot \X[\LpLi]{\tokT[]} \cdot \price[{\tokT[]}]
      - \debtval[\LpLi,\price]{\pmvA}
      && \text{by \eqref{eq:creditVal}}
      \\
      \nonumber
      & = 
      \baseval[\WalWi,\price]{\pmvA} 
      +
      \sum_{\tokT}
      \stateCrediti{\tokT}{\pmvA}
      \cdot \X[\LpL]{\tokT[]} \cdot \price[{\tokT[]}]
      - \debtval[\LpLi,\price]{\pmvA}
      && \text{by Lem. \ref{lem:ER-increasing}}
      \\
      \nonumber
      & = 
      \Big(
      \baseval[\WalW,\price]{\pmvA} 
      - \valV[0] \cdot \price[{\tokT[0]}] 
      \Big)
      +
      \sum_{\tokT\neq \tokT[1]}
      \stateCrediti{\tokT}{\pmvA}
      \cdot \X[\LpL]{\tokT[]} \cdot \price[{\tokT[]}]      
      \\ & \quad
      +
      \Big(
       \stateCredit{\tokT[1]}{\pmvA} 
      + \minted{\valV[1]}
      \Big)
       \cdot \X[\LpL]{\tokT[1]} \cdot \price[{\tokT[1]}]
      - \debtval[\LpL,\price]{\pmvA}
      && \text{by \nrule{[Liq]}}
      \\
      \nonumber
      & = 
      \Big(
          \baseval[\WalW,\price]{\pmvA} 
          - \valV[0] \cdot \price[{\tokT[0]}] 
      \Big)
      \\ & \quad
      +
      \Big(
          \creditval[\LpL,\price]{\pmvA} 
          + \minted{\valV[1]} \cdot \X[\LpL]{\tokT[1]} \cdot \price[{\tokT[1]}]
      \Big)
      - \debtval[\LpL,\price]{\pmvA}
      && \text{by arith.~and \eqref{eq:creditVal}}
      \\
      \nonumber
      & = \baseval[\WalW,\price]{\pmvA} 
      - \valV[0] \cdot \price[{\tokT[0]}] 
      + \creditval[\LpL,\price]{\pmvA} 
    \\ & \quad
            +
        \valV[0] 
          \cdot
        \dfrac{1}{\X[\LpL]{\tokT[1]}} 
          \cdot
        \dfrac{\price({\tokT[0]})}{\price({\tokT[1]})}
        \cdot
        \rLiq
          \cdot
          \X[\LpL]{\tokT[1]}
          \cdot
        \price({\tokT[1]}) 
      - \debtval[\LpL,\price]{\pmvA}    
      && \text{by def. $\minted{\valV[1]}$}
      \\
      \nonumber
      & = \baseval[\WalW,\price]{\pmvA} 
      - \valV[0] \cdot \price[{\tokT[0]}] 
      + \creditval[\LpL,\price]{\pmvA} 
      + 
        \valV[0]  
          \cdot
        \price({\tokT[0]}) 
          \cdot
        \rLiq
      - \debtval[\LpL,\price]{\pmvA}    
      && \text{by arith.}
      \\
      \nonumber
      & = \baseval[\WalW,\price]{\pmvA} 
      + 
      \creditval[\LpL,\price]{\pmvA} 
      - \debtval[\LpL,\price]{\pmvA}
      +
          (\rLiq - 1)
            \cdot
          \valV[0]
            \cdot
          \price({\tokT[0]}) 
    && \text{by arith.}
    \\
      & = \wealth{\confG}{\pmvA} 
    +
    (\rLiq - 1)
    \cdot
    \valV[0]
    \cdot
    \price({\tokT[0]}) 
    && \text{by \eqref{eq:networth}}
    \end{align*}
    
    Recalling that $\valV[0] > 0$, $\price({\tokT[0]}) > 0$ and $\liqReward > 1$, we obtain:
    \[
    \gain{\pmvA}{\confG}{\txT}
    \; = \;  
    (\rLiq - 1)
    \cdot
    \valV[0]
    \cdot
    \price({\tokT[0]}) 
    \; > \; 
    0
    \]

  \item \mbox{$\actLiquidate{\pmvB}{\pmvA}{\valV:\tokT[0]}{\mintedT[1]}$}, where $\pmvB \neq \pmvA$.
  We have that:
    \begin{align*}
      \wealth{\confGi}{\pmvA} 
      \nonumber
      & = \baseval[\WalWi,\price]{\pmvA} 
      + \creditval[\LpLi,\price]{\pmvA} 
      - \debtval[\LpLi,\price]{\pmvA}
      && \eqref{eq:networth}
      \\
      \nonumber
      & = 
      \baseval[\WalWi,\price]{\pmvA} 
      +
      \sum_{\tokT}
      \stateCrediti{\tokT}{\pmvA}
      \cdot \X[\LpLi]{\tokT[]} \cdot \price[{\tokT[]}]
      - \debtval[\LpLi,\price]{\pmvA}
      && \text{by \eqref{eq:creditVal}}
      \\
      \nonumber
      & = 
      \baseval[\WalWi,\price]{\pmvA} 
      +
      \sum_{\tokT}
      \stateCrediti{\tokT}{\pmvA}
      \cdot \X[\LpL]{\tokT[]} \cdot \price[{\tokT[]}]
      - \debtval[\LpLi,\price]{\pmvA}
      && \text{by Lem. \ref{lem:ER-increasing}}
      \\
      \nonumber
      & = 
          \baseval[\WalW,\price]{\pmvA} 
      + \Big(
          \sum_{\tokT \neq \tokT[1]}
          \stateCredit{\tokT}{\pmvA}
          \cdot \X[\LpL]{\tokT[]} \cdot \price[{\tokT[]}]
      \\ & \quad
          +
          (\stateCredit{\tokT[1]}{\pmvA}
          - 
          \minted{\valV[1]}) 
          \cdot \X[\LpL]{\tokT[1]} \cdot \price[{\tokT[1]}]
      \Big)
      - \Big(
        \debtval[\LpL,\price]{\pmvA}
          -
        \valV[0] \cdot \price[{\tokT[0]}]
      \Big)
      && \text{\nrule{[Liq]}}
      \\
      \nonumber
      & = \baseval[\WalW,\price]{\pmvA}
      + \creditval[\LpL,\price]{\pmvA} 
      -      
      \minted{\valV[1]}
      \cdot
      \X[\LpL]{\tokT[1]}
      \cdot
      \price({\tokT[1]}) 
      \\ & \quad
      - \debtval[\LpL,\price]{\pmvA} 
      + \valV[0] \cdot \price[{\tokT[0]}] 
      && \text{(arith. + \eqref{eq:creditVal})}
      \\
      \nonumber
      & = \baseval[\WalW,\price]{\pmvA}
      + \creditval[\LpL,\price]{\pmvA} 
      -
        \valV[0] 
          \cdot
        \dfrac{1}{\X[\LpL]{\tokT[1]}} 
          \cdot
        \dfrac{\price({\tokT[0]})}{\price({\tokT[1]})}
        \cdot
        \rLiq
          \cdot
          \X[\LpL]{\tokT[1]}
          \cdot
        \price({\tokT[1]}) 
    \\ & \quad
             - \debtval[\LpL,\price]{\pmvA} 
      + \valV[0] \cdot \price[{\tokT[0]}] 
   && \text{(def. $\minted{\valV[1]}$)}
      \\
      \nonumber
      & = \baseval[\WalW,\price]{\pmvA} 
      + \creditval[\LpL,\price]{\pmvA} 
      -
        \valV  
          \cdot
        \price({\tokT[0]}) 
          \cdot
        \rLiq
      - \debtval[\LpL,\price]{\pmvA}  
      + \valV[0] \cdot \price[{\tokT[0]}] 
      && \text{(Lemma~\ref{lem:ER-increasing})}
      \\
      \nonumber
      & = \baseval[\WalW,\price]{\pmvA} 
      +
      \creditval[\LpL,\price]{\pmvA} 
      - \debtval[\LpL,\price]{\pmvA}
      +
          (1-\rLiq)
            \cdot
          \valV[0]
            \cdot
          \price({\tokT[0]}) 
    && \text{(arith.)}
    \\
    \nonumber
    & = 
     \wealth{\confG}{\pmvA} +
    (1-\rLiq)
    \cdot
    \valV[0]
    \cdot
    \price({\tokT[0]}) 
    && \eqref{eq:networth}
    \end{align*}  
   
    Recalling that $\valV[0] > 0$, $\price({\tokT[0]}) > 0$ and $\liqReward > 1$, we obtain:
    \[
    \gain{\pmvA}{\confG}{\txT}
    \; = \;  
    (1 - \rLiq)
    \cdot
    \valV[0]
    \cdot
    \price({\tokT[0]}) 
    \; < \; 
    0
    \]

  This case, together with the previous one, proves that, if $\txT = \actLiquidate{\pmvA}{\pmvB}{\valV:\tokT[0]}{\mintedT[1]}$, then
  \(
    \gainSt{\pmvA}{\confG}{\txT} 
    = - \gainSt{\pmvB}{\confG}{\txT} 
    = \valV \cdot \price[{\tokT[0]}] \cdot \rLiq
    \)
    Note that $\gainSt{\pmvB}{\confG}{\txT}$ is obtained from this case by switching $\pmvA$ and $\pmvB$.  
  
  \item  \mbox{$\swapOp(\valV: \tokT[0], \tokT[1])$},
  with $\valVi= \valV \cdot \frac{\statePrice[{\tokT[0]}]{}}{\statePrice[{\tokT[1]}]{}}$.
  We have that:
  
%
\begin{align*}
	\wealth{\confGi}{\pmvA} 
	& = \baseval[\WalWi,\price]{\pmvA} 
	+ \creditval[\LpLi,\price]{\pmvA} 
	- \debtval[\LpLi,\price]{\pmvA}
	&& \text{by \eqref{eq:networth}}
	\\
	& = 
	\Big(
		\baseval[\WalW,\price]{\pmvA} 
		- \valV \cdot \price[{\tokT[0]}] 
		+ \valVi \cdot \price[{\tokT[1]}] 
	\Big)
	+ \creditval[\LpL,\price]{\pmvA} 
	- \debtval[\LpL,\price]{\pmvA}
	&& \text{by \nrule{[swp]}}
	\\
	& = 
	\Big(
	\baseval[\WalW,\price]{\pmvA} 
	- \valV \cdot \price[{\tokT[0]}] 
	+ \valV \cdot \frac{\statePrice[{\tokT[0]}]{}}{\statePrice[{\tokT[1]}]{}}  \cdot \price[{\tokT[1]}] 
	\Big)
	+ \creditval[\LpL,\price]{\pmvA} 
	- \debtval[\LpL,\price]{\pmvA}
	&& \text{by def. $\valVi$} 
	\\
	& = 
	\baseval[\WalW,\price]{\pmvA} 
	+ \creditval[\LpL,\price]{\pmvA} 
	- \debtval[\LpL,\price]{\pmvA}
	&&  \text{by arith.}
	\\     
	 & = \wealth{\confG}{\pmvA} 
	&& \text{by \eqref{eq:networth}}
\end{align*}
  \end{itemize}
  \qed
\end{lemproof}


\begin{lemproof}{lem:gain-px}

Let $\confG = (\WalW,\LpL,\price)$, and let $p = \price[\tokT]$.
Then, given $\txT = \exchUpdateOp{\prIncr:\tokT}$, by~\eqref{eq:wealth-restricted} we have to prove that: 
\[
    \gain{\pmvA}{\confG}{\txT}
    =
    \left(
    \WalW(\tokT,\pmvA) 
    +     
    \LpL(\minted{\tokT},\pmvA) \cdot \X[{\LpL}]{\tokT}
    -
    \LpL(\debt{\tokT},\pmvA)  
    \right)                  
    \cdot 
    \prIncr
\]
Assume that the state transition is given by:
\[
  \confG = (\WalW,\LpL,\price) 
  \; \xrightarrow{\txT} \;
  \confGi = (\WalWi,\LpLi,\pricei)
\]
First, note that, since only the price of $\tokT$ changes, it trivially holds that $\projToki{\wealth{\confGi}{\pmvA}} = \projToki{\wealth{\confG}{\pmvA}}$ for every $\tokTi \neq \tokT$.
Hence, $\gainSt{\pmvA}{\confG}{\txT} = \projTok{\gainSt{\pmvA}{\confG}{\txT}}$, and so we can restrict to analyse 
$\projTok{\gainSt{\pmvA}{\confG}{\txT}} = \projTok{\wealth{\confGi}{\pmvA}} - \projTok{\wealth{\confG}{\pmvA}}$.
The wealth of $\pmvA$ in the old state $\confG$ is given by:
  \begin{align*}  
      \projTok{\wealth{\confG}{\pmvA}} 
      \nonumber
      & = \projTok{\baseval[\WalW,\price]{\pmvA}}
      + \projTok{\creditval[\LpL,\price]{\pmvA}}
      - \projTok{\debtval[\LpL,\price]{\pmvA}}
      && \text{by \eqref{eq:networth}}
      \\
      & \; = \;  
        \stateBase{\tokT}{\pmvA} \cdot \prVal
        +     
            (\stateCredit{\tokT}{\pmvA} \cdot \X[{\LpL}]{\tokT}
            \cdot \prVal)
        -
            (            
            \stateDebt{\tokT}{\pmvA}    
                \cdot 
            \prVal
            )   
      && \text{by \eqref{eq:baseVal},\eqref{eq:creditVal},\eqref{eq:debtVal}} 
      \\   
      & \; = \;  
        \Big(
        \stateBase{\tokT}{\pmvA} 
        +     
            \stateCredit{\tokT}{\pmvA} \cdot \X[{\LpL}]{\tokT}
        -         
            \stateDebt{\tokT}{\pmvA}    
        \Big)   
            \cdot 
        \prVal
      && \text{by arith.}
      \end{align*}
  while for the new state, we have that:
  \begin{align*}  
      \projTok{\wealth{\confGi}{\pmvA}}
      \nonumber
      & = \projTok{\baseval[\WalWi,\price]{\pmvA}}
      + \projTok{\creditval[\LpLi,\price]{\pmvA}}
      - \projTok{\debtval[\LpLi,\price]{\pmvA}}
      && \text{by \eqref{eq:networth}}
      \\
      & =  
        \stateBasei{\tokT}{\pmvA} \cdot (\prVal + \prIncr)
        +     
            \stateCrediti{\tokT}{\pmvA} \cdot \X[{\LpLi}]{\tokT}
            \cdot (\prVal + \prIncr)
        \\
        & \phantom{=}\, 
        - \stateDebti{\tokT}{\pmvA}    
                \cdot 
           (\prVal + \prIncr)
      && \text{by \eqref{eq:baseVal},\eqref{eq:creditVal},\eqref{eq:debtVal}} 
      \\   
      & =  
        \stateBase{\tokT}{\pmvA} \cdot (\prVal + \prIncr)
        +     
            \stateCredit{\tokT}{\pmvA} \cdot \X[{\LpL}]{\tokT}
            \cdot (\prVal + \prIncr)
        \\
        & \phantom{=}\,
        -
            \stateDebt{\tokT}{\pmvA}    
                \cdot 
            (\prVal + \prIncr)
      && \text{by \nrule{[Px]} and ~\eqref{lem:ER-increasing}}
      \\   
      & =  
        \Big(
        \stateBase{\tokT}{\pmvA} 
        +     
            \stateCredit{\tokT}{\pmvA} \cdot \X[{\LpL}]{\tokT}
        -         
            \stateDebt{\tokT}{\pmvA}    
        \Big)   
            \cdot 
        (\prVal + \prIncr)
      && \text{by arith.}
    \end{align*}
    Summing up:
    \begin{align*}
  \gainSt{\pmvA}{\confG}{\txT}
      & \; = \;  
         \projTok{\gainSt{\pmvA}{\confG}{\txT}}
      && \text{by previous observation}
      \\   
      & \; = \;  
          \projTok{\wealth{\confGi}{\pmvA}}
                -
            \projTok{\wealth{\confG}{\pmvA}} 
      && \text{by \eqref{eq:gainSt}}
      \\   
      & \; = \;        
        \Big(
        \stateBase{\tokT}{\pmvA} 
        +     
            \stateCredit{\tokT}{\pmvA} \cdot \X[{\LpL}]{\tokT}
        -         
            \stateDebt{\tokT}{\pmvA}    
        \Big)   
            \cdot 
        \prIncr
      && \text{by arith.}
      \tag*{\qed}
      \end{align*}
\end{lemproof}

\begin{lemproof}{lem:gain-int}
Let 
$\confG = (\WalW,\LpL,\price)$. 
We have to prove that the gain of a user $\pmvA$ upon an $\accrueIntOp$ transaction in $\confG$ \wrt $\tokT$ is given by:
\begin{equation*}
\gainSt{\pmvA}{\confG}{\accrueIntOp}_{|\tokT}
\; = \;
\begin{cases}
    \left(
    \stateCredit{\tokT}{\pmvA}{}
    \cdot 
    \dfrac                
    {\supplyDebt[\LpL]{\tokT}}
    {\supply[{\LpL}]{\tokT}}
    - 
    \LpL(\debt{\tokT},\pmvA)
    \right)
    \cdot 
    \Big(
    \Intr{\LpL}(\tokT)
    \cdot \price[\tokT]
    \Big)
    & \text{if ${\supply[{\LpL}]{\tokT}} > 0$}
    \\ \\
    0
    & \text{otherwise}
\end{cases}
\end{equation*}
Assume that the state transition is given by:
\[
    \confG = (\WalW,\LpL,\price) 
    \; \xrightarrow{\txT} \;
    \confGi = (\WalWi,\LpLi,\price)
\]
By unfolding the definition of $\wealth{}{}$ and its components
with \eqref{eq:baseVal}---\eqref{eq:networth}, we have that:
\begin{align}
    \nonumber
      \projTok{\wealth{\confGi}{\pmvA}}
      & = \projTok{\baseval[\WalWi,\price]{\pmvA}} 
      + \projTok{\creditval[\LpLi,\price]{\pmvA}}
      - \projTok{\debtval[\LpLi,\price]{\pmvA}}
      \\
      \nonumber
      & =
        \stateBasei{\tokT}{\pmvA} \cdot \pricei[\tokT]
        +     
            \stateCrediti{\tokT}{\pmvA} \cdot \X[{\LpLi}]{\tokT}
            \cdot \pricei[\tokT]
        -
            \stateDebti{\tokT}{\pmvA}    
                \cdot 
            \pricei[\tokT]
      \\ 
      \label{eq:gain-int:1}     
      & =  
      \Big(
        \stateBase{\tokT}{\pmvA} 
        +     
            \stateCredit{\tokT}{\pmvA} \cdot \X[{\LpLi}]{\tokT}
        -
            \left(
                \stateDebt{\tokT}{\pmvA}   
                \cdot
                (1 + \Intr{\LpL}(\tokT))
            \right)
        \Big)
                \cdot 
            \price[\tokT]
\end{align}
We have two subcases.
If ${\supply[{\LpL}]{\tokT}} > 0$, by Lemma \ref{lem:ER-increasing} and \eqref{eq:gain-int:1} we obtain:
\begin{equation}
\label{eq:gain-int:2}
\begin{split}
      \projTok{\wealth{\confGi}{\pmvA}}
      =
	\Biggl(
        & 
	\stateBase{\tokT}{\pmvA} 
	+     
	\stateCredit{\tokT}{\pmvA} 
	\cdot
	\left(
	\X[\LpL]{\tokT}
	+   
	\dfrac                
	{\supplyDebt[\LpL]{\tokT}}    
	{\supply[\LpL]{\tokT}}    
	\cdot \Intr{\LpL}(\tokT) 
	\right)
        \\
	& -
	\left(
	\stateDebt{\tokT}{\pmvA}   
	\cdot
	(1 + \Intr{\LpL}(\tokT))
	\right)
	\Biggl)
	\cdot 
	\price[\tokT]
\end{split}
\end{equation}
and so we obtain the following gain restricted to $\tokT$:
\begin{align*}
  \projTok{\gainSt{\pmvA}{\confG}{\txT}}
      & \; = \;  
          \projTok{\wealth{\confGi}{\pmvA}}
                -
            \projTok{\wealth{\confG}{\pmvA}} 
      && \text{by \eqref{eq:gainSt}}
      \\   
      & \; = \;        
        \left(   
            \stateCredit{\tokT}{\pmvA} \cdot 
                \dfrac                
                {\supplyDebt[\LpL]{\tokT}}    
                {\supply[\LpL]{\tokT}}    
                \cdot \Intr{\LpL}(\tokT) 
        -      
            \left(
                \stateDebt{\tokT}{\pmvA}   
                    \cdot
                 \Intr{\LpL}(\tokT)
             \right)
        \right)   
            \cdot 
        \price[\tokT]
      && \text{by \eqref{eq:gain-int:2}}
      \\   
      & \; = \;        
        \left(   
            \stateCredit{\tokT}{\pmvA} \cdot 
                \dfrac                
                {\supplyDebt[\LpL]{\tokT}}
                {\supply[\LpL]{\tokT}}
        -         
            \stateDebt{\tokT}{\pmvA}   
        \right)   
            \cdot
            \Big(
              \Intr{\LpL}(\tokT) 
                \cdot
            \price[\tokT]    
            \Big)
      && \text{by arith.}
\end{align*}

\noindent
Otherwise, if ${\supply[{\LpL}]{\tokT}} = 0$, 
then, by Lemma~\ref{lem:ER-increasing} we have that
$\X[\LpLi]{\tokT} = \X[\LpL]{\tokT} $, and by Lemma~\ref{lem:cred0-imp-debt0} we have that $\stateDebt{\tokT}{\pmvA}  = 0$. 
Hence we have:
\begin{align*}  
      \projTok{\wealth{\confGi}{\pmvA}}
      & =  
      \Big(
        \stateBase{\tokT}{\pmvA} 
        +     
            \stateCredit{\tokT}{\pmvA} \cdot \X[{\LpLi}]{\tokT}
        -
            \left(
                \stateDebt{\tokT}{\pmvA}   
                \cdot
                (1 + \Intr{\LpL}(\tokT))
            \right)
        \Big)
                \cdot 
            \price[\tokT]
      && \text{by~\eqref{eq:gain-int:1}}  
      \\
      & =  
      \Big(
        \stateBase{\tokT}{\pmvA} 
        +     
            \stateCredit{\tokT}{\pmvA} \cdot \X[{\LpLi}]{\tokT}
        \Big)
                \cdot 
            \price[\tokT]
      && \text{by Lem.~\ref{lem:cred0-imp-debt0}}  
      \\
      & =  
      \Big(
        \stateBase{\tokT}{\pmvA} 
        +     
            \stateCredit{\tokT}{\pmvA} \cdot \X[{\LpL}]{\tokT}
        \Big)
                \cdot 
            \price[\tokT]
      && \text{by Lem.~\ref{lem:ER-increasing}}  
      \\
      & =  
      \projTok{\wealth{\confG}{\pmvA}}
      &&  \text{by \eqref{eq:networth}}
\end{align*}   
from which we have the thesis 
$\projTok{\gainSt{\pmvA}{\confG}{\txT}} = 0$.
\qed
\end{lemproof}

\begin{lemproof}{lem:health-tx}
Let $\txT = \pmvA:\ell(\cdots)$, and let 
$\confG \xrightarrow{\txT} \confGi$.
We have to prove that:
\begin{itemize}

\item $\ell \in \setenum{\depositOp,\repayOp, \liquidateOp} \;\implies
		\Health{\confGi}{\pmvA} \geq \Health{\confG}{\pmvA}$

\item $\ell \in \setenum{\borrowOp,\redeemOp} \;\implies
\Health{\confGi}{\pmvA} \leq \Health{\confG}{\pmvA}$ 

\item $\ell \in \setenum{\swapOp} \;\implies
\Health{\confGi}{\pmvA} = \Health{\confG}{\pmvA}
$ 
\end{itemize}
and the inequalities are strict if and only if $\debtval[\confG]{\pmvA}>0$.

Let $\confG = (\WalW,\LpL,\price)$ and let $\confGi = (\WalWi,\LpLi,\pricei)$.
Note that since we are excluding $\pxOp{}$, the prices remain constant, and so $\pricei = \price$.
Recall the the health factor is defined by:
\begin{equation*} 
	\Health{\LpL,\price}{\pmvA}
	\; = \;
	\begin{cases}
		\dfrac{\creditval[\LpL,\price]{\pmvA}}{\debtval[\LpL,\price]{\pmvA}}
		\cdot \liqThreshold
		& \text{if } \debtval[\LpL,\price]{\pmvA} > 0 \\
		+ \infty 
		& \text{otherwise}  
	\end{cases}
\end{equation*}

We start by noting that, if $\debtval[\LpL,\price]{\pmvA} = 0$, the health factor cannot increase, and it decreases if and only if $\pmvA$'s debts  increase (\ie they become strictly positive).
Otherwise, if  $\debtval[\LpL,\price]{\pmvA} = 0$, we have that: 
\begin{enumerate}
\item if the credits increase and the debts do not change, the health factor \emph{increases};
\item if the credits do not change and the debts decrease, the health factor \emph{increases};
\item if the credits decrease and the debts do not change, the health factor \emph{decrease};
\item if the credits do not change and the debts increase, the health factor \emph{decreases};
\item if the credits and the debts do not change, the health factor \emph{remains constant}.
\end{enumerate} 

We then analyse the change in $\pmvA$' credits and debts based on $\ell$:
\begin{itemize}
	
\item \mbox{$\ell = \depositOp$}:
credits increase and debts do not change (case 1).
	
\item \mbox{$\ell = \borrowOp$}: 
debts increase and credits do not change (case 4).
	
\item \mbox{$\ell = \repayOp$}:
debts decreases and credits do not change (case 2).

\item \mbox{$\ell = \redeemOp$}:
credits decrease and debts do not change (case 3).
	
\item \mbox{$\ell = \liquidateOp$}: 
credits increase and debts do not change (case 1).

\item \mbox{$\ell = \swapOp$}: 
the credits  and the debts do not change (case 5).
\qed
\end{itemize}
\end{lemproof}

\begin{lemproof}{lem:health-dep-rep}
	Let
	$\confG[0] \xrightarrow{\actDeposit{\pmvA}{\valV : \tokT}} \confG[1]$
	and
	$\confG[0] \xrightarrow{\actRepay{\pmvA}{\valV : \tokT}} \confGi[1]$.
	We have to prove that:
	\[
	\Health{\confGi[1]}{\pmvA} \geq \Health{\confG[1]}{\pmvA} 
	\text{ if and only if } 
	\valV \cdot \price[\tokT] 
	\geq 
	{\debtval[{\confG[0]}]{\pmvA}}
	-
	{\creditval[{\confG[0]}]{\pmvA}}
	\]
	Let 
	$\confG[0] = (\WalW[0],\LpL[0],\statePrice{0})$, let
	$\confG[1] = (\WalW[1],\LpL[1],\statePrice{1})$, and let
	$\confGi[1] = (\WalWi[1],\LpLi[1],\statePricei{1})$.
	By Lemma \ref{lem:ER-increasing}, we have that both transitions preserve the exchange rate, \ie $\X[{\LpL[1]}]{\tokT} = \X[{\LpL[0]}]{\tokT} = \X[{\LpL[0]}]{\tokT}$.
    
	We first compute the health factor in $\confG[1]$. 
   	First, note that
	$\debtval[{\LpL[0]},\price]{\pmvA} > 0$,
	since the rule~\nrule{[Rep]} is enabled in $\confG[0]$ and its premise requires that the repayer has a strictly positive debit.
	Since $\depositOp$ does not modify the debts, then
        $\debtval[{\LpL[1]},\price]{\pmvA} = \debtval[{\LpL[0]},\price]{\pmvA} > 0$.
        Therefore:
	\begin{align*}
		\Health{\confG[1]}{\pmvA}
		& \; = \;
		\dfrac        
		{\creditval[{\LpL[1]},\price]{\pmvA}}      
		{\debtval[{\LpL[1]},\price]{\pmvA}}
		\cdot 
		{\liqThreshold}
		&& \text{by ~\eqref{eq:healthfactor}}
		\\
		& \; = \;
		\dfrac
		{      
			\sum_{\tokT[i]}
			{\stateCredit[{\LpL[1]}]{\tokT[i]}{\pmvA}} 
			\cdot \X[{\LpL[1]}]{\tokT[i]}
			\cdot \price[{\tokT[i]}]
		}
		{
			\sum_{\tokT[i]} 
			{\stateDebt[{\LpL[1]}]{\tokT}{\pmvA}} 
			\cdot \price[{\tokT[i]}]
		}
		\cdot 
		{\liqThreshold}
		&& \text{by~\eqref{eq:creditVal},\eqref{eq:debtVal}} 
		\\
		& \; = \;
		\dfrac
		{      
			\left(
                \sum_{\tokT[i]}
			{\stateCredit[{\LpL[0]}]{\tokT[i]}{\pmvA}} 
			\cdot \X[{\LpL[1]}]{\tokT[i]}
			\cdot \price[{\tokT[i]}]
                \right)
			+
			\nicefrac{\valV}{\X[{\LpL[0]}]{\tokT}}
			\cdot \X[{\LpL[1]}]{\tokT}
			\cdot \price[{\tokT}]
		}
		{
			\sum_{\tokT[i]} 
			{\stateDebt[{\LpL[0]}]{\tokT}{\pmvA}} 
			\cdot \price[{\tokT[i]}]
		}
		\cdot 
		{\liqThreshold}
		&& \text{by \nrule{[Dep]}} 
		\\
		& \; = \;
		\dfrac
		{    
                \left(
			\sum_{\tokT[i]}
			{\stateCredit[{\LpL[0]}]{\tokT[i]}{\pmvA}} 
			\cdot \X[{\LpL[0]}]{\tokT[i]}
			\cdot \price[{\tokT[i]}]
			\right)
                +
			{\valV}
			\cdot \price[{\tokT}]
		}
		{
			\sum_{\tokT[i]} 
			{\stateDebt[{\LpL[0]}]{\tokT}{\pmvA}} 
			\cdot \price[{\tokT[i]}]
		}
		\cdot 
		{\liqThreshold}
		&& \text{by Lem.~\ref{lem:ER-increasing}} 
		\\
		& \; = \;
		\dfrac        
		{\creditval[{\LpL[0]},\price]{\pmvA}
		+
		{\valV}
		\cdot \price[{\tokT}]
		}      
		{\debtval[{\LpL[0]},\price]{\pmvA}}
		\cdot 
		{\liqThreshold}
		&& \text{by~\eqref{eq:creditVal},\eqref{eq:debtVal}}
	\end{align*}   
	
	We then compute the health factor in $\confGi[1]$.
	There are two subcases, depending on whether $\pmvA$ repays the entirety of her debts or not.
	
	If $\pmvA$ repays all her debts,
	\ie  $\valV = \stateDebt[{\LpL[0]}]{\tokT}{\pmvA}$, then 
	$\valV \cdot \price[\tokT] 
	=
		{\debtval[{\LpL[0]}]{\pmvA}}
	\geq 
	{\debtval[{\LpL[0]}]{\pmvA}}
	-
	{\creditval[{\LpL[0]}]{\pmvA}}$
	 and $\debtval[{\LpLi[1]},\price]{\pmvA} = 0$. 
	 Hence, by the definition of health factor,  $\Health{\LpLi[1]}{\pmvA} = + \infty > \Health{\LpL[1]}{\pmvA}$. 
    	      
	Otherwise, if $\pmvA$ does not repay the entirety of her debts,	we have that $\debtval[{\LpLi[1]},\price]{\pmvA}>0$, and the health factor in $\confGi[1]$ is given by:
	\begin{align*}
		\Health{\confGi[1]}{\pmvA}
		& \; = \;
		\dfrac        
		{\creditval[{\LpLi[1]},\price]{\pmvA}}      
		{\debtval[{\LpLi[1]},\price]{\pmvA}}
		\cdot 
		{\liqThreshold}
		&& \text{by ~\eqref{eq:healthfactor}}
		\\
		& \; = \;
		\dfrac
		{      
			\sum_{\tokT[i]}
			{\stateCredit[{\LpLi[1]}]{\tokT[i]}{\pmvA}} 
			\cdot \X[{\LpLi[1]}]{\tokT[i]}
			\cdot \price[{\tokT[i]}]
		}
		{
			\sum_{\tokT[i]} 
			{\stateDebt[{\LpLi[1]}]{\tokT}{\pmvA}} 
			\cdot \price[{\tokT[i]}]
		}
		\cdot 
		{\liqThreshold}
		&& \text{by~\eqref{eq:creditVal},\eqref{eq:debtVal}} 
		\\
		& \; = \;
		\dfrac
		{      
			\sum_{\tokT[i]}
			{\stateCredit[{\LpL[0]}]{\tokT[i]}{\pmvA}} 
			\cdot \X[{\LpLi[1]}]{\tokT[i]}
			\cdot \price[{\tokT[i]}]
		}
		{
			\sum_{\tokT[i]} 
			{\stateDebt[{\LpL[0]}]{\tokT}{\pmvA}} 
			\cdot \price[{\tokT[i]}]
			-
			{\valV}
			\cdot \price[{\tokT}]
		}
		\cdot 
		{\liqThreshold}
		&& \text{by \nrule{[Rep]}} 
		\\
		& \; = \;
		\dfrac
		{      
			\sum_{\tokT[i]}
			{\stateCredit[{\LpL[0]}]{\tokT[i]}{\pmvA}} 
			\cdot \X[{\LpL[0]}]{\tokT[i]}
			\cdot \price[{\tokT[i]}]
		}
		{
			\sum_{\tokT[i]} 
			{\stateDebt[{\LpL[0]}]{\tokT}{\pmvA}} 
			\cdot \price[{\tokT[i]}]
			-
			{\valV}
			\cdot \price[{\tokT}]
		}
		\cdot 
		{\liqThreshold}
		&& \text{by Lem.~\ref{lem:ER-increasing}} 
		\\
		& \; = \;
		\dfrac        
		{\creditval[{\LpL[0]},\price]{\pmvA}
		}      
		{\debtval[{\LpL[0]},\price]{\pmvA}
		-
		{\valV}
		\cdot \price[{\tokT}]
		}
		\cdot 
		{\liqThreshold}
		&& \text{by  ~\eqref{eq:creditVal} and  ~\eqref{eq:debtVal}}
	\end{align*}   
	
	Let $A = \creditval[{\confG[0]}]{\pmvA}$, 
	let $B = \debtval[{\confG[0]}]{\pmvA}$, 
	and let 
	$C = \valV \cdot \price[{\tokT}]$.
	Note that $B \geq C$ holds by the premise of \nrule{[Rep]}. 
        In particular, in the case that $B = C$, then the health factor is by definition $+\infty$, which is greater than $0$. So from now on, we only consider the case $B>C$.
    The proof follows from the following auxiliary result:
    \begin{equation}
    \label{eq:health-dep-rep:aux}
    \forall A \in \mathbb{R},\; B>C>0 \in \mathbb{R} \quad : \quad
    \frac{A}{B-C} \geq \frac{A+C}{B} \iff C \geq B-A
    \end{equation}
    To prove~\eqref{eq:health-dep-rep:aux}, note that:
    \begin{align*} 
    \frac{A}{B-C} \geq \frac{A+C}{B}         
    \iff         
    & A \cdot B \geq (A+C) \cdot (B-C)           
    \\ 
    \iff        
    & A \cdot B \geq A \cdot B - A \cdot C + B \cdot C - C^2              
    \\ 
    \iff
    & 0 \geq (-A+B-C) \cdot C 
    \\
    \iff  
    & C \geq B-A
    \tag*{\qed}    
    \end{align*}
\end{lemproof}

\section{Proofs for~\Cref{sec:strategic-properties}}
\label{proofs:strategic-properties}


\begin{thmproof}{th:gameLiq}
	Let $\pmvA$ and $\confG$ such that $\Health{\confG}{\pmvA}<1$, and let $\liquidateOp$ be a shorthand for an arbitrary liquidation on $\pmvA$ enabled in $\confG$.
	
	Let $\txT = \pmvA:\ell(\valV: \tokT)$ with $\ell \in \setenum{\depositOp,\repayOp}$, 
	and $\confG \xrightarrow{\txT} \confGi$.
	Then we have to prove that:
	
	\begin{enumerate}
		\item If $\txT = \depositOp$, then $\gain{\pmvA}{\confG}{\txT \, \liquidateOp} 
		\; > \; 
		\gain{\pmvA}{\confG}{\liquidateOp}
		\iff
		\valV 
		\geq 
		\frac{\X[\confG]{\tokT}}
		{\statePrice[\tokT]{\confG}}
		\cdot \left(
		\dfrac
		{\debtval[{\confG}]{\pmvA}}
		{\liqThreshold}
		-
		\creditval[\confG]{\pmvA}
		\right)
		$
		\item If $\txT = \repayOp$, then $\gain{\pmvA}{\confG}{\txT \, \liquidateOp} 
		\; > \; 
		\gain{\pmvA}{\confG}{\liquidateOp}
		\iff
		\valV 
		\geq 
		\frac{1}
		{\statePrice[\tokT]{\confG}}
		\cdot \left(
		{\debtval[{\confG}]{\pmvA}}
		-
		\creditval[\confG]{\pmvA}
		\cdot
		{\liqThreshold}
		\right)
		$
	\end{enumerate}

Note that we have to cases:
\begin{itemize}
	
	\item If after firing $\txT$ the $\liquidateOp$ is still enabled, we have that $\gain{\pmvA}{\confG}{\txT \, \liquidateOp} = \gain{\pmvA}{\confG}{\txT } + \gain{\pmvA}{\confGi}{\liquidateOp} = 0 +  \gain{\pmvA}{\confGi}{\liquidateOp}  <0 $ 	 by Lemma \ref{lem:gain-base}.
	\item If after firing $\txT$ the $\liquidateOp$ is not enabled anymore, we have that $\gain{\pmvA}{\confG}{\txT \, \liquidateOp} = \gain{\pmvA}{\confG}{\txT }  = 0 $ 	 by Lemma \ref{lem:gain-base}.
\end{itemize}
	Hence, we now compute the threshold values of $\valV$ for which the $\liquidateOp$ gets disabled, \ie for which $\Health{\confGi}{\pmvA} = 1$ (since by Lemma \ref{lem:health-tx} both $\depositOp$ and $\repayOp$ increase the health factor, for higher values of $\valV$ we have that $\Health{\confGi}{\pmvA} \geq 1$). 
	Note that, if we consider a fixed liquidation on $\pmvA$ with parameter $\valV[l]$, then, in the case of $\ell = \repayOp$, it would be possible to disable the $\liquidateOp$ also by making the condition $\stateDebt[\LpL]{\tokT}{\pmvA}\geq \valV[l]$ false, by repaying a $\valV > \stateDebt[\LpL]{\tokT}{\pmvA}-\valV[l]$.
	However, in general, a user  in danger of being liquidated does not precisely know  the amount of debt that a liquidator will try to liquidate. 
	For arbitrary values of $\valV[l]$ indeed, as it is in the hypothesis of the theorem, the only way to avoid being liquidated is making the health factor greater or equal to $1$.
	
\begin{itemize}
	\item If $\txT = \depositOp$, then
\[
\Health{\confGi}{\pmvA} = 
\frac
{
\creditval[\confG]{\pmvA}
+
\nicefrac{\valV}{\X[\confG]{\tokT}}
\cdot 
\statePrice[\tokT]{\confG}
}
{
	\debtval[\confG]{\pmvA}
}
\cdot
\liqThreshold
\]
hence we have that 	$\Health{\confGi}{\pmvA}\geq 1$ if and only if	
	\[
	\valV 
	\geq 
	\frac{\X[\confG]{\tokT}}
	{\statePrice[\tokT]{\confG}}
	\cdot \left(
	\dfrac
	{\debtval[{\confG}]{\pmvA}}
	{\liqThreshold}
	-
	\creditval[\confG]{\pmvA}
	\right)
	\]
	
	\item If $\txT = \repayOp$, then
	\[
	\Health{\confGi}{\pmvA} = 
	\frac
	{
		\creditval[\confG]{\pmvA}
	}
	{
		\debtval[\confG]{\pmvA}
		-
		{\valV}
		\cdot 
		\statePrice[\tokT]{\confG}
	}
	\cdot
	\liqThreshold
	\]
	hence we have that 	$\Health{\confGi}{\pmvA}\geq 1$ if and only if	
	\[ 
	\gain{\pmvA}{\confG}{\liquidateOp}
	\iff
	\valV 
	\geq 
	\frac{1}
	{\statePrice[\tokT]{\confG}}
	\cdot \left(
	{\debtval[{\confG}]{\pmvA}}
	-
	\creditval[\confG]{\pmvA}
		\cdot
	{\liqThreshold}
	\right)
	\]

\end{itemize}

\end{thmproof}


\begin{thmproof}{th:gamePx}
Let $\pxOp{}$ be a shorthand for $\exchUpdateOp{\prIncr:\tokT}$.
Let $\confG \xrightarrow{\txT} \confGi$
with $\txT = \pmvA:\ell(\cdots)$ mentioning token $\tokT$.
We have to prove that:
\begin{equation}
\label{eq:gamePx:0}
\gain{\pmvA}{\confG}{\txT \, \exchUpdateOp{}} 
\; \circ \;
\gain{\pmvA}{\confG}{\exchUpdateOp{}}
\; = \;
\gain{\pmvA}{\confG}{\exchUpdateOp{} \, \txT} 
\end{equation}
where the relation $\circ$ is given by:
\[
\circ
\; = \;
\begin{cases}
=
& \text{if $\ell \in \setenum{\depositOp, \repayOp, \borrowOp, \redeemOp}$}
\\
\,> & \text{if ($\prIncr > 0$ and $\ell = \swapOp(\valV: \tokTi, \tokT)$) or ($\prIncr < 0$ and $\ell = \swapOp(\valV: \tokT, \tokTi)$) }
\\
\,< & \text{if ($\prIncr < 0$ and $\ell = \swapOp(\valV: \tokTi, \tokT)$) or ($\prIncr > 0$ and $\ell = \swapOp(\valV: \tokT, \tokTi)$) }
\end{cases}
\]
More precisely, we have to prove that if $\confG$ has price function $\price$, then:
\[
\gain{\pmvA}{\confG}{\txT \, \exchUpdateOp{}} 
\; = \;
\gain{\pmvA}{\confG}{\exchUpdateOp{}} +    
    \sigma \cdot
    \valV \cdot \prIncr \cdot 
    \left(
    \frac{\price(\tokTi)}{\price(\tokT)}
    \right)^{\sigma}
    \qquad
    \sigma
        \; = \; 
    \begin{cases}
    1
    & \text{$\ell = \swapOp(\valV: \tokTi, \tokT)$}
    \\
    -1
    & \text{$\ell = \swapOp(\valV: \tokT, \tokTi)$}
\end{cases}
\]







Note that the rightmost equality in~\eqref{eq:gamePx:0} is given by Lemma~\ref{lem:gain-base}. 
Hence, we only have to prove the leftmost equality/inequality.
Under the hypotheses of the theorem, we have:
\begin{align*}
\gain{\pmvA}{\confG}{\txT \, \pxOp{}}
& =
\gain{\pmvA}{\confG}{\txT}
+
\gain{\pmvA}{\confGi}{\pxOp{}}
&& \text{by~\eqref{eq:gainSt}}
\\
& = \gain{\pmvA}{\confGi}{\pxOp{}}
&& \text{by Lem.~\ref{lem:gain-base}}
\\
& =
\left(
\stateBase[{\WalWi}]{\tokT}{\pmvA}
+     
\stateCredit[{\LpLi}]{\tokT}{\pmvA}
\cdot \X[{\LpLi}]{\tokT}
- \stateDebt[{\LpLi}]{\tokT}{\pmvA}
\right)                  
\cdot 
\prIncr
&& \text{by Lem.~\ref{lem:gain-px}}
\end{align*}
We proceed from here by cases on $\ell$:
\begin{itemize}


\item  \mbox{$\actDeposit{\pmvA}{\valV:\tokT}$}.
\begin{align*}
\gain{\pmvA}{\confG}{\txT \, \pxOp{}}
& =
\left(
(
\stateBase[{\WalW}]{\tokT}{\pmvA}
-
\valV
)
+    
( 
\stateCredit[{\LpL}]{\tokT}{\pmvA}
+
\nicefrac{\valV}{ \X[{\LpL}]{\tokT}}
)
\cdot \X[{\LpLi}]{\tokT}
-		
\stateDebt[{\LpL}]{\tokT}{\pmvA}
\right)                  
\cdot 
\prIncr
&& \text{by \nrule{[Dep]}}	
\\
& =
\left(
(
\stateBase[{\WalW}]{\tokT}{\pmvA}
-
\valV
)
+    
( 
\stateCredit[{\LpL}]{\tokT}{\pmvA}
+
\nicefrac{\valV}{ \X[{\LpL}]{\tokT}}
)
\cdot \X[{\LpL}]{\tokT}
-		
\stateDebt[{\LpL}]{\tokT}{\pmvA}
\right)                  
\cdot 
\prIncr
&& \text{by Lem~\ref{lem:ER-increasing}}	
\\
& =
\left(
\stateBase[{\WalW}]{\tokT}{\pmvA}
+    
\stateCredit[{\LpL}]{\tokT}{\pmvA}
\cdot \X[{\LpL}]{\tokT}
-		
\stateDebt[{\LpL}]{\tokT}{\pmvA}
\right)                  
\cdot 
\prIncr
&& \text{by arith.}
\\
& =
\gain{\pmvA}{\confG}{\pxOp{}}
&& \text{by Lem.~\ref{lem:gain-px}}
\end{align*}


\item  \mbox{$\actBorrow{\pmvA}{\valV:\tokT}$}.
\begin{align*}
\gain{\pmvA}{\confG}{\txT \, \pxOp{}}
& =
\left(
(
\stateBase[{\WalW}]{\tokT}{\pmvA}
+
\valV
)
+     
\stateCredit[{\LpL}]{\tokT}{\pmvA}
\cdot \X[{\LpLi}]{\tokT}
- 
(
\stateDebt[{\LpL}]{\tokT}{\pmvA}
+ 
\valV
)
\right)                  
\cdot 
\prIncr
&& \text{by \nrule{[Bor]}}	
\\
& =
\left(
(
\stateBase[{\WalW}]{\tokT}{\pmvA}
+
\valV
)
+     
\stateCredit[{\LpL}]{\tokT}{\pmvA}
\cdot \X[{\LpL}]{\tokT}
- 
(
\stateDebt[{\LpL}]{\tokT}{\pmvA}
+ 
\valV
)
\right)                  
\cdot 
\prIncr
&& \text{by Lem.~\ref{lem:ER-increasing}}	
\\
& =
\left(
\stateBase[{\WalW}]{\tokT}{\pmvA}
+    
\stateCredit[{\LpL}]{\tokT}{\pmvA}
\cdot \X[{\LpL}]{\tokT}
-		
\stateDebt[{\LpL}]{\tokT}{\pmvA}
\right)                  
\cdot 
\prIncr
&& \text{by arith.}
\\
& =
\gain{\pmvA}{\confG}{\pxOp{}}
&& \text{by Lem.~\ref{lem:gain-px}}
\end{align*}


\item  \mbox{$\actRepay{\pmvA}{\valV:\tokT}$}.
\begin{align*}
\gain{\pmvA}{\confG}{\txT \, \pxOp{}}
& =
\left(
(
\stateBase[{\WalW}]{\tokT}{\pmvA}
-
\valV
)
+     
\stateCredit[{\LpL}]{\tokT}{\pmvA}
\cdot \X[{\LpLi}]{\tokT}
- 
(
\stateDebt[{\LpL}]{\tokT}{\pmvA}
- 
\valV
)
\right)                  
\cdot 
\prIncr
&& \text{by \nrule{[Rep]}}	
\\
& =
\left(
(
\stateBase[{\WalW}]{\tokT}{\pmvA}
-
\valV
)
+     
\stateCredit[{\LpL}]{\tokT}{\pmvA}
\cdot \X[{\LpL}]{\tokT}
- 
(
\stateDebt[{\LpL}]{\tokT}{\pmvA}
- 
\valV
)
\right)                  
\cdot 
\prIncr
&& \text{by Lem.~\ref{lem:ER-increasing}}	
\\
& =
\left(
\stateBase[{\WalW}]{\tokT}{\pmvA}
+    
\stateCredit[{\LpL}]{\tokT}{\pmvA}
\cdot \X[{\LpL}]{\tokT}
-		
\stateDebt[{\LpL}]{\tokT}{\pmvA}
\right)                  
\cdot 
\prIncr
&& \text{by arith.}
\\
& =
\gain{\pmvA}{\confG}{\pxOp{}}
&& \text{by Lem.~\ref{lem:gain-px}}
\end{align*}


\item  \mbox{$\actRedeem{\pmvA}{\minted{\valV}:\minted{\tokT}}$}.
Let $\valV = \minted{\valV} \cdot \X[\LpL]{\tokT}$.
We have two subcases.
If $\supply[\LpLi]{\tokT} > 0$, then:
\begin{align*}
\gain{\pmvA}{\confG}{\txT \, \pxOp{}}
& =
\left(
(
\stateBase[{\WalW}]{\tokT}{\pmvA}
+
\valV
)
+     
(
\stateCredit[{\LpL}]{\tokT}{\pmvA}
- \minted{\valV}
)
\cdot \X[{\LpLi}]{\tokT}
- 
\stateDebt[{\LpL}]{\tokT}{\pmvA}
\right)                  
\cdot 
\prIncr
&& \text{by \nrule{[Rdm]}}	
\\
& =
\left(
(
\stateBase[{\WalW}]{\tokT}{\pmvA}
+
\valV
)
+     
(
\stateCredit[{\LpL}]{\tokT}{\pmvA}
- \minted{\valV}
)
\cdot \X[{\LpL}]{\tokT}
- 
\stateDebt[{\LpL}]{\tokT}{\pmvA}
\right)                  
\cdot 
\prIncr
&& \text{by Lem.~\ref{lem:ER-increasing}}	
\\
& =
\left(
\stateBase[{\WalW}]{\tokT}{\pmvA}
+    
\stateCredit[{\LpL}]{\tokT}{\pmvA}
\cdot \X[{\LpL}]{\tokT}
-		
\stateDebt[{\LpL}]{\tokT}{\pmvA}
\right)                  
\cdot 
\prIncr
&& \text{by arith.}
\\
& =
\gain{\pmvA}{\confG}{\pxOp{}}
&& \text{by Lem.~\ref{lem:gain-px}}
\end{align*}

\noindent
Otherwise, if $\supply[\LpLi]{\tokT} = 0$, then
$\minted{\valV} = \LpL(\minted{\tokT},\pmvA)$, and so:
\begin{align*}
\gain{\pmvA}{\confG}{\txT \, \pxOp{}}
& =
\left(
(
\stateBase[{\WalW}]{\tokT}{\pmvA}
+
\valV
)
+     
(
\stateCredit[{\LpL}]{\tokT}{\pmvA}
- \minted{\valV}
)
\cdot \X[{\LpLi}]{\tokT}
- 
\stateDebt[{\LpL}]{\tokT}{\pmvA}
\right)                  
\cdot 
\prIncr
&& \text{by \nrule{[Rdm]}}	
\\
& =
\left(
(
\stateBase[{\WalW}]{\tokT}{\pmvA}
+
\valV
)
+     
(
\stateCredit[{\LpL}]{\tokT}{\pmvA}
- \minted{\valV}
)
\cdot \X[{\LpL}]{\tokT}
- 
\stateDebt[{\LpL}]{\tokT}{\pmvA}
\right)                  
\cdot 
\prIncr
&& \text{by arith.}	
\\
& =
\left(
\stateBase[{\WalW}]{\tokT}{\pmvA}
+    
\stateCredit[{\LpL}]{\tokT}{\pmvA}
\cdot \X[{\LpL}]{\tokT}
-		
\stateDebt[{\LpL}]{\tokT}{\pmvA}
\right)                  
\cdot 
\prIncr
&& \text{by arith.}
\\
& =
\gain{\pmvA}{\confG}{\pxOp{}}
&& \text{by Lem.~\ref{lem:gain-px}}
\end{align*}


\item \mbox{$\swapOp(\valV: \tokTi, \tokT)$}.
Let $\valVi= \valV \cdot \nicefrac{\price(\tokTi)}{\price(\tokT)}$.
We have that:
\begin{align*}
\gain{\pmvA}{\confG}{\txT \, \pxOp{}}
& =
\left(
(
\stateBase[{\WalW}]{\tokT}{\pmvA}
+
\valVi
)
+     
\stateCredit[{\LpL}]{\tokT}{\pmvA}
\cdot \X[{\LpLi}]{\tokT}
- 
\stateDebt[{\LpL}]{\tokT}{\pmvA}
\right)                  
\cdot 
\prIncr
&& \text{by \nrule{[Swp]}}	
\\
& =
\left(
(
\stateBase[{\WalW}]{\tokT}{\pmvA}
+
\valVi
)
+     
\stateCredit[{\LpL}]{\tokT}{\pmvA}
\cdot \X[{\LpL}]{\tokT}
- 
\stateDebt[{\LpL}]{\tokT}{\pmvA}
\right)                  
\cdot 
\prIncr
&& \text{by Lem.~\ref{lem:ER-increasing}}	
\\
& =
\left(
\stateBase[{\WalW}]{\tokT}{\pmvA}
+    
\stateCredit[{\LpL}]{\tokT}{\pmvA}
\cdot \X[{\LpL}]{\tokT}
-		
\stateDebt[{\LpL}]{\tokT}{\pmvA}
\right)                  
\cdot 
\prIncr
+
\valVi \cdot \prIncr
&& \text{by arith.}
\\
& =
\gain{\pmvA}{\confG}{\pxOp{}}
+
\valVi \cdot \prIncr
&& \text{by Lem.~\ref{lem:gain-px}}
\end{align*}

\noindent
Since $\valVi > 0$, we have that
$\gain{\pmvA}{\confG}{\txT \pxOp{}} \circ \gain{\pmvA}{\confG}{\pxOp{}}$
whenever $\prIncr \circ 0$ for $\circ \in \setenum{<,>}$.


\item \mbox{$\swapOp(\valV: \tokT, \tokTi)$}.
Let $\valVi= \valV \cdot \nicefrac{\price(\tokT)}{\price(\tokTi)}$.
We have that:
\begin{align*}
\gain{\pmvA}{\confG}{\txT \, \pxOp{}}
& =
\left(
(
\stateBase[{\WalW}]{\tokT}{\pmvA}
-
\valVi
)
+     
\stateCredit[{\LpL}]{\tokT}{\pmvA}
\cdot \X[{\LpLi}]{\tokT}
- 
\stateDebt[{\LpL}]{\tokT}{\pmvA}
\right)                  
\cdot 
\prIncr
&& \text{by \nrule{[Swp]}}	
\\
& =
\left(
(
\stateBase[{\WalW}]{\tokT}{\pmvA}
-
\valVi
)
+     
\stateCredit[{\LpL}]{\tokT}{\pmvA}
\cdot \X[{\LpL}]{\tokT}
- 
\stateDebt[{\LpL}]{\tokT}{\pmvA}
\right)                  
\cdot 
\prIncr
&& \text{by Lem.~\ref{lem:ER-increasing}}	
\\
& =
\left(
\stateBase[{\WalW}]{\tokT}{\pmvA}
+    
\stateCredit[{\LpL}]{\tokT}{\pmvA}
\cdot \X[{\LpL}]{\tokT}
-		
\stateDebt[{\LpL}]{\tokT}{\pmvA}
\right)                  
\cdot 
\prIncr
-
\valVi \cdot \prIncr
&& \text{by arith.}
\\
& =
\gain{\pmvA}{\confG}{\pxOp{}}
-
\valVi \cdot \prIncr
&& \text{by Lem.~\ref{lem:gain-px}}
\end{align*}

\noindent
Since $\valVi > 0$, we have that
$\gain{\pmvA}{\confG}{\txT \pxOp{}} \circ \gain{\pmvA}{\confG}{\pxOp{}}$
whenever $0 \circ \prIncr$ for $\circ \in \setenum{<,>}$.
\qed
\end{itemize}
\end{thmproof}


\begin{lemproof}{lem:gameOptions}
Let $\pxOp{}$ be a shorthand for $\pxOp{\prIncr:\tokT[1]}$, with $\prIncr > 0$, 
and consider the following sequence of transactions:
\[
\TxTS 
\; = \;
\pmvA:\depositOp(\valV[1]:\tokT[1]) \;
\pmvA:\borrowOp(\valV[2]:\tokT[2]) \;
\pmvA:\swapOp(\valV[2]:\tokT[2],  \tokT[1])
\]
For all $\confG$ such that $\TxTS$ is enabled in $\confG$, we prove that:
\[
\gain{\pmvA}{\confG}{\TxTS[] \exchUpdateOp{}} 
>
\gain{\pmvA}{\confG}{ \exchUpdateOp{}} 
\]	
We start by giving names to the intermediate states reached during the execution of $\TxTS$. Let:
\begin{align*}
\confG = (\WalW[0],\LpL[0],\price)
& \xrightarrow{\; \pmvA:\depositOp(\valV[1]:\tokT[1]) \;\;\;\;} 
\confG[1] = (\WalW[1],\LpL[1],\price)
\\
& \xrightarrow{\; \pmvA:\borrowOp(\valV[2]:\tokT[2]) \;\;\;\;} 
\confG[2] = (\WalW[2],\LpL[2],\price)
\\
& \xrightarrow{\pmvA:\swapOp(\valV[2]:\tokT[2], \tokT[1])} 
\confG[3] = (\WalW[3],\LpL[3],\price)
\end{align*}


\noindent
Let $\valVi[1] = \valV[2] \cdot \nicefrac{\price(\tokT[2])}{\price(\tokT[1])}$.
By the rules \nrule{[Dep]}, \nrule{[Bor]} and \nrule{[Swp]}, we have that:	
\begin{align*}
& \WalW[1] = \WalW[0] - \setenum{(\tokT[1],\pmvA) \mapsto \valV[1]}
&& \LpL[1] = \LpL[0] + \setenum{\tokT[1] \mapsto \valV[1]} + \setenum{(\mintedT[1],\pmvA) \mapsto \nicefrac{\valV[1]}{\X[{\LpL[0]}]{\tokT[1]}}}
\\
& \WalW[2] = \WalW[1] + \setenum{(\tokT[2],\pmvA) \mapsto \valV[2]}
&& \LpL[2] = \LpL[1] - \setenum{\tokT[2] \mapsto \valV[2]} + \setenum{(\debtT[2],\pmvA) \mapsto \valV[2]}
\\
& \WalW[3] = \WalW[2] - \setenum{(\tokT[2],\pmvA) \mapsto \valV[2]} + \setenum{(\tokT[1],\pmvA) \mapsto \valVi[1]}
&& \LpL[3] = \LpL[2]
\end{align*}

\noindent
We then estimate $\pmvA$'s gain as follows:
\begin{align*}
\gain{\pmvA}{\confG}{\TxTS \pxOp{}}
& =
\gain{\pmvA}{\confG}{\TxTS} + \gain{\pmvA}{\confG[3]}{\pxOp{}}
&& \text{by~\eqref{eq:gainSt}}
\\
& =
\gain{\pmvA}{\confG[3]}{\pxOp{}}
&& \text{by Lem.~\ref{lem:gain-base}}
\\
& = 
\left(
\stateBase[{\WalW[3]}]{\tokT[1]}{\pmvA}
+     
\stateCredit[{\LpL[3]}]{\tokT[1]}{\pmvA}
\cdot \X[{\LpL[3]}]{\tokT[1]}
-
\stateDebt[{\LpL[3]}]{\tokT[1]}{\pmvA}
\right)                  
\cdot 
\prIncr
&& \text{by Lem.~\ref{lem:gain-px}}
\\
& = 
\left(
\stateBase[{\WalW[3]}]{\tokT[1]}{\pmvA}
+     
\stateCredit[{\LpL[3]}]{\tokT[1]}{\pmvA}
\cdot \X[{\LpL[0]}]{\tokT[1]}
-
\stateDebt[{\LpL[3]}]{\tokT[1]}{\pmvA}
\right)                  
\cdot 
\prIncr
&& \text{by Lem.~\ref{lem:ER-increasing}}
\\
& = 
\Big(
(
\stateBase[{\WalW[0]}]{\tokT[1]}{\pmvA}
- \valV[1] + \valVi[1]
)
\\
& \phantom{=} \, +
(
\stateCredit[{\LpL[0]}]{\tokT[1]}{\pmvA}
+ 
\nicefrac{\valV[1]}{\X[{\LpL[0]}]{\tokT[1]}}
)
\cdot \X[{\LpL[0]}]{\tokT[1]}
- \stateDebt[{\LpL[3]}]{\tokT[1]}{\pmvA}
\Big)                  
\cdot 
\prIncr
&& \text{by~\nrule{[Dep,Bor,Swp]}}
\\
& = 
\Big(
(
\stateBase[{\WalW[0]}]{\tokT[1]}{\pmvA}
+ \valVi[1]
)
+
\stateCredit[{\LpL[0]}]{\tokT[1]}{\pmvA}
\cdot \X[{\LpL[0]}]{\tokT[1]}
- \stateDebt[{\LpL[3]}]{\tokT[1]}{\pmvA}
\Big)                  
\cdot 
\prIncr
&& \text{by arith.}
\\
& = \gain{\pmvA}{\confG}{\pxOp{}} + \valVi[1] \cdot \prIncr
\end{align*}

\noindent
Therefore, $\gain{\pmvA}{\confG}{\TxTS \, \pxOp{}} - \gain{\pmvA}{\confG}{\pxOp{}} = \valVi[1] \cdot \prIncr > 0$.
\qed
\end{lemproof}


\begin{thmproof}{th:X-int-vs-X}
Let $\txT = \pmvA:\ell(\cdots)$ mentioning token $\tokT$ with transaction parameter $\valV$.
We have to prove that:
\begin{enumerate}
	\item If we do not make assumptions on the interest rate function $\Intr{\LpL}(\tokT)$, then we have that, for every $\ell \in \setenum{\depositOp,\borrowOp,\repayOp,\redeemOp,\liquidateOp}$ and for every $\circ \in \{>,  =, <\}$, there exists $\confG[0]$ and 
	$\valV$
	such that 
	$\gain{\pmvA}{\confG[0]}{\txT \, \accrueIntOp} \circ \gain{\pmvA}{\confG[0]}{\accrueIntOp}$.
	
	\item If we assume a constant interest rate function, \ie
	$
	\Intr{\LpL}(\tokT) 
	\; = \;
	r_{\tokT}$,
	then:
	\begin{enumerate}
		\item $\ell \in \setenum{\depositOp,\repayOp} \;\implies
		\text{ for all } \confG[0] \text{ and } \valV, \ 
		\gain{\pmvA}{\confG[0]}{\txT \, \accrueIntOp} \geq \gain{\pmvA}{\confG[0]}{\accrueIntOp} 
		$
		\item $\ell \in \setenum{\borrowOp, \redeemOp} \implies
		\text{ for all } \confG[0] \text{ and } \valV,
		\gain{\pmvA}{\confG[0]}{\txT \, \accrueIntOp} \leq \gain{\pmvA}{\confG[0]}{\accrueIntOp} 
		$
		\item $\ell \in \setenum{\liquidateOp} \implies$ for all $\circ \in \{\geq, \leq\}$, there exists $\confG[0]$ and $\valV$  such that $\gain{\pmvA}{\confG[0]}{\txT \, \accrueIntOp} \circ \gain{\pmvA}{\confG[0]}{\accrueIntOp}$
	\end{enumerate}
	
\end{enumerate}

\noindent
In order to prove the theorem, we proceed as follows. 
We explicitly compute the formula of 
the gain for each $\ell \in \setenum{\depositOp,\borrowOp,\repayOp,\redeemOp,\liquidateOp}$.
We then directly prove points (2.a) and (2.b), \ie we show that the inequalities for the cases in which $\ell \in \setenum{\depositOp,\repayOp, \borrowOp, \redeemOp}$ and $
\Intr{\LpL}(\tokT) 
\; = \;
r_{\tokT}$ hold; moreover, we  precisely determinate when the inequalities are strict or not.
To prove the rest of the theorem, have to provide a counter-examples for each case.
A counter-example consists of an interest rate function $\Intr{\LpL}(\tokT)$, a reachable state $\confG[0]$, and a choice of parameter
$\valV$
such that 
$\gain{\pmvA}{\confG[0]}{\txT \, \accrueIntOp} \circ \gain{\pmvA}{\confG[0]}{\accrueIntOp}$ (for $\circ$ being one of $ \{>,  =, <\}$).
Note that the points (2.a) and (2.b) already gives us 8 counter-example (for $\ell \in \setenum{\depositOp,\repayOp}$ we have the cases $\circ \in \{>,  =\}$, and for $\ell \in \setenum{\borrowOp,\redeemOp}$ we have the cases $\circ \in \{<,  =\}$).
For the remaining 7 cases, in order not to overload the proof with simple yet long computations, we provide the following link to the counter-examples: \url{https://github.com/bitbart/lp-model/tree/main/examples-lmcs/frontrun-int}.

\noindent
Let
$\confG[0] \xrightarrow{\accrueIntOp} \confG[1]$
and
$\confG[0] \xrightarrow{\txT} \confGi[0] \xrightarrow{\accrueIntOp} \confGi[1]$,
and assume that the states are deconstructed as follows:	
\[
    \confG[0] = (\WalW[0],\LpL[0],\statePrice{0})
    \qquad
    \confGi[0] = (\WalWi[0],\LpLi[0],\statePricei{0})
    \qquad
    \confG[1] = (\WalW[1],\LpL[1],\statePrice{1})
    \qquad
    \confGi[1] = (\WalWi[1],\LpLi[1],\statePricei{1})
\]

First, we note that, for every $\tokTi\neq \tokT$, $\projTok[\tokTi]{\gain{\pmvA}{\confG[0]}{\accrueIntOp}} =  \projTok[\tokTi]{\gain{\pmvA}{\confG[0]}{\txT \, \accrueIntOp}}$.
Hence we will focus only on   $\projTok[\tokT]{\gain{\pmvA}{\confG[0]}{\accrueIntOp}}$ and $ \projTok[\tokT]{\gain{\pmvA}{\confG[0]}{\txT \, \accrueIntOp}}$.

By Lemma~\ref{lem:gain-int} we have that:
\[
    \projTok[\tokT]{\gain{\pmvA}{\confG[0]}{\accrueIntOp}}
    =   
    \begin{cases}
	\left(
    \stateCredit[{\LpL[0]}]{\tokT[i]}{\pmvA}
    \cdot 
    \dfrac                
    {\supplyDebt[{\LpL[0]}]{\tokT[i]}}
    {\supply[{{\LpL[0]}}]{\tokT[i]}}
    - 
    \LpL[0](\debt{\tokT[i]},\pmvA)
	\right)
    \cdot 
    \Intr{{\LpL[0]}}(\tokT[i])
    \cdot \statePrice{0}(\tokT[i])
    & \text{ if  ${\supply[{\LpL}]{\tokT}} > 0$}\\
    0 & \text{otherwise}
	\end{cases}
\]




We now compute $\pmvA$'s net worth in  $\confGi[1]$. We proceed by cases on $\txT$.
\begin{itemize}

\item  \mbox{$\actDeposit{\pmvA}{\valV:\tokT}$}.
Note that we have that ${\supply[{\LpLi}]{\tokT}} > 0$, since a successful $\depositOp$ generates a positive amount of credits.

We have that:
\begin{align*}
	\projTok[\tokT]{\gain{\pmvA}{\confG[0]}{\txT \, \accrueIntOp}}
	& =
	\projTok[\tokT]{\gain{\pmvA}{\confG[0]}{\txT}}
	+
	\projTok[\tokT]{\gain{\pmvA}{\confGi[0]}{\accrueIntOp}}
	\\
	& =   
	0 
	+
	\left(
	\stateCredit[{\LpLi[0]}]{\tokT[]}{\pmvA}
	\cdot 
	\dfrac                
	{\supplyDebt[{\LpLi[0]}]{\tokT[]}}
	{\supply[{{\LpLi[0]}}]{\tokT[]}}
	- 
	\LpLi[0](\debt{\tokT[]},\pmvA)
	\right)
	\cdot 
	\Big(
	\Intr{{\LpLi[0]}}(\tokT[])
	\cdot \statePricei{0}{\tokT[]}
	\Big)		
	&& \text{(\ref{lem:gain-base} + \ref{lem:gain-int})}
	\\ 
	& =      
	\Biggl(
	\Big(
	\stateCredit[{\LpL[0]}]{\tokT}{\pmvA}
	 	+
	 	\nicefrac{\valV}{\X[{\LpL[0]}]{\tokT}}
	\Big)
	\cdot 
	\dfrac                
	{\supplyDebt[{\LpL[0]}]{\tokT}}
	{\supply[{{\LpL[0]}}]{\tokT}
		+
		\nicefrac{\valV}{\X[{\LpL[0]}]{\tokT}}
	}
	- 
	\LpL[0](\debt{\tokT},\pmvA)
	\Biggl)
	\cdot 
	\Big(
	\Intr{{\LpLi[0]}}(\tokT)
	\cdot \statePrice{0}{\tokT}
	\Big)		
		&& \text{by \nrule{[Dep]}}				
\end{align*}

We have two cases, depending on whether  ${\supply[{\LpL}]{\tokT}} = 0$ or not.

If ${\supply[{\LpL}]{\tokT}} = 0$, we have that $\projTok[\tokT]{\gain{\pmvA}{\confG[0]}{\accrueIntOp}=0}$, and, from Lemma \ref{lem:cred0-imp-debt0}, that $\supplyDebt[{\LpL[0]}]{\tokT}=0$ (and so also $\LpL[0](\debt{\tokT},\pmvA)=0$). 
Hence we obtain that also $	\projTok[\tokT]{\gain{\pmvA}{\confG[0]}{\txT \, \accrueIntOp}}=0$.

Otherwise, we have that:
\begin{align*}	
   	\projTok[\tokT]{\gain{\pmvA}{\confG[0]}{\txT \, \accrueIntOp}}
    	-
 		\projTok[\tokT]{\gain{\pmvA}{\confG[0]}{\accrueIntOp} }
    & =
    \Bigg(
	     \Big(
	    \stateCredit[{\LpL[0]}]{\tokT}{\pmvA}
	    +
	    \nicefrac{\valV}{\X[{\LpL[0]}]{\tokT}}
	    \Big)
	    \cdot 
	    \dfrac                
	    {\supplyDebt[{\LpL[0]}]{\tokT}}
	    {\supply[{{\LpL[0]}}]{\tokT}
	        +
	    \nicefrac{\valV}{\X[{\LpL[0]}]{\tokT}}
	}
	-
	 \LpL[0](\debt{\tokT},\pmvA) 
	\Bigg)
	\cdot \Intr{{\LpLi[0]}}(\tokT)
	\cdot \statePrice{0}{\tokT}
	\\ &
		-
	\Bigg(
    \stateCredit[{\LpL[0]}]{\tokT}{\pmvA}
    \cdot 
    \dfrac                
    {\supplyDebt[{\LpL[0]}]{\tokT}}
    {\supply[{{\LpL[0]}}]{\tokT}
    }	
	-
	\LpL[0](\debt{\tokT},\pmvA) 
	\Bigg)
	\cdot \Intr{{\LpL[0]}}(\tokT)
	\cdot \statePrice{0}{\tokT}
\end{align*}


\noindent
If we consider a constant interest rate, \ie $\Intr{{\LpLi[0]}}(\tokT)= \Intr{{\LpL[0]}}(\tokT)$, then

\begin{align*}	
	\projTok[\tokT]{\gain{\pmvA}{\confG[0]}{\txT \, \accrueIntOp}}
	-
	\projTok[\tokT]{\gain{\pmvA}{\confG[0]}{\accrueIntOp} } 
	& =
	\supplyDebt[{\LpL[0]}]{\tokT}
	\cdot
	\Bigg(
	\dfrac                
	{\stateCredit[{\LpL[0]}]{\tokT}{\pmvA}
		+
		\nicefrac{\valV}{\X[{\LpL[0]}]{\tokT}}}
	{\supply[{{\LpL[0]}}]{\tokT}
		+
		\nicefrac{\valV}{\X[{\LpL[0]}]{\tokT}}
	}
	-
	\dfrac                
	{\stateCredit[{\LpL[0]}]{\tokT}{\pmvA}}
	{\supply[{{\LpL[0]}}]{\tokT}
	}
	\Bigg)
	\cdot \Intr{{\LpL[0]}}(\tokT)
	\cdot \statePrice{0}{\tokT}
\end{align*}

This amount is positive, and it is equal to zero if and only if $\supplyDebt[{\LpL[0]}]{\tokT}=0$. Indeed, $\Intr{{\LpL[0]}}(\tokT)>0$ and $\statePrice{0}{\tokT}>0$ by definition,
and 
$\dfrac                
{\stateCredit[{\LpL[0]}]{\tokT}{\pmvA}
	+
	\nicefrac{\valV}{\X[{\LpL[0]}]{\tokT}}}
{\supply[{{\LpL[0]}}]{\tokT}
	+
	\nicefrac{\valV}{\X[{\LpL[0]}]{\tokT}}
}
-
\dfrac                
{\stateCredit[{\LpL[0]}]{\tokT}{\pmvA}}
{\supply[{{\LpL[0]}}]{\tokT}
}
>
0$
follows from the simple mathematical fact that, given $A, B\in \mathbb{R}_{\geq 0}$ with $A<B$, and $C\in \mathbb{R}_{> 0}$, then $\frac{A+C}{B+C} > \frac{A}{B}$. 

\ \\

\item  \mbox{$\actBorrow{\pmvA}{\valV:\tokT}$}.
		Note that a $\borrowOp$ can only be fired if the reserves are non-zero, \ie $\reserves[\LpL]{\tokT}>0$. By Lemma \ref{lem:cred0-imp-debt0}, this implies that   ${\supply[{\LpL}]{\tokT}} > 0$.
		Moreover, after a successful $\borrowOp$, we have  ${\supplyDebt[{\LpLi}]{\tokT}} > 0$.
		By Lemma \ref{lem:cred0-imp-debt0}, this implies that   ${\supply[{\LpLi}]{\tokT}} > 0$.

		We have that:
		\begin{align*}
	\projTok[\tokT]{\gain{\pmvA}{\confG[0]}{\txT \, \accrueIntOp}}
	& =
	\projTok[\tokT]{\gain{\pmvA}{\confG[0]}{\txT}}
	+
	\projTok[\tokT]{\gain{\pmvA}{\confGi[0]}{\accrueIntOp}}
	\\
			& = 
			0
			+
			\left(
			\stateCredit[{\LpLi[0]}]{\tokT[]}{\pmvA}
			\cdot 
			\dfrac                
			{\supplyDebt[{\LpLi[0]}]{\tokT[]}}
			{\supply[{{\LpLi[0]}}]{\tokT[]}}
			- 
			\LpLi[0](\debt{\tokT[]},\pmvA)
			\right)
			\cdot 
			\Big(
			\Intr{{\LpLi[0]}}(\tokT[])
			\cdot \statePricei{0}{\tokT[]}
			\Big)
			&& \text{(\ref{lem:gain-base} + \ref{lem:gain-int})}
			\\ 
			& =   
			\Biggl(
			\stateCredit[{\LpL[0]}]{\tokT}{\pmvA}
			\cdot 
			\dfrac                
			{\supplyDebt[{\LpL[0]}]{\tokT}
			+
			\valV
			}
			{\supply[{{\LpL[0]}}]{\tokT}			}
			- 
			\Big(
				\LpL[0](\debt{\tokT},\pmvA)
					+
				\valV
			\Big)
			\Biggl)
			\cdot 
			\Big(
			\Intr{{\LpLi[0]}}(\tokT)
			\cdot \statePrice{0}{\tokT}
			\Big)		
			&& \text{by \nrule{[Bor]}}				
		\end{align*}
		
	Hence we have that

\begin{align*}	
	\projTok[\tokT]{\gain{\pmvA}{\confG[0]}{\txT \, \accrueIntOp}}
	-
	\projTok[\tokT]{\gain{\pmvA}{\confG[0]}{\accrueIntOp} }
	& =	
	\Biggl(
	\stateCredit[{\LpL[0]}]{\tokT}{\pmvA}
	\cdot 
	\dfrac                
	{\supplyDebt[{\LpL[0]}]{\tokT}
		+
		\valV
	}
	{\supply[{{\LpL[0]}}]{\tokT}			}
	- 
	\Big(
	\LpL[0](\debt{\tokT},\pmvA)
	+
	\valV
	\Big)
	\Biggl)
	\cdot 
	\Big(
	\Intr{{\LpLi[0]}}(\tokT)
	\cdot \statePrice{0}{\tokT}
	\Big)	
	\\	&
	 - 
	\Bigg(
	\stateCredit[{\LpL[0]}]{\tokT}{\pmvA}
	\cdot 
	\dfrac                
	{\supplyDebt[{\LpL[0]}]{\tokT}}
	{\supply[{{\LpL[0]}}]{\tokT}
	}	
	-
	\LpL[0](\debt{\tokT},\pmvA) 
	\Bigg)
	\cdot \Intr{{\LpL[0]}}(\tokT)
	\cdot \statePrice{0}{\tokT}
\end{align*}


\noindent
If we consider a constant interest rate, \ie $\Intr{{\LpLi[0]}}(\tokT)= \Intr{{\LpL[0]}}(\tokT)$, then

\begin{align*}	
	\projTok[\tokT]{\gain{\pmvA}{\confG[0]}{\txT \, \accrueIntOp}}
	-
	\projTok[\tokT]{\gain{\pmvA}{\confG[0]}{\accrueIntOp} }
	& =	
	\valV
	\cdot
	\Biggl(
	\dfrac                
	{
		\stateCredit[{\LpL[0]}]{\tokT}{\pmvA}
	}
	{\supply[{{\LpL[0]}}]{\tokT}			}
	- 
	1
	\Biggl)
	\cdot 
	\Big(
	\Intr{{\LpLi[0]}}(\tokT)
	\cdot \statePrice{0}{\tokT}
	\Big)
\end{align*}

which is $\leq0$, since $               
	\stateCredit[{\LpL[0]}]{\tokT}{\pmvA}
\leq
\supply[{{\LpL[0]}}]{\tokT}$.

\ \\
		
\item  \mbox{$\actRepay{\pmvA}{\valV:\tokT}$}.
	Note that a $\repayOp$ can only be fired if the exist some debts, \ie ${\supplyDebt[{\LpL}]{\tokT}} > 0$, which, by Lemma \ref{lem:cred0-imp-debt0}, implies  ${\supply[{\LpL}]{\tokT}} > 0$.
	Moreover, since $\repayOp$ does not impact credits, we have that ${\supply[{\LpLi}]{\tokT}} = {\supply[{\LpL}]{\tokT}}  > 0$.
		
		We have that:
		\begin{align*}
			\projTok[\tokT]{\gain{\pmvA}{\confG[0]}{\txT \, \accrueIntOp}}
			& =
			\projTok[\tokT]{\gain{\pmvA}{\confG[0]}{\txT}}
			+
			\projTok[\tokT]{\gain{\pmvA}{\confGi[0]}{\accrueIntOp}}
			\\
			& = 
			0
			+
			\left(
			\stateCredit[{\LpLi[0]}]{\tokT[]}{\pmvA}
			\cdot 
			\dfrac                
			{\supplyDebt[{\LpLi[0]}]{\tokT[]}}
			{\supply[{{\LpLi[0]}}]{\tokT[]}}
			- 
			\LpLi[0](\debt{\tokT[]},\pmvA)
			\right)
			\cdot 
			\Big(
			\Intr{{\LpLi[0]}}(\tokT[])
			\cdot \statePricei{0}{\tokT[]}
			\Big)
			&& \text{(\ref{lem:gain-base} + \ref{lem:gain-int})}
			\\ 
			& =   
			\Biggl(
			\stateCredit[{\LpL[0]}]{\tokT}{\pmvA}
			\cdot 
			\dfrac                
			{\supplyDebt[{\LpL[0]}]{\tokT}
				-
				\valV
			}
			{\supply[{{\LpL[0]}}]{\tokT}			}
			- 
			\Big(
			\LpL[0](\debt{\tokT},\pmvA)
			-
			\valV
			\Big)
			\Biggl)
			\cdot 
			\Big(
			\Intr{{\LpLi[0]}}(\tokT)
			\cdot \statePrice{0}{\tokT}
			\Big)		
			&& \text{by \nrule{[Rep]}}				
		\end{align*}
		
		Hence we have that

		\begin{align*}	
			\projTok[\tokT]{\gain{\pmvA}{\confG[0]}{\txT \, \accrueIntOp}}
			-
			\projTok[\tokT]{\gain{\pmvA}{\confG[0]}{\accrueIntOp} }
			& =	
			\Biggl(
			\stateCredit[{\LpL[0]}]{\tokT}{\pmvA}
			\cdot 
			\dfrac                
			{\supplyDebt[{\LpL[0]}]{\tokT}
				-
				\valV
			}
			{\supply[{{\LpL[0]}}]{\tokT}			}
			- 
			\Big(
			\LpL[0](\debt{\tokT},\pmvA)
			-
			\valV
			\Big)
			\Biggl)
			\cdot 
			\Big(
			\Intr{{\LpLi[0]}}(\tokT)
			\cdot \statePrice{0}{\tokT}
			\Big)	
			\\	&
			- 
			\Bigg(
			\stateCredit[{\LpL[0]}]{\tokT}{\pmvA}
			\cdot 
			\dfrac                
			{\supplyDebt[{\LpL[0]}]{\tokT}}
			{\supply[{{\LpL[0]}}]{\tokT}
			}	
			-
			\LpL[0](\debt{\tokT},\pmvA) 
			\Bigg)
			\cdot \Intr{{\LpL[0]}}(\tokT)
			\cdot \statePrice{0}{\tokT}
		\end{align*}

		\noindent
		If we consider a constant interest rate, \ie $\Intr{{\LpLi[0]}}(\tokT)= \Intr{{\LpL[0]}}(\tokT)$, then

		\begin{align*}	
			\projTok[\tokT]{\gain{\pmvA}{\confG[0]}{\txT \, \accrueIntOp}}
			-
			\projTok[\tokT]{\gain{\pmvA}{\confG[0]}{\accrueIntOp} }
			& =	
			-\valV
			\cdot
			\Biggl(
			\dfrac                
			{
				\stateCredit[{\LpL[0]}]{\tokT}{\pmvA}
			}
			{\supply[{{\LpL[0]}}]{\tokT}			}
			- 
			1
			\Biggl)
			\cdot 
			\Big(
			\Intr{{\LpLi[0]}}(\tokT)
			\cdot \statePrice{0}{\tokT}
			\Big)
		\end{align*}
		
		which is $\geq0$, since $               
		\stateCredit[{\LpL[0]}]{\tokT}{\pmvA}
		\leq
		\supply[{{\LpL[0]}}]{\tokT}$.

		\ \\

		\item  \mbox{$\actRedeem{\pmvA}{\valV:\tokT}$}.
		Note that a $\redeemOp$ can only be fired if the credits are non-zero, \ie  ${\supply[{\LpL}]{\tokT}} > 0$.
		There are two cases.

		In the first case, $\pmvA$ redeems all the $\tokT$-credits available in the lending pool.
		Then, ${\supply[{\LpLi}]{\tokT}} = 0$, and hence, by Lemma \ref{lem:cred0-imp-debt0}, also ${\supplyDebt[{\LpLi}]{\tokT}} = 0$.
		Since $\redeemOp$ does not affect debts, this implies that also ${\supplyDebt[{\LpL}]{\tokT}} = 0$.
		Hence we have that
		\begin{align*}
			\projTok[\tokT]{\gain{\pmvA}{\confG[0]}{\txT \, \accrueIntOp}}
			& =
			\projTok[\tokT]{\gain{\pmvA}{\confG[0]}{\txT}}
			+
			\projTok[\tokT]{\gain{\pmvA}{\confGi[0]}{\accrueIntOp}}
			\\
			& = 
			0
			+0
			&& \text{(\ref{lem:gain-base} + \ref{lem:gain-int})}		
		\end{align*}
		which implies that
		$
		\projTok[\tokT]{\gain{\pmvA}{\confG[0]}{\txT \, \accrueIntOp}}
		-
		\projTok[\tokT]{\gain{\pmvA}{\confG[0]}{\accrueIntOp} }
		\leq 
		0$,
		since
		$\projTok[\tokT]{\gain{\pmvA}{\confG[0]}{\accrueIntOp} } \geq 0$
		(recall that, by hypothesis,
		${\supplyDebt[{\LpL}]{\tokT}} = 0$%
		).
		
		\
		
		\noindent
		In the second case, $\pmvA$ does not redeem all the  $\tokT$-credits available in the lending pool, \ie  ${\supply[{\LpLi}]{\tokT}} > 0$.

		We have that:
		\begin{align*}
			\projTok[\tokT]{\gain{\pmvA}{\confG[0]}{\txT \, \accrueIntOp}}
			& =
			\projTok[\tokT]{\gain{\pmvA}{\confG[0]}{\txT}}
			+
			\projTok[\tokT]{\gain{\pmvA}{\confGi[0]}{\accrueIntOp}}
			\\
			& =   
			0 
			+
			\left(
			\stateCredit[{\LpLi[0]}]{\tokT[]}{\pmvA}
			\cdot 
			\dfrac                
			{\supplyDebt[{\LpLi[0]}]{\tokT[]}}
			{\supply[{{\LpLi[0]}}]{\tokT[]}}
			- 
			\LpLi[0](\debt{\tokT[]},\pmvA)
			\right)
			\cdot 
			\Big(
			\Intr{{\LpLi[0]}}(\tokT[])
			\cdot \statePricei{0}{\tokT[]}
			\Big)		
			&& \text{(\ref{lem:gain-base} + \ref{lem:gain-int})}
			\\ 
			& =      
			\Biggl(
			\Big(
			\stateCredit[{\LpL[0]}]{\tokT}{\pmvA}
			-
			\nicefrac{\valV}{\X[{\LpL[0]}]{\tokT}}
			\Big)
			\cdot 
			\dfrac                
			{\supplyDebt[{\LpL[0]}]{\tokT}}
			{\supply[{{\LpL[0]}}]{\tokT}
				-
				\nicefrac{\valV}{\X[{\LpL[0]}]{\tokT}}
			}
			- 
			\LpL[0](\debt{\tokT},\pmvA)
			\Biggl)
			\cdot 
			\Big(
			\Intr{{\LpLi[0]}}(\tokT)
			\cdot \statePrice{0}{\tokT}
			\Big)		
			&& \text{by \nrule{[Dep]}}				
		\end{align*}
		and hence
		\begin{align*}	
			\projTok[\tokT]{\gain{\pmvA}{\confG[0]}{\txT \, \accrueIntOp}}
			-
			\projTok[\tokT]{\gain{\pmvA}{\confG[0]}{\accrueIntOp} }
			& =
			\Bigg(
			\Big(
			\stateCredit[{\LpL[0]}]{\tokT}{\pmvA}
			-
			\nicefrac{\valV}{\X[{\LpL[0]}]{\tokT}}
			\Big)
			\cdot 
			\dfrac                
			{\supplyDebt[{\LpL[0]}]{\tokT}}
			{\supply[{{\LpL[0]}}]{\tokT}
				-
				\nicefrac{\valV}{\X[{\LpL[0]}]{\tokT}}
			}
			-
			\LpL[0](\debt{\tokT},\pmvA) 
			\Bigg)
			\cdot \Intr{{\LpLi[0]}}(\tokT)
			\cdot \statePrice{0}{\tokT}
			\\ &
			-
			\Bigg(
			\stateCredit[{\LpL[0]}]{\tokT}{\pmvA}
			\cdot 
			\dfrac                
			{\supplyDebt[{\LpL[0]}]{\tokT}}
			{\supply[{{\LpL[0]}}]{\tokT}
			}	
			-
			\LpL[0](\debt{\tokT},\pmvA) 
			\Bigg)
			\cdot \Intr{{\LpL[0]}}(\tokT)
			\cdot \statePrice{0}{\tokT}
		\end{align*}

		\noindent
		If we consider a constant interest rate, \ie $\Intr{{\LpLi[0]}}(\tokT)= \Intr{{\LpL[0]}}(\tokT)$, then

		\begin{align*}	
			\projTok[\tokT]{\gain{\pmvA}{\confG[0]}{\txT \, \accrueIntOp}}
			-
			\projTok[\tokT]{\gain{\pmvA}{\confG[0]}{\accrueIntOp} } 
			& =
			\supplyDebt[{\LpL[0]}]{\tokT}
			\cdot
			\Bigg(
			\dfrac                
			{\stateCredit[{\LpL[0]}]{\tokT}{\pmvA}
				-
				\nicefrac{\valV}{\X[{\LpL[0]}]{\tokT}}}
			{\supply[{{\LpL[0]}}]{\tokT}
				-
				\nicefrac{\valV}{\X[{\LpL[0]}]{\tokT}}
			}
			-
			\dfrac                
			{\stateCredit[{\LpL[0]}]{\tokT}{\pmvA}}
			{\supply[{{\LpL[0]}}]{\tokT}
			}
			\Bigg)
			\cdot \Intr{{\LpL[0]}}(\tokT)
			\cdot \statePrice{0}{\tokT}
		\end{align*}
		
		This amount is negative, and it is equal to zero if and only if $\supplyDebt[{\LpL[0]}]{\tokT}=0$. Indeed, $\Intr{{\LpL[0]}}(\tokT)>0$ and $\statePrice{0}{\tokT}>0$ by definition,
		and 
		$\dfrac                
		{\stateCredit[{\LpL[0]}]{\tokT}{\pmvA}
			-
			\nicefrac{\valV}{\X[{\LpL[0]}]{\tokT}}}
		{\supply[{{\LpL[0]}}]{\tokT}
			-
			\nicefrac{\valV}{\X[{\LpL[0]}]{\tokT}}
		}
		-
		\dfrac                
		{\stateCredit[{\LpL[0]}]{\tokT}{\pmvA}}
		{\supply[{{\LpL[0]}}]{\tokT}
		}
		<
		0$
		follows from the simple mathematical fact that, given $A, B\in \mathbb{R}_{\geq 0}$ with $A<B$, and $C\in \mathbb{R}_{> 0}$, then $\frac{A-C}{B-C} < \frac{A}{B}$. 
		
	\end{itemize}	
\end{thmproof}




%








\section{Proofs for~\Cref{sec:attacks}}
\label{proofs:attacks}

\begin{thmproof}{th:gameUndercollLoanAttck}
Let $\confG =  (\WalW,\LpL,\statePrice{})$, and assume that $\pmvA$ has no credits or debts with the LP, \ie, ${\creditval[{\confG}]{\pmvA}=\debtval[\confG]{\pmvA}=0}$.
\noindent
Consider the following sequence of transactions:
\[
\TxTS \; = \;
\pmvA:\depositOp(\valV[1]:\tokT[1]) \;\; 
\pxOp{-\prIncr:\tokT[2]}  \;\;
\pmvA:\borrowOp(\valV[2]:\tokT[2]) \;\;	
\pxOp{\prIncr:\tokT[2]}
\]
where 
$0  < \prIncr < \price(\tokT[2])$
and
$\valV[2] = \frac{\valV[1]}{\X[{\LpL}]{\tokT[1]}} \cdot \frac{\price[{\tokT[1]}]}{\price[{\tokT[2]}]-\prIncr} \cdot \liqThreshold$.
We have to prove that:
\begin{enumerate}
\item \label{eq:gameUndercollLoanAttck:1} 
$\gain{\pmvA}{\confG}{\TxTS} = 0$ 
\item \label{eq:gameUndercollLoanAttck:2} 
$\netCredit{\confGi}{\pmvA} < 0$ if $\confG \xrightarrow{\TxTS} \confGi$.
\end{enumerate}	

\noindent
To prove Item~\eqref{eq:gameUndercollLoanAttck:1}, consider the sequence of transitions: 
\[
\confG
\xrightarrow{\txT[1] = \pmvA:\depositOp(\valV[1]:\tokT[1])}
\confG[1]
\xrightarrow{\txT[2] = \pxOp{-\prIncr:\tokT[2]}}
\confG[2]
\xrightarrow{\txT[3] = \pmvA:\borrowOp(\valV[2]:\tokT[2])}
\confG[3]
\xrightarrow{\txT[4] = \pxOp{\prIncr:\tokT[2]}}
\confGi
\]
Note that if $\txT[1]$ is not enabled in $\confG$, then by the hypothesis that $\pmvA$ has no credits in $\confG$ then also $\txT[3]$ will not be enabled, and so $\gain{\pmvA}{\confG}{\TxTS} = 0$.
Otherwise, if $\txT[1]$ is enabled in $\confG$ but $\txT[3]$ is not enabled in $\confG[2]$, then the effects of $\txT[2]$ and $\txT[4]$ cancel out, and then $\gain{\pmvA}{\confG}{\TxTS} = \gain{\pmvA}{\confG}{\txT[1]} = 0$ by Lemma~\ref{lem:gain-base}.
Finally, in case all the transactions in $\TxTS$ are enabled, we have that:
\begin{align*}
\gain{\pmvA}{\confG}{\TxTS} 
& =
\gain{\pmvA}{\confG}{\txT[1]} + 
\gain{\pmvA}{\confG[1]}{\txT[2]} +
\gain{\pmvA}{\confG[2]}{\txT[3]} +
\gain{\pmvA}{\confG[3]}{\txT[4]}
&& \text{by \eqref{eq:gainSt}}
\\
& =
\gain{\pmvA}{\confG[1]}{\txT[2]} +
\gain{\pmvA}{\confG[3]}{\txT[4]}
&& \text{by Lem.~\ref{lem:gain-base}}
\\
& =
\projTok[{\tokT[2]}]{\wealth{\confG[1]}{\pmvA}} \cdot \frac{-\prIncr}{\price(\tokT[2])}
+
\projTok[{\tokT[2]}]{\wealth{\confG[3]}{\pmvA}} \cdot \frac{\prIncr}{\price(\tokT[2])-\prIncr}
&& \text{by Lem~\ref{lem:gain-px}}
\\
& =
    \left(
    \WalW[1](\tokT[2],\pmvA) 
    +     
    \LpL[1](\mintedT[2],\pmvA) \cdot \X[{\LpL[1]}]{\tokT[2]}
    -
    \LpL[1](\debtT[2],\pmvA)  
    \right)                  
    \cdot 
    -\prIncr
\\
& + 
    \left(
    \WalW[3](\tokT[2],\pmvA) 
    +     
    \LpL[3](\mintedT[2],\pmvA) \cdot \X[{\LpL[3]}]{\tokT[2]}
    -
    \LpL[3](\debtT[2],\pmvA)  
    \right)                  
    \cdot 
    \prIncr
&& \text{by~\eqref{eq:wealth-restricted}}
\\
& =
    \left(
    \WalW[1](\tokT[2],\pmvA) 
    +     
    \LpL[1](\mintedT[2],\pmvA) \cdot \X[{\LpL[1]}]{\tokT[2]}
    -
    \LpL[1](\debtT[2],\pmvA)  
    \right)                  
    \cdot 
    -\prIncr
\\
& + 
    \left(
    \WalW[1](\tokT[2],\pmvA) + \valV[2] 
    +     
    \LpL[1](\mintedT[2],\pmvA) \cdot \X[{\LpL[3]}]{\tokT[2]}
    -
    \LpL[1](\debtT[2],\pmvA)  
    - \valV[2]
    \right)                  
    \cdot 
    \prIncr
&& \text{by \nrule{[Bor]}}
\\
& =
    \left(
    \WalW[1](\tokT[2],\pmvA) 
    +     
    \LpL[1](\mintedT[2],\pmvA) \cdot \X[{\LpL[1]}]{\tokT[2]}
    -
    \LpL[1](\debtT[2],\pmvA)  
    \right)                  
    \cdot 
    -\prIncr
\\
& + 
    \left(
    \WalW[1](\tokT[2],\pmvA) 
    +     
    \LpL[1](\mintedT[2],\pmvA) \cdot \X[{\LpL[3]}]{\tokT[2]}
    -
    \LpL[1](\debtT[2],\pmvA) 
    \right)                  
    \cdot 
    \prIncr
&& \text{by arith.}
\\
& =
    \left(
    \WalW[1](\tokT[2],\pmvA) 
    +     
    \LpL[1](\mintedT[2],\pmvA) \cdot \X[{\LpL[1]}]{\tokT[2]}
    -
    \LpL[1](\debtT[2],\pmvA)  
    \right)                  
    \cdot 
    -\prIncr
\\
& + 
    \left(
    \WalW[1](\tokT[2],\pmvA) 
    +     
    \LpL[1](\mintedT[2],\pmvA) \cdot \X[{\LpL[1]}]{\tokT[2]}
    -
    \LpL[1](\debtT[2],\pmvA) 
    \right)                  
    \cdot 
    \prIncr
&& \text{by Lem.~\ref{lem:ER-increasing}}
\\
& = 0
&& \text{by arith.}
\end{align*}

\medskip\noindent
To prove Item~\eqref{eq:gameUndercollLoanAttck:2}, observe that since by hypothesis $\creditval[{\confG}]{\pmvA} = \debtval[\confG]{\pmvA} = 0$, then
in $\confGi$ we have that:
\begin{align*}
\creditval[{\confGi}]{\pmvA} 
& = 
\frac{\valV[1]}{\X[{\LpL}]{\tokT[1]}} \cdot \price(\tokT[1])
\\
\debtval[{\confGi}]{\pmvA} 
& =
\valV[2]  \cdot \price(\tokT[2]) = 
\frac{\valV[1]}{\X[{\LpL}]{\tokT[1]}} \cdot \frac{\price(\tokT[1])}{\price(\tokT[2])-\prIncr} \cdot \liqThreshold	
\cdot \price(\tokT[2])
\end{align*}
Therefore:
\begin{align*}
\netCredit{\confGi}{\pmvA}
& =
\creditval[{\confGi}]{\pmvA} 
-
\debtval[{\confGi}]{\pmvA}
=  
\frac{\valV[1]}{\X[{\LpL}]{\tokT[1]}}
\cdot
\price(\tokT[1])
\cdot
\Big(
    1
    - \frac
    {\liqThreshold	
    \cdot \price(\tokT[2])
    }
    {\price(\tokT[2])-\prIncr} 
    \Big)
\end{align*}
This amount is negative if and only if 
\[
\frac
	{\liqThreshold	
		\cdot \price(\tokT[2])
	}
	{\price(\tokT[2])-\prIncr} > 
	1
\]
or, equivalently, since $0 < \prIncr < \price(\tokT[2])$:
\[
	\prIncr
	 >
	  \price(\tokT[2])
	  \cdot
	  \Big(
	  1
	  -
	  \liqThreshold	
	  \Big)
\tag*{\qed}
\]
\end{thmproof}

\begin{thmproof}{th:gameLiqAttck}
Let $\confG =  (\WalW,\LpL,\price{})$ and
let $\pmvA$ and $\pmvB$ such that:
\begin{enumerate}

\item $\LpL(\mintedT[1],\pmvB) = \valV[c]$, and $\stateCredit{\tokT}{\pmvB} = 0$ for all $\tokT \neq \tokT[1]$ (\ie the collateral of $\pmvB$ relies on a single token type $\tokT[1]$)

\item $\LpL(\debtT[2],\pmvB) = \valV[d]$,
and $\stateDebt{\tokT}{\pmvB} = 0$ for all $\tokT \neq \tokT[2]$ (\ie $\pmvB$ has debts only in $\tokT[2]$)

\item $\stateBase{\tokT[2]}{\pmvA} > 0$ 

\item $\Health{\confG[]}{\pmvB} \geq 1$ (\ie $\pmvB$ cannot be liquidated in $\confG$)

\end{enumerate}

\noindent
Then,
for every $\prIncr>0$ sufficiently small,
and 
for every $\valV[l]>0$
 such that $\valV[l] \leq \stateBase{\tokT[2]}{\pmvA}$, $\valV[l] \leq \valV[d]$
 and 
$\valV[l]<
\valV[c] 
\cdot
\frac{
	\X[\LpL]{\tokT[1]}
}
{\rLiq}
\cdot
\frac{\prIncr}{\price(\tokT[2])}$,
given
the following sequence of transactions:
\[
\TxTS 
\; = \; 
\pxOp{(-\price(\tokT[1])+\prIncr):\tokT[1]} \;\;
\pmvA:\liquidateOp(\pmvB, \valV[l]: \tokT[2], \mintedT[1]) \;\;
\pxOp{(\price(\tokT[1])-\prIncr):\tokT[1]}
\]
it holds that:
\begin{enumerate}

\item \label{eq:gameLiqAttck:1}
$\TxTS$ is enabled in $\confG$ 

\item \label{eq:gameLiqAttck:2}
$\gainSt{\pmvA}{\confG}{\TxTS} > 0$

\end{enumerate}	

\noindent
We start by giving names to the intermediate states reached during the execution of $\TxTS$:
\begin{align*}
\confG = (\WalW[],\LpL[],\price)
& 
\xrightarrow{\; \txT[1] \; = \; \pxOp{(-\price(\tokT[1])+\prIncr):\tokT[1]}\;}
\confG[1] = (\WalW[1],\LpL[1],\statePrice{1})
&& \WalW[1] = \WalW, \LpL[1] = \LpL
\\
& \xrightarrow{\; \txT[2] \; = \; \pmvA:\liquidateOp(\pmvB, \valV[l]: \tokT[2], \mintedT[1])\;\;\;\;} 
\confG[2] = (\WalW[2],\LpL[2],\statePrice{2})
&& \statePrice{2} = \statePrice{1}
\\
& \xrightarrow{\; \txT[3] \; = \; \pxOp{(\price(\tokT[1])-\prIncr):\tokT[1]}\;\;\;} 
\confG[3] = (\WalW[3],\LpL[3],\statePrice{3})
&& \WalW[3] = \WalW[2], \LpL[3] = \LpL[2]
\end{align*}

We first prove Item~\eqref{eq:gameLiqAttck:1}.
Of course, both price updates are enabled whenever $\prIncr > 0$, so in order to prove that $\TxTS$ is enabled in $\confG$ we only need to prove that the liquidation is enabled in $\confG[1]$.
We check that all the premises of \nrule{[Liq]} hold:
\begin{itemize}

\item $\WalW[1](\tokT[2],\pmvA) \geq \valV[l] > 0$ is given by the conditions on $\valV[l]$

\item $\LpL[1](\debtT[2],\pmvB) \geq \valV[l]$ is given by the conditions on $\valV[l]$

\item $\LpL[1](\mintedT[1],\pmvB) \geq \frac{\valV[l]}{\X[{\LpL[1]}]{\tokT[1]}} \cdot \frac{\statePrice{1}(\tokT[2])}{\statePrice{1}(\tokT[1])} \cdot \rLiq$ is given by the fact that $\statePrice{1}(\tokT[2])=\prIncr$, $\statePrice{1}(\tokT[1])=\statePrice{}(\tokT[1])$, and 
	$\valV[l]<
	\valV[c] 
	\cdot
	\frac{
		\X[\LpL]{\tokT[1]}
	}
	{\rLiq}
	\cdot
	\frac{\prIncr}{\price(\tokT[2])}$.
	
\item $\Health{\confG[1]}{\pmvB} < 1$ is given by the fact that
	\[
	\Health{\confG[1]}{\pmvB} 
	= 
	\frac{\valV[c] \cdot \statePrice{1}(\tokT[1])}
	{\valV[d] \cdot \statePrice{1}(\tokT[2])}
	\cdot \rLiq
	=
	\frac{\valV[c] \cdot \prIncr}
	{\valV[d] \cdot \price(\tokT[2])}
	\cdot \rLiq
	\]
which, for $\prIncr$ sufficiently small, is strictly less than $1$. 
    
\item $\Health{\confG[2]}{\pmvB} \leq 1$ is given by the fact that 
\begin{align*}	
		\Health{\confG[2]}{\pmvB} 
		& =  
		\frac{
			\Big(
			\valV[c]
			-
			\frac
			{\valV[l]}{\X[{\LpL[1]}]{\tokT[1]}}
			\cdot \frac{\statePrice{1}(\tokT[2])}{\statePrice{1}(\tokT[1])}
			\cdot \rLiq
			\Big)
			\cdot \statePrice{1}(\tokT[1])
		}
		{
			\big(	
			\valV[d] - \valV[l]
			\big)
			\cdot \statePrice{1}(\tokT[2])
		}
		\cdot \rLiq
		&& \text{by~\nrule{[Liq]}}
		\\
		& =  
		\frac
		{
			\Big(
			\valV[c]
			-
			\frac{\valV[l]}{\X[\LpL]{\tokT[1]}}
			\cdot
			\frac{\price(\tokT[2])}{\prIncr}
			\cdot
			\rLiq
			\Big)
			\cdot \prIncr
		}
		{
			\big(	
			\valV[d] 
			-
			\valV[l]
			\big)
			\cdot 
			\price(\tokT[2])}
		\cdot \rLiq
		&& \text{by Lem.~\ref{lem:ER-increasing}}
		\\
		& =  
		\frac
		{
			\valV[c] \cdot \prIncr
			-
			\frac{\valV[l]}{\X[\LpL]{\tokT[1]}}
			\cdot
			\price(\tokT[2])
			\cdot
			\rLiq
		}
		{
			\big(	
			\valV[d] - \valV[l]
			\big)
			\cdot 
			\price(\tokT[2])
		}
		\cdot \rLiq
		&& \text{by arith.}
	\end{align*}
	which,
	for $\prIncr$ sufficiently small, is less or equal to $1$ (note that $\prIncr$ bounds $\valV[l]$). 
\end{itemize}


\medskip

We now prove Item~\eqref{eq:gameLiqAttck:2}. Since by the previous point all transactions are enabled, we have that:
\begin{align*}
	\gain{\pmvA}{\confG}{\TxTS} 
	& =
	\gain{\pmvA}{\confG}{\txT[1]} + 
	\gain{\pmvA}{\confG[1]}{\txT[2]} +
	\gain{\pmvA}{\confG[2]}{\txT[3]} 
	&& \text{by \eqref{eq:gainSt}}
	\\
	& =
	\projTok[{\tokT[1]}]{\wealth{\confG[]}{\pmvA}} \cdot \frac{-\price(\tokT[1])+\prIncr}{\price(\tokT[1])}
	+	
	\gain{\pmvA}{\confG[1]}{\txT[2]} 
	+
	\projTok[{\tokT[1]}]{\wealth{\confG[2]}{\pmvA}} \cdot \frac{\price(\tokT[1])-\prIncr}{\prIncr}
	&& \text{by Lem.~\ref{lem:gain-px}}
	\\
	& =
	\left(
	\WalW[](\tokT[1],\pmvA) 
	+     
	\LpL[](\mintedT[1],\pmvA) \cdot \X[{\LpL[]}]{\tokT[1]}
	-
	\LpL[](\debtT[1],\pmvA)  
	\right)                  
	\cdot 
	(-\price(\tokT[1])+\prIncr)
	\\
	& \quad
	+		\gain{\pmvA}{\confG[1]}{\txT[2]} 
	\\
	& \quad + 
	\left(
	\WalW[2](\tokT[1],\pmvA) 
	+     
	\LpL[2](\mintedT[1],\pmvA) \cdot \X[{\LpL[2]}]{\tokT[1]}
	-
	\LpL[2](\debtT[1],\pmvA)  
	\right)                  
	\cdot 
	(\price(\tokT[1])-\prIncr)
	&& \text{by~\eqref{eq:wealth-restricted}}
	\\
	& >
	\left(
	\WalW[](\tokT[1],\pmvA) 
	+     
	\LpL[](\mintedT[1],\pmvA) \cdot \X[{\LpL[]}]{\tokT[1]}
	-
	\LpL[](\debtT[1],\pmvA)  
	\right)                  
	\cdot 
	(-\price(\tokT[1])+\prIncr)
	\\
	& \quad
	+		\gain{\pmvA}{\confG[1]}{\txT[2]} 
	\\
	& \quad + 
	\left(
	\WalW[1](\tokT[1],\pmvA) 
	+     
	\LpL[1](\mintedT[1],\pmvA) \cdot \X[{\LpL[2]}]{\tokT[1]}
	-
	\LpL[1](\debtT[1],\pmvA)  
	\right)                  
	\cdot 
	(\price(\tokT[1])-\prIncr)
	&& \text{by \nrule{[Liq]}}
	\\
	& =
	\left(
	\WalW[](\tokT[1],\pmvA) 
	+     
	\LpL[](\mintedT[1],\pmvA) \cdot \X[{\LpL[]}]{\tokT[1]}
	-
	\LpL[](\debtT[1],\pmvA)  
	\right)                  
	\cdot 
	(-\price(\tokT[1])+\prIncr)
	\\
	& \quad
	+		\gain{\pmvA}{\confG[1]}{\txT[2]} 
	\\
	& \quad + 
	\left(
	\WalW[](\tokT[1],\pmvA) 
	+     
	\LpL[](\mintedT[1],\pmvA) \cdot \X[{\LpL[2]}]{\tokT[1]}
	-
	\LpL[](\debtT[1],\pmvA)  
	\right)                  
	\cdot 
	(\price(\tokT[1])-\prIncr)
	&& \text{by \nrule{[Px]}}	
	\\
	& =
	\left(
	\WalW[](\tokT[1],\pmvA) 
	+     
	\LpL[](\mintedT[1],\pmvA) \cdot \X[{\LpL[]}]{\tokT[1]}
	-
	\LpL[](\debtT[1],\pmvA)  
	\right)                  
	\cdot 
	(-\price(\tokT[1])+\prIncr)
	\\
	& \quad
	+		\gain{\pmvA}{\confG[1]}{\txT[2]} 
	\\
	& \quad + 
	\left(
	\WalW[](\tokT[1],\pmvA) 
	+     
	\LpL[](\mintedT[1],\pmvA) \cdot \X[{\LpL[]}]{\tokT[1]}
	-
	\LpL[](\debtT[1],\pmvA)  
	\right)                  
	\cdot 
	(\price(\tokT[1])-\prIncr)
	&& \text{by Lem.~\ref{lem:ER-increasing}}	
	\\
	& =	\gain{\pmvA}{\confG[1]}{\txT[2]} 
	&& \text{by arith.}	
	\\
	& > 0
	&& \text{by Lem.~\ref{lem:gain-base}}	
\end{align*}
\qed
\end{thmproof}

\begin{thmproof}{th:underutilizationAttack}
Let $\confG[] = (\WalW,\LpL,\price{})$, and let  $\pmvA, \pmvB$ and $\tokT$ be such that:
\begin{enumerate}
\item $\stateCredit{\tokT}{\pmvA}=0$ and   
\item $\stateCredit{\tokT}{\pmvB}>0$ and  $\stateDebt{\tokT}{\pmvB}=0$
\end{enumerate}
Then, let $\TxTS$ be the following sequence of transactions:
\[
\TxTS 
\; = \;
\pmvA:\depositOp{(\valV: \tokT)}
\;\;	
\accrueIntOp
\;\;
\pmvA:\redeemOp{(\minted{\valV}: \tokT)}
\]
where $\minted{\valV}$ is the amount of credits held by $\pmvA$ in the intermediate state before $\redeemOp$.

Assuming $\TxTS$ is enabled in $\confG$, and that the lending protocol uses the linear utility interest rate function in~\eqref{eq:intr-util} with $\alpha > 0$,  we have  to prove that:
\begin{itemize}
\item $\gainSt{\pmvA}{\confG[]}{\TxTS} > \gainSt{\pmvA}{\confG[]}{\accrueIntOp}$, 
\item $\gainSt{\pmvB}{\confG[]}{\TxTS} < \gainSt{\pmvB}{\confG[]}{\accrueIntOp}$
\end{itemize}

\noindent
We start by giving names to the intermediate states reached during the execution of $\TxTS$:
\begin{align*}
\confG = (\WalW[],\LpL[],\price)
& 
\xrightarrow{\; \txT[1] \; = \; \pmvA:\depositOp{(\valV: \tokT)}\;}
\confG[1] = (\WalW[1],\LpL[1],\price)
\\
& \xrightarrow{\; \txT[2] \; = \; \accrueIntOp\;\;\;\;\;\;\;\;\;\;\;\;\;\;} 
\confG[2] = (\WalW[2],\LpL[2],\price)
\\
& \xrightarrow{\; \txT[3] \; = \; \pmvA:\redeemOp{(\minted{\valV}: \tokT)}\;} 
\confG[3] = (\WalW[3],\LpL[3],\price)
\end{align*}

\noindent
First note that, for every $\tokTi \neq \tokT$, 
it trivially holds that 
$\projTok[\tokTi]{\gain{\pmvA}{\confG}{\TxTS}} = 
\projTok[\tokTi]{\gain{\pmvA}{\confG}{\accrueIntOp}}$
and
$\projTok[\tokTi]{\gain{\pmvB}{\confG}{\TxTS}} = 
\projTok[\tokTi]{\gain{\pmvB}{\confG}{\accrueIntOp}}$.
Hence, we only focus on the gains restricted to $\tokT$.

By hypothesis, we have that $\supply[{\LpL[]}]{\tokT}>0$ and $\supplyDebt[{\LpL[]}]{\tokT}>0$. 
Since $\depositOp$ does not decrease the supply of credits nor that of debts, we also have that $\supply[{\LpL[1]}]{\tokT}>0$ and $\supplyDebt[{\LpL[1]}]{\tokT}>0$.
Moreover, since $\stateDebt{\tokT}{\pmvB}=0$ and the deposit fired by $\pmvA$ does not impact the credits of $\pmvB$, we also have  $\stateDebt[{\LpL[1]}]{\tokT}{\pmvB}=0$

\noindent
The gain of $\pmvA$ is given by:
\begin{align*}
		\projTok{\gain{\pmvA}{\confG}{\TxTS}} 
		& =
		\projTok{\gain{\pmvA}{\confG}{\txT[1]}} +  
		\projTok{\gain{\pmvA}{\confG[1]}{\txT[2]}} + 
		\projTok{\gain{\pmvA}{\confG[2]}{\txT[3]}} 
		&& \text{by \eqref{eq:gainSt}}
		\\
		& =
		0 + 
		\projTok{\gain{\pmvA}{\confG[1]}{\txT[2]}} + \
		0 
		&& \text{by Lem. \ref{lem:gain-base}}
		\\
		& =
		\left(
		\dfrac
		{\stateCredit[{\LpL[1]}]{\tokT}{\pmvA}{}}
		{\supply[{\LpL[1]}]{\tokT}}
		\cdot 
		{\supplyDebt[{\LpL[1]}]{\tokT}}
		- 
		\stateDebt[{\LpL[1]}]{\tokT}{\pmvA}{}
		\right)
		\cdot 
		\Intr{\LpL[1]}(\tokT)
		\cdot \price[\tokT]	
		&&  \text{by Lem. \ref{lem:gain-int}}
		\\
		& =
		\left(
		\dfrac
		{\stateCredit[{\LpL}]{\tokT}{\pmvA}{} + \nicefrac{\valV}{\X[{\LpL}]{\tokT}
		}}
		{\supply[{\LpL}]{\tokT} 
		+
		\nicefrac{\valV}{\X[{\LpL}]{\tokT}}	}
		\cdot 
		{\supplyDebt[{\LpL}]{\tokT}}
		- 
		\stateDebt[{\LpL}]{\tokT}{\pmvA}{}
		\right)
		\cdot 
		\Intr{\LpL[1]}(\tokT)
		\cdot \price[\tokT]	
		&&  \text{by $\nrule{[Dep]}$}
		\\
		& =
		\left(
		\dfrac
		{ \nicefrac{\valV}{\X[{\LpL}]{\tokT}
		}}
		{\supply[{\LpL}]{\tokT} 
			+
			\nicefrac{\valV}{\X[{\LpL}]{\tokT}}	}
		\cdot 
		{\supplyDebt[{\LpL}]{\tokT}}
		- 
		\stateDebt[{\LpL}]{\tokT}{\pmvA}{}
		\right)
		\cdot 
		\Intr{\LpL[1]}(\tokT)
		\cdot \price[\tokT]	
		&&  \text{by hyp.}
		\\	
		& > 
		- 
		\stateDebt[{\LpL}]{\tokT}{\pmvA}{}
		\cdot 
		\Intr{\LpL[1]}(\tokT)
		\cdot \price[\tokT]	
		&&  \text{by arith.}
		\\
		& =
		- 
		\stateDebt[{\LpL}]{\tokT}{\pmvA}{}
		\cdot 
		\left(
		\alpha \cdot
		\frac
		{\supplyDebt[{\LpL}]{\tokT}}
		{	(\reserves[{\LpL}]{\tokT}
			+\valV
			)
			+
			\supplyDebt[{\LpL}]{\tokT}}
		+\beta		
		\right)
		\cdot \price[\tokT]	
		&&  \text{by \eqref{eq:intr-util} +  $\nrule{[Dep]}$}
		\\	
		& >
		- 
		\stateDebt[{\LpL}]{\tokT}{\pmvA}{}
		\cdot 
		\left(
		\alpha \cdot
		\frac
		{\supplyDebt[{\LpL}]{\tokT}}
		{	\reserves[{\LpL}]{\tokT}
			+
			\supplyDebt[{\LpL}]{\tokT}}
		+\beta		
		\right)
		\cdot \price[\tokT]	
		&&  \text{by arith.}
		\\	
		& =	\projTok{\gain{\pmvA}{\confG}{\accrueIntOp}}
		&&  \text{Lem. \ref{lem:gain-int} + hyp. 
		 }
\end{align*}
	
\medskip\noindent	
The gain of $\pmvB$ is given by:
\begin{align*}
		\projTok{\gain{\pmvB}{\confG}{\TxTS}} 
		& =
		\projTok{\gain{\pmvB}{\confG}{\txT[1]}} + 
		\projTok{\gain{\pmvB}{\confG[1]}{\txT[2]}} +
		\projTok{\gain{\pmvB}{\confG[2]}{\txT[3]}} 
		&& \text{by \eqref{eq:gainSt}}
		\\
		& =
		0 + 
		\projTok{\gain{\pmvB}{\confG[1]}{\txT[2]}} +
		0 
		&& \text{by Lem. \ref{lem:gain-base}}
		\\
		& =
		\left(
		\dfrac
		{\stateCredit[{\LpL[1]}]{\tokT}{\pmvB}{}}
		{\supply[{\LpL[1]}]{\tokT}}
		\cdot 
		{\supplyDebt[{\LpL[1]}]{\tokT}}
		- 
		\stateDebt[{\LpL[1]}]{\tokT}{\pmvB}{}
		\right)
		\cdot 
		\Intr{\LpL[1]}(\tokT)
		\cdot \price[\tokT]	
		&&  \text{by Lem. \ref{lem:gain-int}}
		\\
		& =
		\dfrac
		{\stateCredit[{\LpL[1]}]{\tokT}{\pmvB}{}}
		{\supply[{\LpL[1]}]{\tokT}}
		\cdot 
		{\supplyDebt[{\LpL[1]}]{\tokT}}
		\cdot 
		\Intr{\LpL[1]}(\tokT)
		\cdot \price[\tokT]	
		&&  \text{by hyp. 
		}
		\\
		& =
		\dfrac
		{\stateCredit[{\LpL}]{\tokT}{\pmvB}{}}
		{\supply[{\LpL}]{\tokT} 
			+
		\nicefrac{\valV}{\X[{\LpL}]{\tokT}
		}}
		\cdot 
		{\supplyDebt[{\LpL}]{\tokT}}
		\cdot 
		\left(
		\alpha \cdot
		\frac
		{\supplyDebt[{\LpL}]{\tokT}}
		{	(\reserves[{\LpL}]{\tokT}
			+\valV
			)
			+
			\supplyDebt[{\LpL}]{\tokT}}
		+\beta		
		\right)
		\cdot \price[\tokT]	
		&&  \text{by $\nrule{[Dep]}$}
		\\
		& <
		\dfrac
		{\stateCredit[{\LpL}]{\tokT}{\pmvB}{}}
		{\supply[{\LpL}]{\tokT}}
		\cdot 
		{\supplyDebt[{\LpL}]{\tokT}}
		\cdot 
		\left(
		\alpha \cdot
		\frac
		{\supplyDebt[{\LpL}]{\tokT}}
		{	\reserves[{\LpL}]{\tokT}
			+
			\supplyDebt[{\LpL}]{\tokT}}
		+\beta		
		\right)
		\cdot \price[\tokT]	
		&&  \text{by arith}
		\\
		& =		\projTok{\gain{\pmvB}{\confG}{\accrueIntOp}}
		&&  \text{by Lem. \ref{lem:gain-int}}
\end{align*}

\qed
\end{thmproof}

\begin{thmproof}{th:overutilizationAttack}
Let $\confG[] = (\WalW,\LpL,\price{})$, and let  $\pmvA, \pmvB$ and $\tokT$ be such that:
\begin{enumerate}

\item $\stateCredit{\tokT}{\pmvA}=\supply[\LpL]{\tokT}$ and  $\stateDebt{\tokT}{\pmvA} < \supplyDebt[\LpL]{\tokT}$

\item $\stateDebt{\tokT}{\pmvB}>0$

\end{enumerate}

\noindent
Then, let the following sequence of transactions:
\[
	\TxTS 
	\; = \;
	\pmvA:\borrowOp{(\valV: \tokT)}
	\;\;	
	\accrueIntOp
	\;\;
	\pmvA:\repayOp{(\valV: \tokT)}
\]
be enabled in $\confG$.
If $\Intr{\LpL}(\tokT)$ is a linear utility interest rate function with $\alpha > 0$, then:
	\begin{itemize}
		\item $\gainSt{\pmvA}{\confG[]}{\TxTS} > \gainSt{\pmvA}{\confG[]}{\accrueIntOp}$, 
		\item $\gainSt{\pmvB}{\confG[]}{\TxTS} < \gainSt{\pmvB}{\confG[]}{\accrueIntOp}$
\end{itemize}

\noindent
We start by giving names to the intermediate states reached during the execution of $\TxTS$:
\begin{align*}
\confG = (\WalW[],\LpL[],\price)
& 
\xrightarrow{\; \txT[1] \; = \; \pmvA:\borrowOp{(\valV: \tokT)}\;}
\confG[1] = (\WalW[1],\LpL[1],\price)
\\
& \xrightarrow{\; \txT[2] \; = \; \accrueIntOp\;\;\;\;\;\;\;\;\;\;\;\;\;\;} 
\confG[2] = (\WalW[2],\LpL[2],\price)
\\
& \xrightarrow{\; \txT[3] \; = \; \pmvA:\repayOp{(\valV: \tokT)}\;} 
\confG[3] = (\WalW[3],\LpL[3],\price)
\end{align*}

\noindent
First note that, for every $\tokTi \neq \tokT$, 
it trivially holds that 
$\projTok[\tokTi]{\gain{\pmvA}{\confG}{\TxTS}} = 
\projTok[\tokTi]{\gain{\pmvA}{\confG}{\accrueIntOp}}$
and
$\projTok[\tokTi]{\gain{\pmvB}{\confG}{\TxTS}} = 
\projTok[\tokTi]{\gain{\pmvB}{\confG}{\accrueIntOp}}$.
Hence, we only focus on the gains restricted to $\tokT$.

By hypothesis, we have that  $\supplyDebt[{\LpL[]}]{\tokT}>0$, and, by Lemma \ref{lem:cred0-imp-debt0}, also $\supply[{\LpL[]}]{\tokT}>0$. 
Since $\borrowOp$ does not decrease the supply of credits nor that of debit tokens, we also have that $\supply[{\LpL[1]}]{\tokT}>0$ and $\supplyDebt[{\LpL[1]}]{\tokT}>0$.

The gain of $\pmvA$ is given by:
\begin{align*}
	\projTok{\gain{\pmvA}{\confG}{\TxTS}} 
		& =
		\projTok{\gain{\pmvA}{\confG}{\txT[1]}} + 
		\projTok{\gain{\pmvA}{\confG[1]}{\txT[2]}} +
		\projTok{\gain{\pmvA}{\confG[2]}{\txT[3]}} 
		&& \text{by \eqref{eq:gainSt}}
		\\
		& =
		0 + 
		\projTok{\gain{\pmvA}{\confG[1]}{\txT[2]}} + \
		0 
		&& \text{by Lem. \ref{lem:gain-base}}
		\\
		& =
		\left(
		\dfrac
		{\stateCredit[{\LpL[1]}]{\tokT}{\pmvA}{}}
		{\supply[{\LpL[1]}]{\tokT}}
		\cdot 
		{\supplyDebt[{\LpL[1]}]{\tokT}}
		- 
		\stateDebt[{\LpL[1]}]{\tokT}{\pmvA}{}
		\right)
		\cdot 
		\Intr{\LpL[1]}(\tokT)
		\cdot \price[\tokT]	
		&&  \text{by Lem. \ref{lem:gain-int}}
		\\
		& =
		\left(
		\dfrac
		{\stateCredit[{\LpL}]{\tokT}{\pmvA}{}}
		{\supply[{\LpL}]{\tokT}}
		\cdot 
		\left(
		{\supplyDebt[{\LpL}]{\tokT}}
		+\valV
		\right)
		- 
		\left(
		\stateDebt[{\LpL}]{\tokT}{\pmvA}
		+ \valV
		\right)
		\right)
		\cdot 
		\Intr{\LpL[1]}(\tokT)
		\cdot \price[\tokT]	
		&&  \text{by $\nrule{[bor]}$}
		\\
		& =
		\left(
		{\supplyDebt[{\LpL}]{\tokT}}
		+\valV
		- 
		\stateDebt[{\LpL}]{\tokT}{\pmvA}
		- \valV
		\right)
		\cdot 
		\left(
		\alpha \cdot
		\frac
		{\supplyDebt[{\LpL}]{\tokT}+\valV}
		{	\reserves[{\LpL}]{\tokT}
			+
			\supplyDebt[{\LpL}]{\tokT}}
		+\beta		
		\right)
		\cdot \price[\tokT]	
		&&  \text{by hyp.}
		\\
		& >
		\left(
		{\supplyDebt[{\LpL}]{\tokT}}
		- 
		\stateDebt[{\LpL}]{\tokT}{\pmvA}
		\right)
		\cdot 
		\left(
		\alpha \cdot
		\frac
		{\supplyDebt[{\LpL}]{\tokT}}
		{	\reserves[{\LpL}]{\tokT}
			+
			\supplyDebt[{\LpL}]{\tokT}}
		+\beta		
		\right)
		\cdot \price[\tokT]	
		&&  \text{by arith.}
		\\	
		& =
		\left(
		{\supplyDebt[{\LpL}]{\tokT}}
		- 
		\stateDebt[{\LpL}]{\tokT}{\pmvA}
		\right)
		\cdot 
            \Intr{\LpL}(\tokT)
		\cdot \price[\tokT]	
		&&  \text{by~\eqref{eq:intr-util}}
            \\
		& =	\projTok{\gain{\pmvA}{\confG}{\accrueIntOp}}
		&&  \text{Lem. \ref{lem:gain-int} + hyp. 
		}
\end{align*}
	
Note that the hypothesis that $
{\supplyDebt[{\LpL}]{\tokT}}
>
\stateDebt[{\LpL}]{\tokT}{\pmvA}$
is necessary to show the strict inequality. 
Otherwise, we would have $
\projTok{\gain{\pmvA}{\confG}{\TxTS}}  = 0 = 
\projTok{\gain{\pmvA}{\confG}{\accrueIntOp}}$.
	
\medskip\noindent
We now compute the gain of $\pmvB$. 
Note that, since by hypothesis  $\stateCredit{\tokT}{\pmvA}=\supply[\LpL]{\tokT}$, then  $\stateCredit{\tokT}{\pmvB}=0$, and, since the borrow fired by $\pmvA$ does not affect the credits of $\pmvB$, then we also have  $\stateCredit[{\LpL[1]}]{\tokT}{\pmvB}=0$.
The gain:
\begin{align*}
\projTok{\gain{\pmvB}{\confG}{\TxTS}} 
& =
\projTok{\gain{\pmvB}{\confG}{\txT[1]}} + 
\projTok{\gain{\pmvB}{\confG[1]}{\txT[2]}} +
\projTok{\gain{\pmvB}{\confG[2]}{\txT[3]}} 
&& \text{by \eqref{eq:gainSt}}
\\
& =
0 + 
\projTok{\gain{\pmvB}{\confG[1]}{\txT[2]}} + \
0 
&& \text{by Lem. \ref{lem:gain-base}}
		\\
		& =
		\left(
		\dfrac
		{\stateCredit[{\LpL[1]}]{\tokT}{\pmvB}{}}
		{\supply[{\LpL[1]}]{\tokT}}
		\cdot 
		{\supplyDebt[{\LpL[1]}]{\tokT}}
		- 
		\stateDebt[{\LpL[1]}]{\tokT}{\pmvB}{}
		\right)
		\cdot 
		\Intr{\LpL[1]}(\tokT)
		\cdot \price[\tokT]	
		&&  \text{by Lem. \ref{lem:gain-int}}
		\\
		& =
		- 
		\stateDebt[{\LpL[1]}]{\tokT}{\pmvB}
		\cdot 
		\left(
		\alpha \cdot
		\frac
		{\supplyDebt[{\LpL[1]}]{\tokT} }
		{	\reserves[{\LpL[1]}]{\tokT}
			+
			\supplyDebt[{\LpL[1]}]{\tokT}
		}
		+\beta		
		\right)
		\cdot \price[\tokT]	
		&&  \text{by hyp.}
		\\
		& =
		-\stateDebt[{\LpL}]{\tokT}{\pmvB}
		\cdot 
		\left(
		\alpha \cdot
		\frac
		{\supplyDebt[{\LpL}]{\tokT} + \valV}
		{	\reserves[{\LpL}]{\tokT}
			+
			\supplyDebt[{\LpL}]{\tokT}
		}
		+\beta		
		\right)
		\cdot \price[\tokT]	
		&&  \text{by $\nrule{[Bor]}$}
		\\
		& <
		-\stateDebt[{\LpL}]{\tokT}{\pmvB}
		\cdot 
		\left(
		\alpha \cdot
		\frac
		{\supplyDebt[{\LpL}]{\tokT} }
		{	\reserves[{\LpL}]{\tokT}
			+
			\supplyDebt[{\LpL}]{\tokT}
		}
		+\beta		
		\right)
		\cdot \price[\tokT]	
		&&  \text{by arith.}
		\\
		& =	\projTok{\gain{\pmvB}{\confG}{\accrueIntOp}}
		&&  \text{by Lem.~\ref{lem:gain-int}}
	\end{align*}
\qed
\end{thmproof}

\end{document}